\documentclass[aps,prd,showkeys,superscriptaddress,twocolumn,nofootinbib,floatfix]{revtex4-1}

\pdfoutput=1

\usepackage{amsmath}
\usepackage{amssymb}
\usepackage{amsthm}
\usepackage{mathrsfs}
\usepackage{graphicx}
\usepackage{epstopdf}
\usepackage{fancyhdr}
\usepackage{array}
\usepackage[all]{xy}
\usepackage{eufrak}
\usepackage{euscript}
\usepackage{enumerate}
\usepackage{slashed}
\usepackage{hyperref}
\usepackage{subfigure} 
\usepackage{epstopdf} 

\hypersetup{pdftex,colorlinks=true,linkcolor=blue,citecolor=blue,menucolor=black,urlcolor=blue,filecolor=blue}

\begin{document}

\title{Dynamical vs. equilibrium properties of the QCD phase transition: a holographic perspective}

\author{Romulo Rougemont}
\email{rrougemont@iip.ufrn.br}
\affiliation{International Institute of Physics, Federal University of Rio Grande do Norte,
Campus Universit\'{a}rio - Lagoa Nova, CEP 59078-970, Natal, Rio Grande do Norte, Brazil}

\author{Renato Critelli}
\email{renato.critelli@usp.br}
\affiliation{Instituto de F\'{i}sica, Universidade de S\~{a}o Paulo, Rua do Mat\~{a}o, 1371, Butant\~{a}, CEP 05508-090, S\~{a}o Paulo, S\~{a}o Paulo, Brazil}

\author{Jacquelyn Noronha-Hostler}
\email{jakinoronhahostler@gmail.com}
\affiliation{Department of Physics, University of Houston, Houston TX 77204, USA}

\author{Jorge Noronha}
\email{noronha@if.usp.br}
\affiliation{Instituto de F\'{i}sica, Universidade de S\~{a}o Paulo, Rua do Mat\~{a}o, 1371, Butant\~{a}, CEP 05508-090, S\~{a}o Paulo, S\~{a}o Paulo, Brazil}

\author{Claudia Ratti}
\email{cratti@uh.edu}
\affiliation{Department of Physics, University of Houston, Houston TX 77204, USA}

\begin{abstract}
We employ an Einstein-Maxwell-Dilaton (EMD) holographic model, which is known to be in good agreement with lattice results for the QCD equation of state with $(2+1)$ flavors and physical quark masses, to investigate the temperature and baryon chemical potential dependence of the susceptibilities, conductivities, and diffusion coefficients associated with baryon, electric, and strangeness conserved charges. We also determine how the bulk and shear viscosities of the plasma vary in the plane of temperature and baryon chemical potential. The diffusion of conserved charges and the hydrodynamic viscosities in a baryon rich quark-gluon plasma are found to be suppressed with respect to the zero net baryon case. The transition temperatures associated with equilibrium and non-equilibrium quantities are determined as a function of the baryon chemical potential for the first time. Because of the crossover nature of the QCD phase transition even at moderately large values of the chemical potential, we find that the transition temperatures associated with different quantities are spread in the interval between $130-200$ MeV and they all decrease with increasing baryon chemical potential.
\end{abstract}



\maketitle
\tableofcontents

\section{Introduction}
\label{sec:intro}

The quark-gluon plasma (QGP) \cite{Gyulassy:2004zy} produced in ultrarelativistic heavy ion collisions \cite{Arsene:2004fa,Adcox:2004mh,Back:2004je,Adams:2005dq,Aad:2013xma} has been the focus of intensive experimental and theoretical efforts in the last years \cite{Heinz:2013th,Shuryak:2014zxa,Lappi:2016gmk}.  Due to the highly explosive nature of heavy ion collisions, not only are the thermodynamical properties in equilibrium relevant but one must also understand the Quark Gluon Plasma's response to perturbations around equilibrium, which is encoded in the behavior of transport coefficients.

By varying the experimental conditions under which heavy ion collisions take place it is possible to probe different aspects of the QGP. For instance, at the LHC the large center of mass energy $\sqrt{s_{\textrm{NN}}}= 2.76\,\textrm{\textemdash}\,5.02$ TeV of the collisions produce a medium with very large temperature ($T$) and small baryon chemical potential ($\mu_B$) such that $\mu_B/T \ll 1$. On the other hand, the Beam Energy Scan (BES) program at RHIC \cite{Aggarwal:2010cw} scans collision energies spanning the interval $\sqrt{s_{\textrm{NN}}}= 7.7\,\textrm{\textemdash}\,200$ GeV, where it is expected that $\mu_B/T\sim 1-3$, and effects due to finite baryon density become relevant. Indeed, one of the main goals of the BES program at RHIC, and of other future projects such as the CBM experiment at FAIR \cite{Staszel:2010zza}, the possible Fixed Target Experiments at RHIC \cite{Meehan:2016qon}, and experiments at the NICA facility \cite{NICA}, is to explore the baryon dense regime of the QGP looking for possible experimental signatures of the long-sought critical end point (CEP) of the QCD phase diagram in the $(T,\mu_B)$ plane \cite{Stephanov:1998dy,Stephanov:1999zu,Rischke:2003mt,Kogut:2004su,Stephanov:2004wx,Stephanov:2007fk,Alford:2007xm}.

On the theoretical side, the main non-perturbative tool available to study QCD thermodynamics in the deconfinement/hadronization crossover region \cite{Aoki:2006we,Borsanyi:2016ksw} at $\mu_B=0$ is lattice QCD. Very recently, it has been shown by lattice QCD simulations \cite{Borsanyi:2016ksw} that the contribution of charm quarks to the QCD equation of state only begins to play a role for temperatures $T\gtrsim 300$ MeV, while bottom quarks only become relevant at much higher temperatures, $T\gtrsim 600$ MeV. Therefore, for the temperatures probed at RHIC, FAIR, and NICA it is  reasonable to consider just the contribution of up, down, and strange quarks to QCD thermodynamics. In such a scenario there are three $U(1)$ global symmetries associated with three conserved charges: baryon charge, electric charge, and strangeness. In equilibrium each one of these conserved charges is associated with a corresponding chemical potential, $\mu_B$, $\mu_Q$, and $\mu_S$, whose gradients in the plasma control the diffusion of these conserved charges from higher density regions towards regions where the density is lower. The chemical potentials associated with each of the three lighter quarks, $\mu_u$, $\mu_d$, and $\mu_s$, are related to the $\mu_B$, $\mu_Q$, and $\mu_S$ chemical potentials as follows\footnote{Note that the coefficient in each $\mu_{B,Q,S}$ in Eq.\ \eqref{eq:murel} is given by the corresponding value of each conserved charge for a given flavor.} (see, for instance, Ref.\ \cite{Bazavov:2017dus}),
\begin{align}
\mu_u&=\frac{1}{3}\mu_B+\frac{2}{3}\mu_Q, \quad \mu_d=\frac{1}{3}\mu_B-\frac{1}{3}\mu_Q,\nonumber\\
\mu_s&=\frac{1}{3}\mu_B-\frac{1}{3}\mu_Q-\mu_S.
\label{eq:murel}
\end{align}

Under the experimental conditions realized in heavy ion collisions, $\mu_B>\mu_S>\mu_Q$ \cite{Bazavov:2017dus,Karsch:2010ck,Bazavov:2012vg,Borsanyi:2013hza,Borsanyi:2014ewa}, one may consider for the sake of simplicity that $\mu_B\neq 0$ while $\mu_S=\mu_Q=0$ (this approximation should be valid as long $\mu_B$ is not very large).  In fact, this was one of the configurations considered in Ref.\ \cite{Bazavov:2017dus} to compute the properties of $(2+1)$-flavor QCD equation of state at $\mathcal{O}\left[(\mu_B/T)^6\right]$ in a lattice setup using a Taylor expansion of the pressure, whose expansion coefficients are conserved charge susceptibilities. This assumption appears to be quite reasonable considering that it was previously shown that a finite $\mu_S$ has only a very minimal effect on the charge susceptibilities and the location of the critical point in a Nambu-Jona-Lasinio (NJL) model \cite{Fan:2016ovc}.

A previous estimate of the QCD equation of state also at $\mathcal{O}\left[(\mu_B/T)^6\right]$ was presented in Ref.\ \cite{Gunther:2016vcp},\footnote{Ref.\ \cite{Gunther:2016vcp} presented results for the equation of state along isentropic trajectories in the $(T,\mu_B)$ plane. Here we are interested in comparing lattice results with holographic predictions at some fixed values of $\mu_B$ and/or $\mu_B/T$, instead of fixed $S/N_B$ (where $S$ is the entropy and $N_B$ is the baryon number) because $S/N_B$ is not conserved in the presence of viscosity. Therefore, the analysis made in Ref.\ \cite{Bazavov:2017dus} is especially suited for the study done here.} while the seminal lattice work regarding the finite baryon density $(2+1)$-flavor QCD equation of state at $\mathcal{O}\left[(\mu_B/T)^2\right]$ was first published in Ref.\ \cite{Borsanyi:2012cr}. Regarding conserved charge susceptibilities, results for the second order susceptibility $\chi_2^B$ for $(2+1)$-flavor QCD in the continuum limit were first presented in Ref.\ \cite{Borsanyi:2011sw} and, more recently, results for $\chi_4^B$ were given in Ref.\ \cite{Bellwied:2015lba} while $\chi_8^B$ was computed in Ref.\ \cite{DElia:2016jqh}. According to Ref.\ \cite{Bazavov:2017dus}, the lattice QCD equation of state at finite $\mu_B$, determined on the lattice via the Taylor series procedure, is under control up to $\mu_B/T\sim 2.2$ with no signatures of a CEP being found in the scanned window corresponding to $T\gtrsim 135$ MeV and $\mu_B\lesssim 600$ MeV.

A key question concerning lattice QCD at finite temperature and density is the determination of the hadronic freeze-out line as a function of $\mu_B$. In recent years a strong connection \cite{Gupta:2011wh,Karsch:2012wm} between charge susceptibilities calculated using Lattice QCD and experimental measurements of the moments of the distribution of net-charge, net-protons \cite{Borsanyi:2014ewa,Bazavov:2017dus}, and net-kaons \cite{Noronha-Hostler:2016rpd} has been explored in order to determine the freeze-out line directly from first principles.  Even though experimental effects \cite{Bzdak:2012ab} such as efficiencies, acceptance cuts, and inability to experimentally measure neutral baryons are expected to play a role, recent works have been exploring new methods to take them into account in effective models \cite{Alba:2014eba,Hippert:2017xoj,Petersen:2015pcy,Jiang:2015hri,Zhou:2017jfk}. 

Despite the enormous progress achieved using ab initio lattice QCD calculations in the last decade, it is not clear if reliable signatures of the QCD CEP in the $(T,\mu_B)$ plane will be found using lattice simulations in the near future. Furthermore, another severe difficulty faced by lattice calculations concerns the computation of real-time, transport observables \cite{Meyer:2011gj}. Relativistic hydrodynamics has been enormously successful in describing experimental observables with a small shear viscosity to entropy density ratio ($\eta/s$) and more recently with a bulk viscosity to entropy density ratio ($\zeta/s$) exhibiting a peak at some characteristic temperature. At finite baryon densities diffusion coefficients of baryon number, electric charge, and strangeness begin to play a role although at this point there have not yet been any model calculations that included a nontrivial temperature dependence of the electric charge and strangeness diffusion coefficients. Even at zero baryon density, most of these transport coefficients have not yet been calculated within the framework of Lattice QCD with physical quark masses in the continuum limit and, yet, they are vital input parameters in dynamical models of the QGP.

Since there are no ab initio QCD calculations of these transport coefficients, it is not known at the moment if there is a clear hierarchy among the inflection points (or some other characteristic value of temperature) of each respective transport coefficient. Furthermore, it is not clear if the minimum of $\eta/s(T)$ should correspond to the peak in $\zeta/s(T)$ or if they occur at different temperatures. In fact, due to the crossover nature of the QCD phase transition it is more likely that each transport coefficient experiences the change in degrees of freedom (i.e., from hadrons to quarks and gluons) at different temperatures. This is the case for several quantities computed in equilibrium at zero chemical potential, such as the inflection point of the second order baryon susceptibility and the peak of the trace anomaly, though it is not known how these transition temperatures change as a function of $\mu_B$.  

By focusing on the tasks of computing finite $\mu_B$ observables and transport coefficients, one finds oneself in a corner of theory space that is currently incredibly challenging for first principles lattice QCD simulations. Consequently, complementary theoretical approaches, such as effective models and the holographic gauge-gravity duality \cite{Maldacena:1997re,Gubser:1998bc,Witten:1998qj,Witten:1998zw} may be useful to give physical insight into such problems.

While initially focused on studying the top-down conformal plasma of $\mathcal{N}=4$ super Yang-Mills (SYM) theory \cite{CasalderreySolana:2011us,Adams:2012th}, which is profoundly different from the real-world non-conformal QGP (see \cite{Rougemont:2016etk} for a  discussion), recently the gauge-gravity duality has been employed to ``engineer"\, bottom-up holographic models that closely mimic the physics of the QGP around the crossover \cite{Gubser:2008ny,Gubser:2008yx,Gubser:2008sz,Noronha:2009ud,DeWolfe:2010he,DeWolfe:2011ts,Ficnar:2011yj,Ficnar:2012yu,Finazzo:2014cna,Rougemont:2015wca,Rougemont:2015ona,Finazzo:2015xwa,Rougemont:2015oea,Finazzo:2016mhm,Critelli:2016cvq}. The main idea of Ref.\ \cite{Gubser:2008ny} (see also \cite{Gursoy:2007cb,Gursoy:2007er,Gursoy:2010fj} for the case of pure glue gauge theories) consists in deforming the strongly coupled quantum field theory, defined at the boundary of a 5-dimensional asymptotic anti-de Sitter (AdS) spacetime, by considering a relevant operator in the gauge theory dual to a massive dilaton-like field in the bulk. The dilaton potential is then engineered in such a way to emulate equilibrium properties of (2+1)-flavor QCD at $\mu_B=0$. According to the holographic dictionary, at finite temperature the spacetime backgrounds that are solutions of Einstein's equations comprise a black hole in the bulk and the free parameters of the gravity action may be dynamically fixed by solving the equations of motion for the bulk fields with the requirement that the holographic equation of state at $\mu_B=0$ matches the corresponding lattice QCD results. 

Furthermore, chemical potentials associated with different globally conserved Abelian charges may be included in the holographic model by adding to the gravity action different Maxwell-like vector fields, which then define an Einstein-Maxwell-Dilaton (EMD) model. The coupling between the Maxwell and the dilaton fields may be then dynamically fixed by matching some holographically computed second order susceptibility to the corresponding lattice QCD result (for baryon, electric charge, and/or strangeness susceptibilities) at $\mu_B=0$ \cite{DeWolfe:2010he}. Recently, an anisotropic version of the holographic EMD model comprising an external magnetic field ($B$) at zero chemical potential has also been studied in \cite{Rougemont:2015oea,Finazzo:2016mhm}. Both the isotropic EMD model at finite $T$ and $\mu_B$ from Refs.\ \cite{Rougemont:2015wca,Rougemont:2015ona,Finazzo:2015xwa} and the anisotropic EMD model at finite $T$ and $B$ from Refs.\ \cite{Rougemont:2015oea,Finazzo:2016mhm,Critelli:2016cvq}, were found to give results that are in good quantitative agreement with a large set of physical observables calculated within Lattice QCD in Refs.\ \cite{Borsanyi:2012cr,Borsanyi:2011sw,Bellwied:2015lba,Bonati:2015bha,Bellwied:2015rza,Aarts:2014nba} and \cite{Borsanyi:2013bia,Bonati:2013vba,Bali:2014kia,Bruckmann:2013oba,Endrodi:2015oba,Bazavov:2016uvm}, respectively.

In the present work, we focus on the calculation of several transport coefficients relevant for the physics of the QGP across the $(T,\mu_B)$ plane. In Section \ref{sec:thermo}, we briefly review the main features of the holographic EMD model used in Ref.\ \cite{Rougemont:2015wca} and its thermodynamic properties. We also present in this section the comparison between the holographic equation of state at finite $\mu_B$ and the recent lattice results from Ref.\ \cite{Bazavov:2017dus} at $\mathcal{O}\left[(\mu_B/T)^6\right]$ with $\mu_B\neq 0$ and $\mu_S=\mu_Q=0$. In Section \ref{sec:transport} we discuss the transport of conserved charges in the hot and baryon rich QGP by computing the susceptibilities, conductivities, and diffusion rates for the baryon (subsection \ref{sec:baryon}), electric charge (subsection \ref{sec:electric}), and strangeness (subsection \ref{sec:strangeness}) sectors. The results in \ref{sec:strangeness} are the first calculations in the literature concerning the transport of strangeness and how it is affected by finite baryon density effects across the crossover phase transition. The results reviewed in subsections \ref{sec:baryon} and \ref{sec:electric} were originally obtained in Refs.\ \cite{Rougemont:2015ona} and \cite{Finazzo:2015xwa}, respectively, and are only included here for completeness. In Section \ref{sec:bulk} we evaluate the temperature and baryon chemical potential dependence of the bulk viscosity in the hot and baryon dense QGP, while a hybrid quasiparticle-holographic estimate providing a temperature dependent profile for the shear viscosity coefficient is presented in Appendix \ref{sec:shear}. In Section \ref{sec:Tcs} we obtain the $\mu_B$ dependence of the transition temperatures extracted from characteristic points of different equilibrium and transport observables calculated in the aforementioned sections. We close the paper in Section \ref{sec:conclusion} with a discussion of the main results derived in this work.

We use in this paper natural units where $\hbar = k_B = c = 1$ and a mostly plus metric signature.

\section{Brief review of the EMD model and its thermodynamics}
\label{sec:thermo}

The holographic EMD model at finite temperature and baryon density used here is discussed in detail in Ref.\ \cite{Rougemont:2015wca} (we invite the interested reader to check this reference for the technical aspects and details about the numerics). The bulk EMD action reads,
\begin{align}
S=\frac{1}{2\kappa_5^2}\int d^5x\sqrt{-g}\left[R-\frac{(\partial_\mu\phi)^2}{2}-V(\phi) -\frac{f_B(\phi)(F_{\mu\nu}^B)^2}{4}\right],
\label{eq:EMDaction}
\end{align}
where $\kappa_5^2\equiv 8\pi G_5$ is the 5-dimensional gravitational constant, $\phi$ is the dilaton field with potential $V(\phi)$, $A_\mu^B$ is the Maxwell field associated with the baryon sector, and $f_B(\phi)$ is the coupling function between $\phi$ and $A_\mu^B$. The charged, spatially isotropic and rotationally invariant black hole Ansatz we employ for the bulk EMD fields is given by
\begin{align}
ds^2&=e^{2A(r)}\left[-h(r)dt^2+d\vec{x}^2\right]+\frac{dr^2}{h(r)},\nonumber\\
\phi&=\phi(r), \quad A^B=A_\mu^B dx^\mu=\Phi(r)dt,
\label{eq:EMDansatz}
\end{align}
with the boundary of the asymptotically AdS$_5$ spacetime placed at $r\to\infty$. Also, the black hole horizon is given by the largest root of $h(r_H)=0$ and the radius of the asymptotically AdS$_5$ background is set to unity for simplicity.

\begin{figure}[htp!]
\center
\subfigure[]{\includegraphics[width=0.8\linewidth]{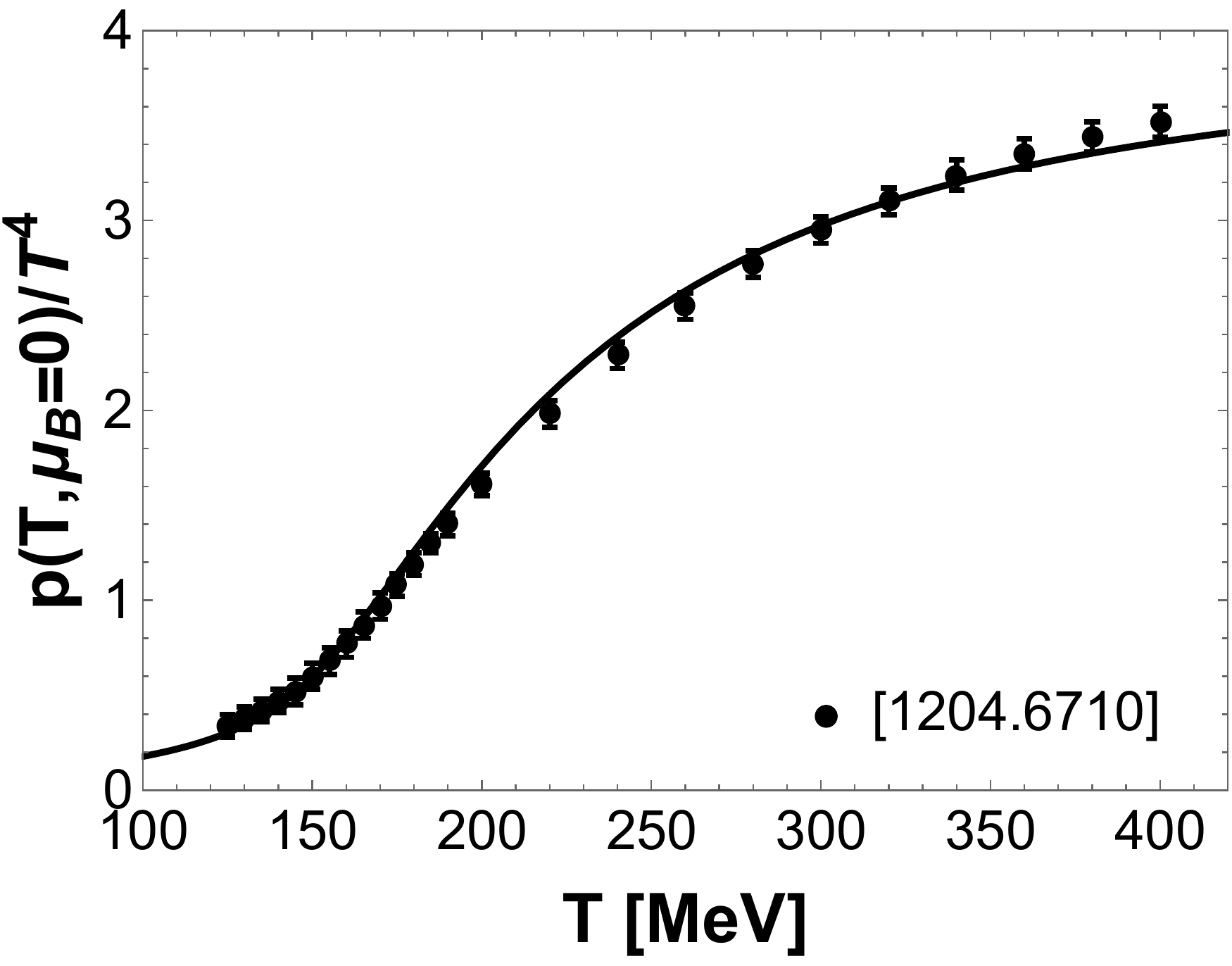}}
\qquad
\subfigure[]{\includegraphics[width=0.8\linewidth]{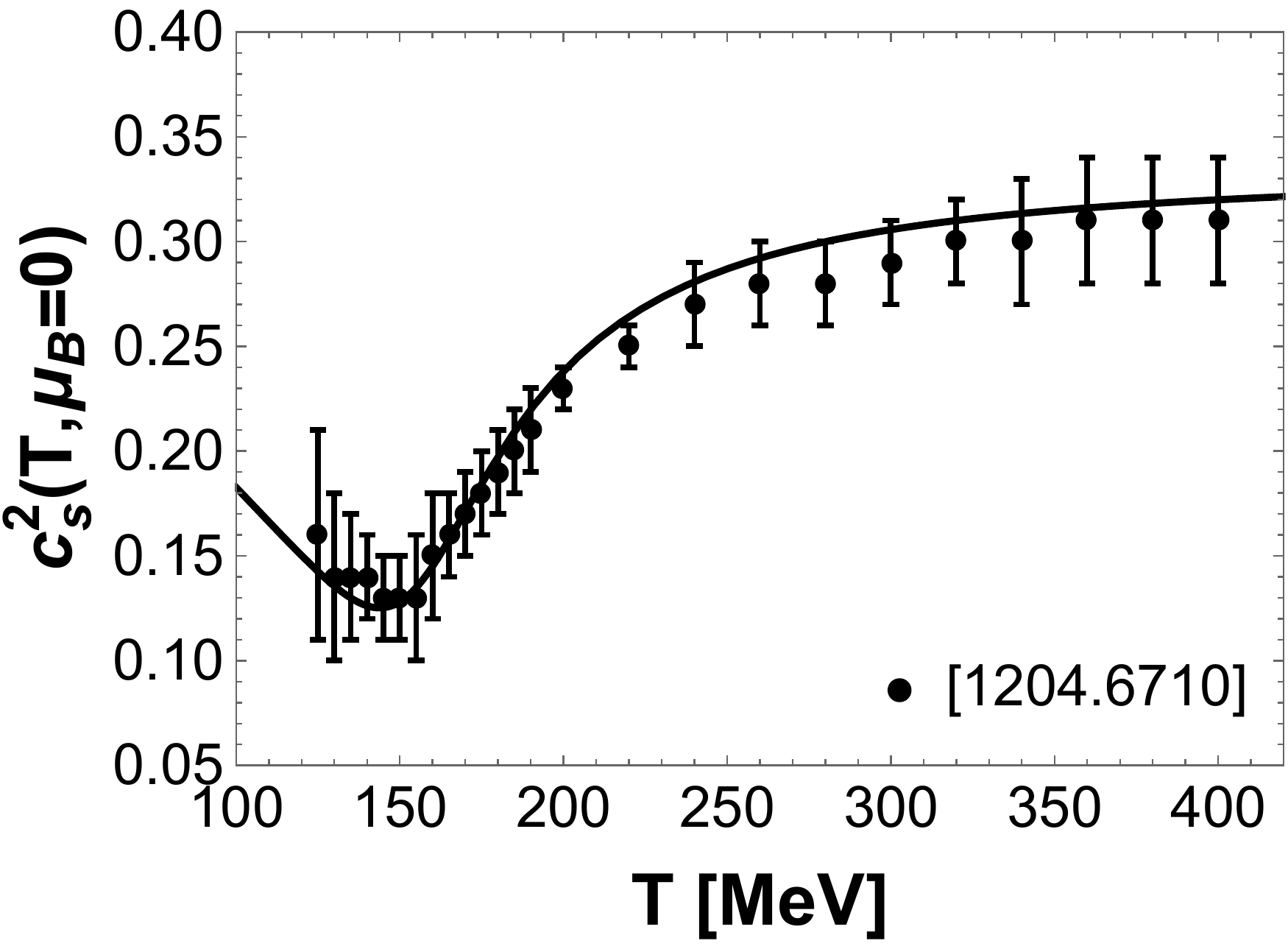}}
\qquad
\subfigure[]{\includegraphics[width=0.8\linewidth]{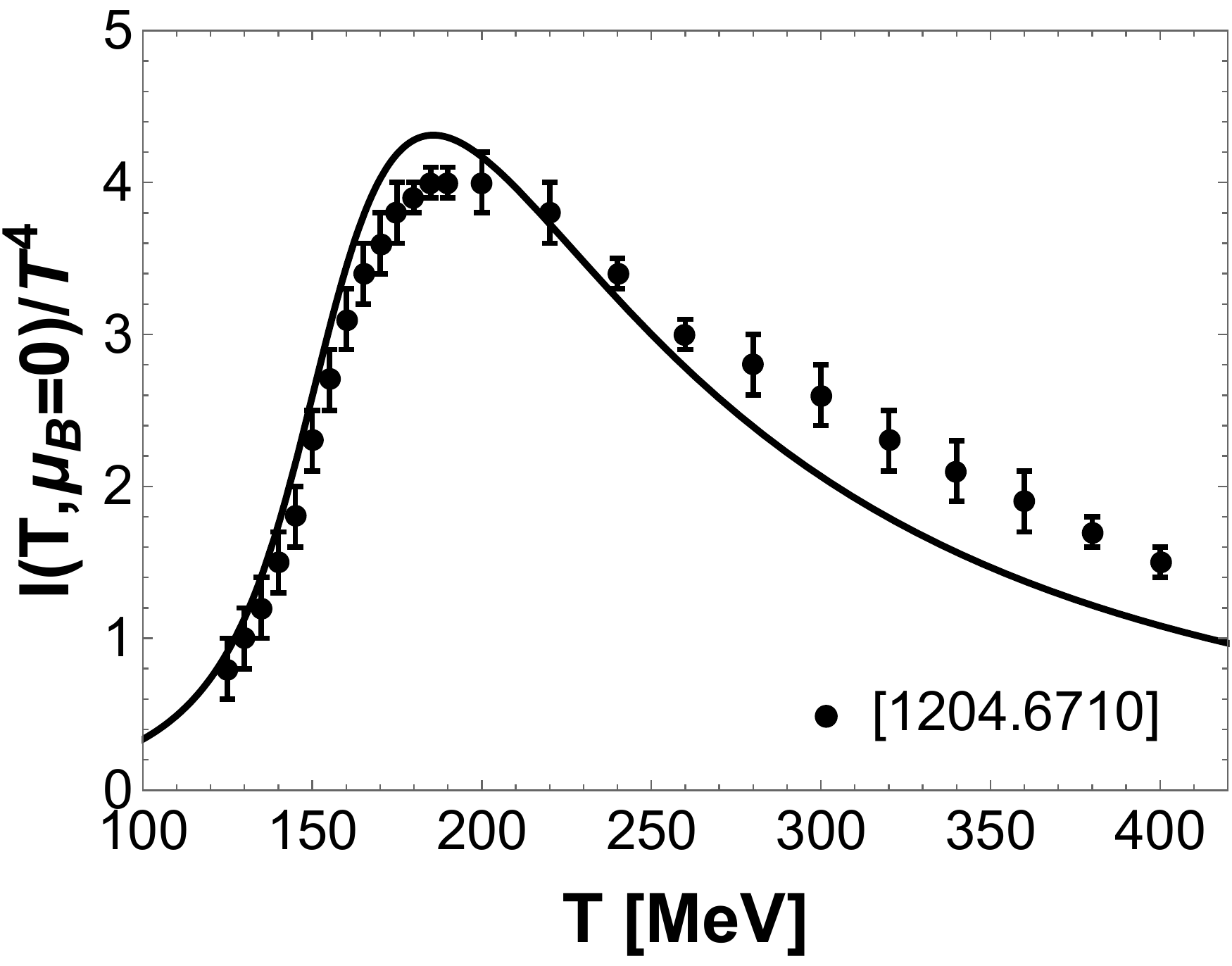}}
\qquad
\subfigure[]{\includegraphics[width=0.8\linewidth]{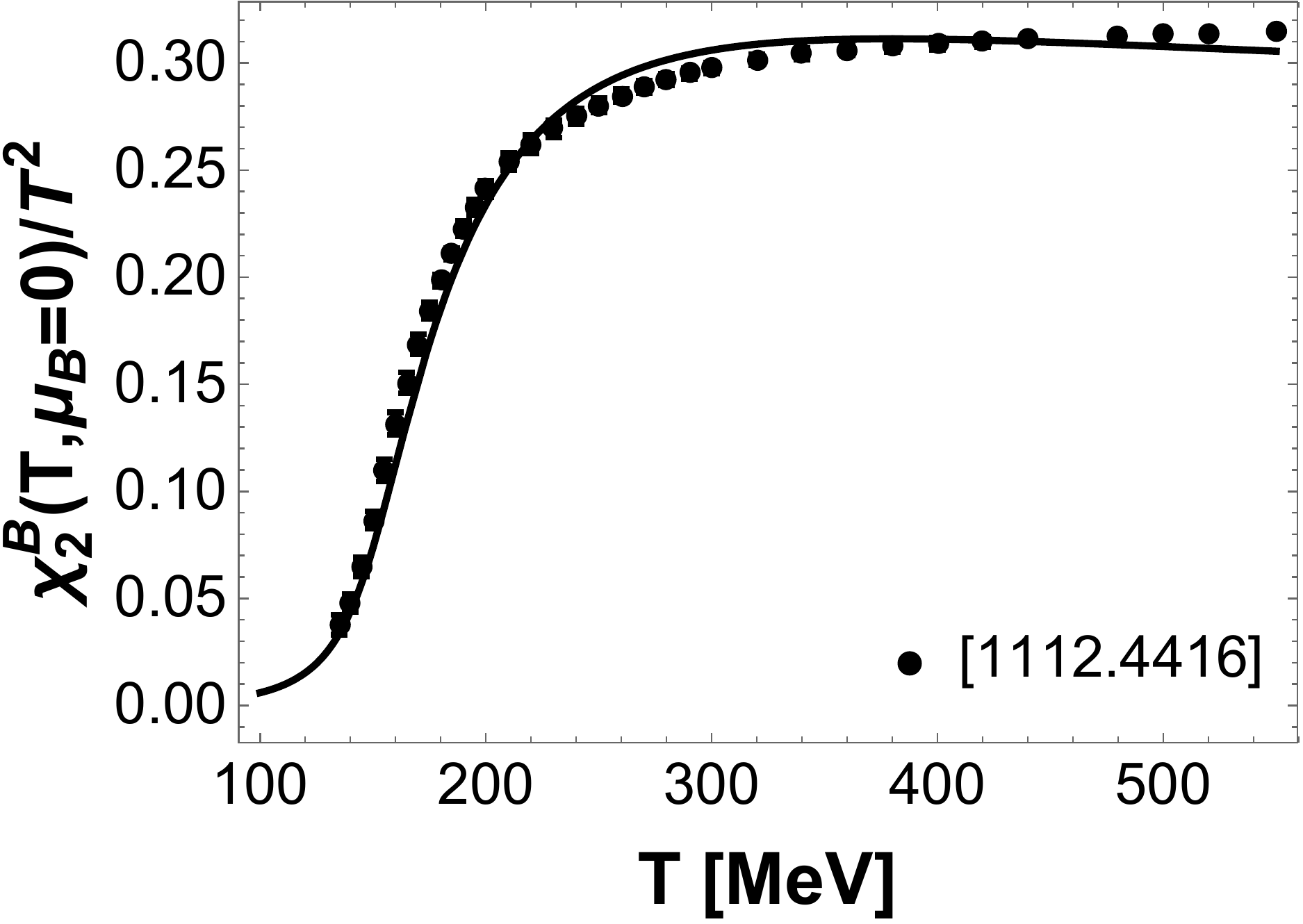}}
\caption{Holographic thermodynamics at $\mu_B=0$ compared to lattice results from Refs.\ \cite{Borsanyi:2011sw,Borsanyi:2012cr}. (a) Pressure. (b) Speed of sound squared. (c) Trace anomaly. (d) Baryon susceptibility.}
\label{fig:thermomuB0}
\end{figure}

Once $V(\phi)$ and $f_B(\phi)$ are specified, one may obtain numerical solutions for the background functions $A(r)$, $h(r)$, $\phi(r)$, and $\Phi(r)$ by following the steps discussed in Ref.\ \cite{Rougemont:2015wca}. Each possible solution of the EMD equations of motion is generated by choosing different values for the pair of initial conditions $(\phi(r_H),\Phi'(r_H))$, specified at the black hole horizon, and then numerically integrating the coupled system of differential equations for the EMD model towards the boundary. From the far from the horizon, near-boundary behavior of these background functions one may extract the needed ultraviolet expansion coefficients used in the holographic calculation of physical observables of the gauge theory at the boundary. Each black hole solution numerically generated corresponds to a different physical state of the gauge theory with definite values of temperature, baryon chemical potential, entropy density ($s$), and baryon charge density ($\rho_B$), obtained according to the following relations,
\begin{align}
T&=\frac{\Lambda}{4\pi\phi_A^{1/\nu}\sqrt{h_0^{\textrm{far}}}}, \quad
\mu_B=\frac{\Phi_0^{\textrm{far}}\,\Lambda}{\phi_A^{1/\nu}\sqrt{h_0^{\textrm{far}}}},\nonumber\\
s&=\frac{2\pi\,\Lambda^3}{\kappa_5^2\,\phi_A^{3/\nu}}, \quad
\rho_B=-\frac{\Phi_2^{\textrm{far}}\,\Lambda^3}{\kappa_5^2\,\phi_A^{3/\nu}\sqrt{h_0^{\textrm{far}}}}.
\label{eq:thermovariables}
\end{align}
In Eq.\ \eqref{eq:thermovariables}, the ultraviolet expansion coefficients $h_0^{\textrm{far}}$, $\Phi_0^{\textrm{far}}$, $\Phi_2^{\textrm{far}}$, and $\phi_A$ are obtained from the near-boundary asymptotics of the EMD fields \cite{DeWolfe:2010he}, $A(r)\approx\alpha(r)$, $h(r)\approx h_0^{\textrm{far}}$, $\phi(r)\approx\phi_A e^{-\nu\alpha(r)}$, and $\Phi(r)\approx\Phi_0^{\textrm{far}}+\Phi_2^{\textrm{far}}e^{-2\alpha(r)}$, where $\alpha(r)=r/\sqrt{h_0^{\textrm{far}}}+A_0^{\textrm{far}}$ and $\nu=4-\Delta$, with $\Delta$ being the scaling dimension of the gauge field theory operator dual to the bulk dilaton field. The quantity $\Lambda$ is a scaling factor with dimensions of energy needed to convert observables computed on the gravity side to field theory units in MeV \cite{Finazzo:2014cna,Rougemont:2015wca}.

As mentioned in the introduction, the free parameters of the holographic model are dynamically fixed by solving the EMD equations of motion at $\mu_B=0$ (corresponding to the initial condition $\Phi'(r_H)=0$) with the constraint that the holographic equation of state and baryon susceptibility match the corresponding lattice results for QCD with $(2+1)$ flavors and physical quark masses from Refs.\ \cite{Borsanyi:2012cr} and \cite{Borsanyi:2011sw}, respectively, which gives \cite{Rougemont:2015wca},
\begin{align}
V(\phi)&=-12\cosh(0.606\phi)+0.703\phi^2-0.1\phi^4+0.0034\phi^6,\nonumber\\
f_B(\phi)&=\frac{\textrm{sech}(1.2\,\phi-0.69)}{3\,\textrm{sech}(0.69)}+\frac{2e^{-100\,\phi}}{3},\nonumber\\
\kappa_5^2&=12.5, \quad \Lambda=831\,\textrm{MeV}.
\label{eq:EMDparameters}
\end{align}
The corresponding results are shown in Fig.\ \ref{fig:thermomuB0}.

\begin{figure}[htp!]
\center
\subfigure[]{\includegraphics[width=0.8\linewidth]{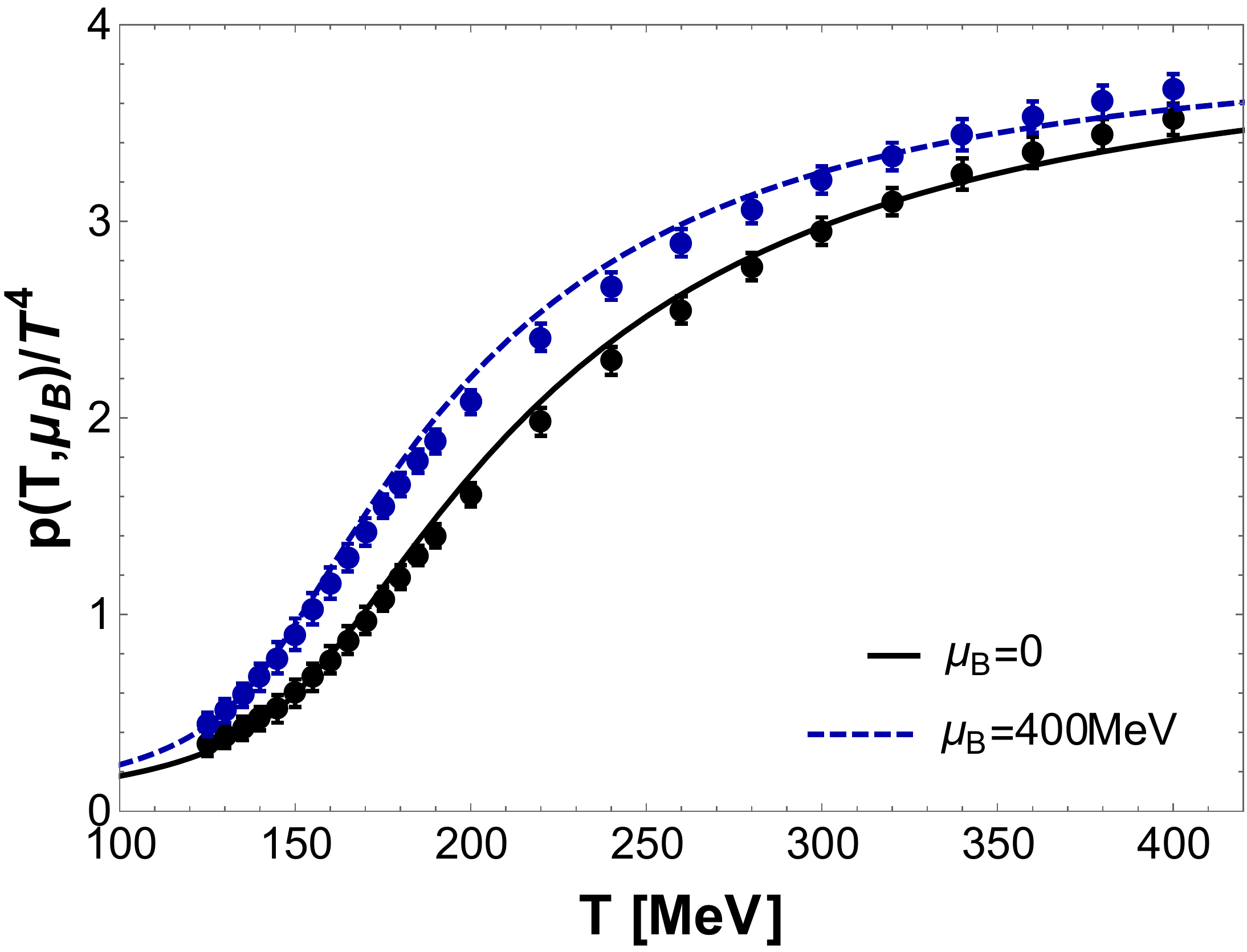}}
\qquad
\subfigure[]{\includegraphics[width=0.8\linewidth]{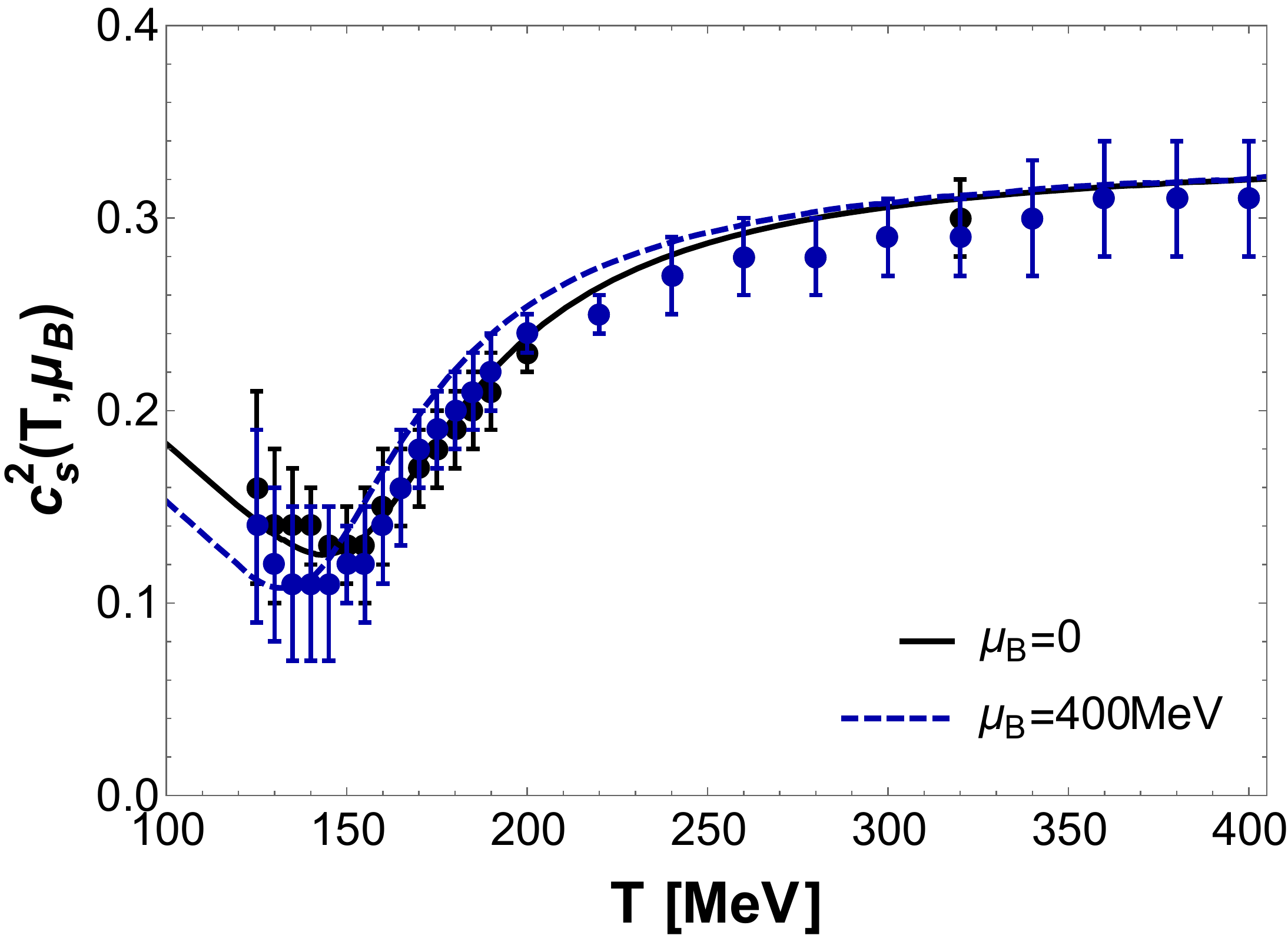}}
\qquad
\subfigure[]{\includegraphics[width=0.8\linewidth]{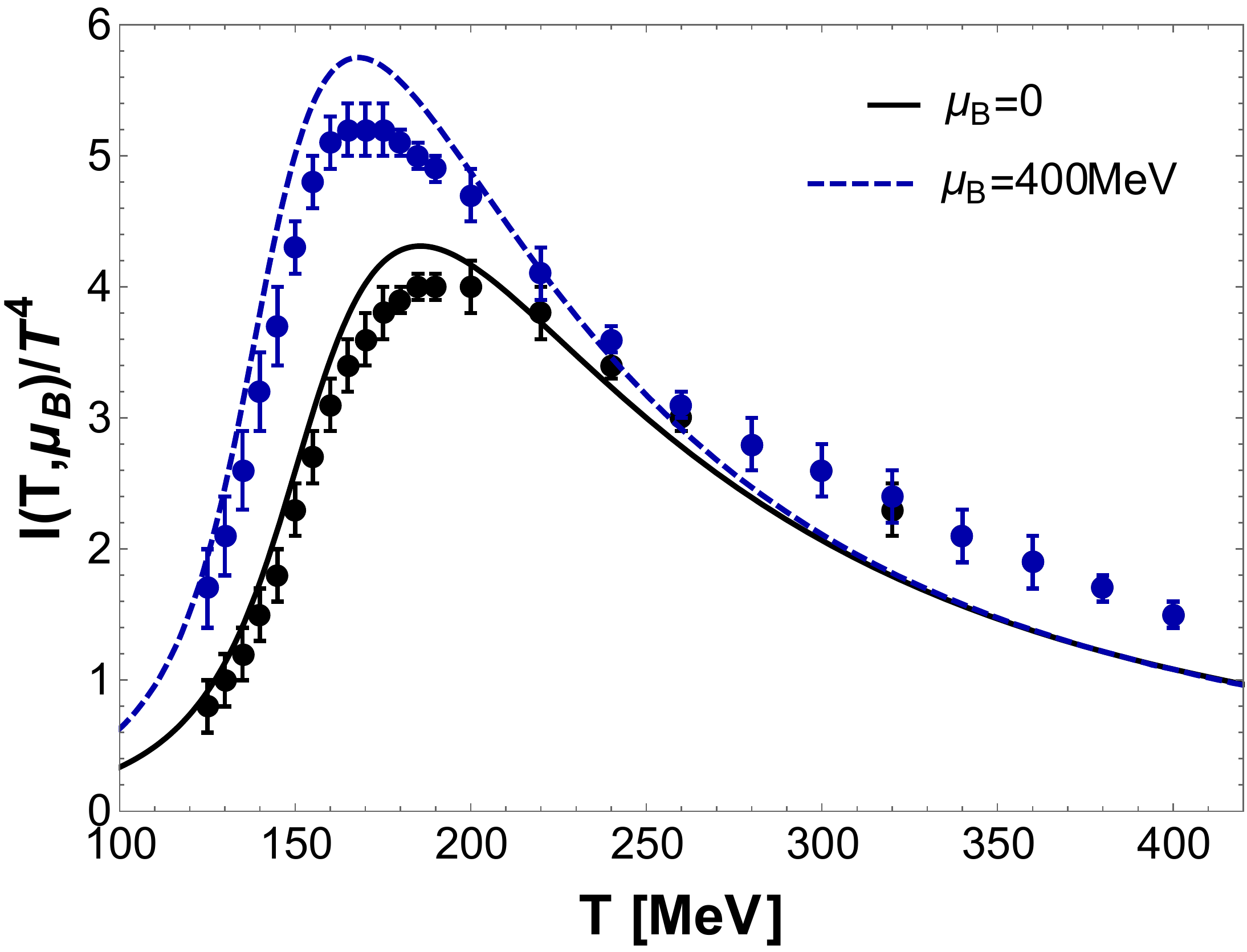}}
\caption{(Color online) Holographic equation of state at finite baryon chemical potential compared to lattice results computed at $\mathcal{O}\left[(\mu_B/T)^2\right]$ from Ref.\ \cite{Borsanyi:2012cr}. (a) Pressure. (b) Speed of sound squared. (c) Trace anomaly.}
\label{fig:thermofinitemuB2012}
\end{figure}

\begin{figure}[htp!]
\center
\subfigure[]{\includegraphics[width=0.8\linewidth]{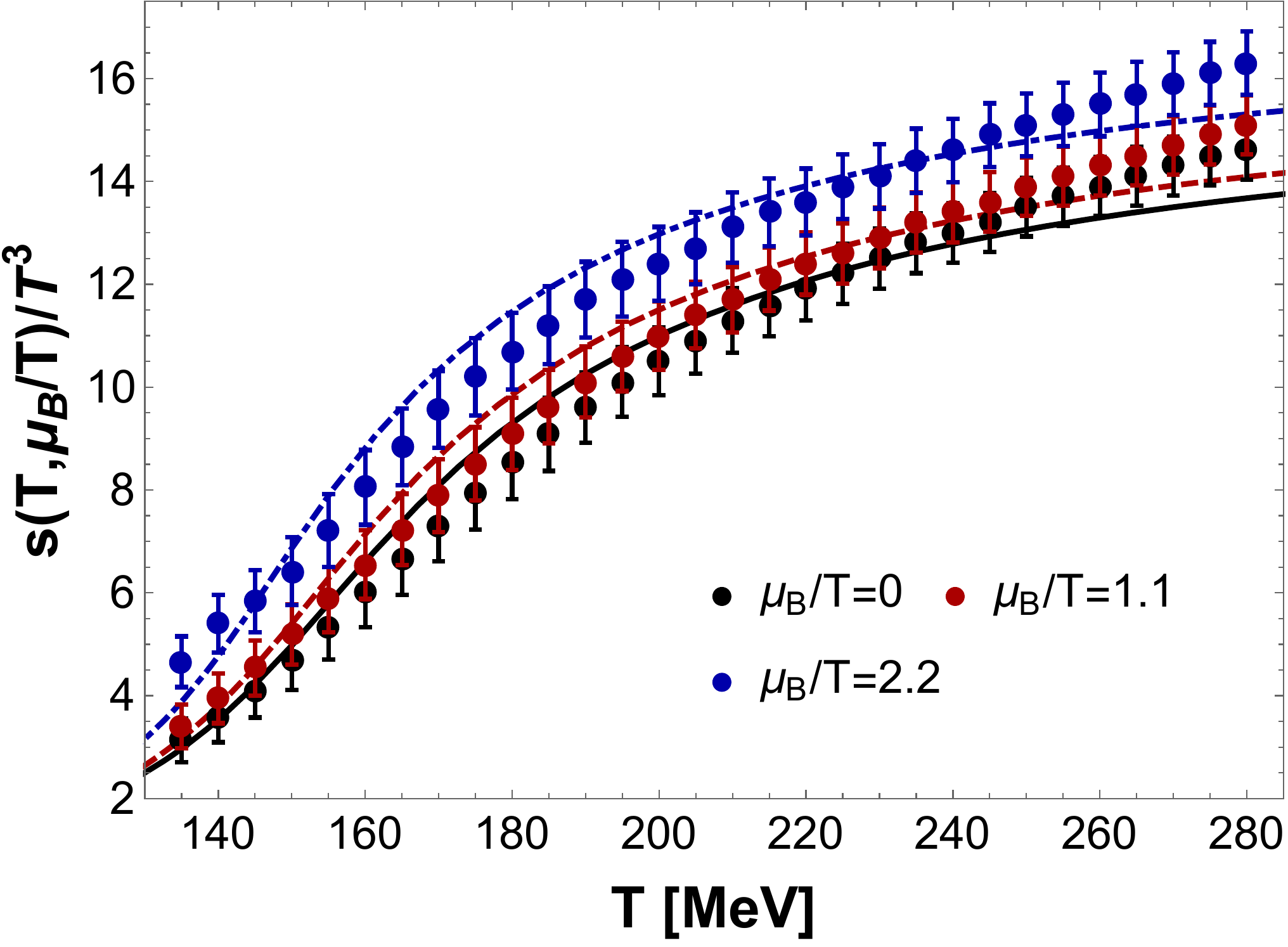}}
\qquad
\subfigure[]{\includegraphics[width=0.8\linewidth]{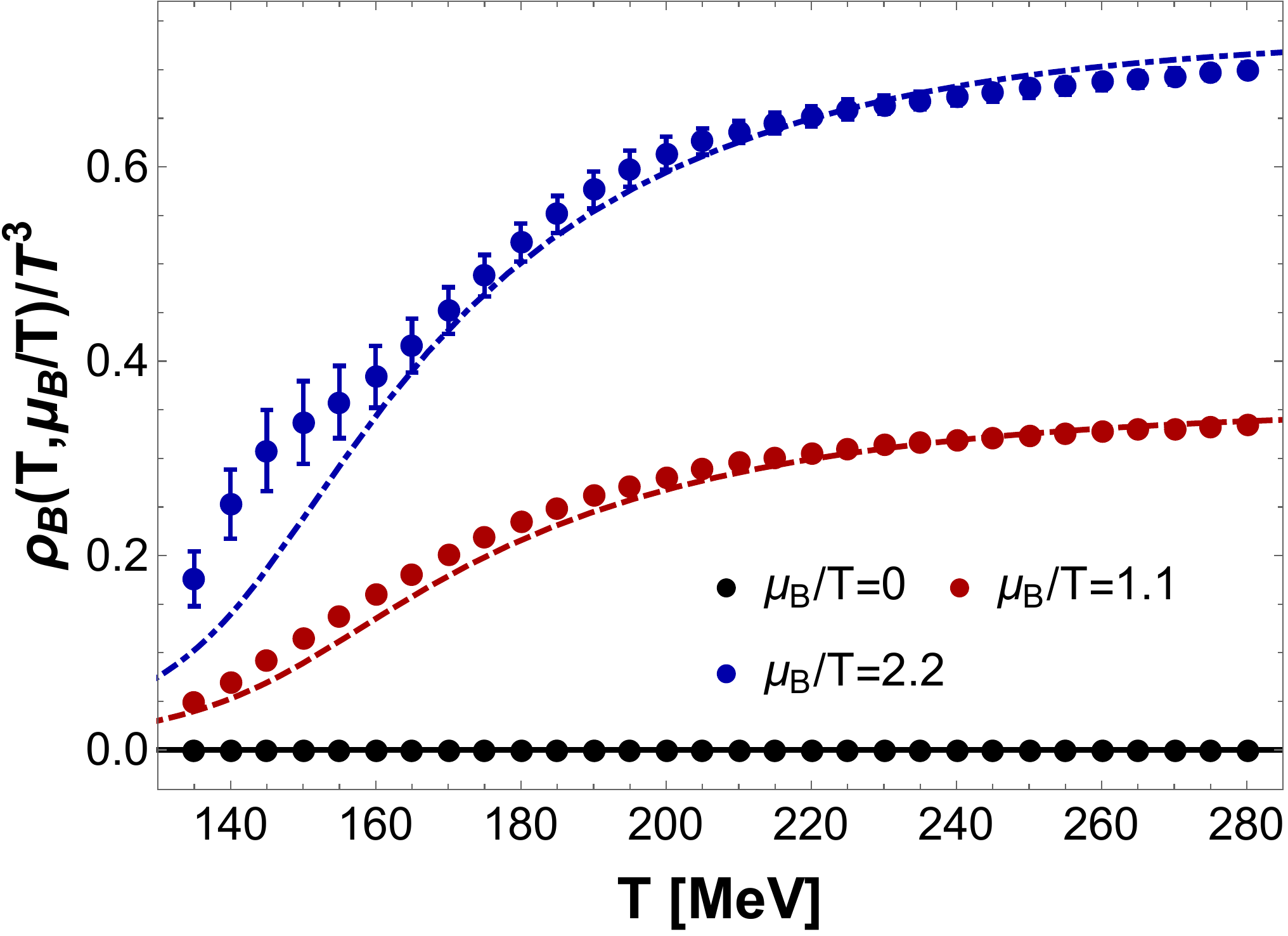}}
\caption{(Color online) Holographic equation of state at finite baryon chemical potential compared to lattice results computed at $\mathcal{O}\left[(\mu_B/T)^6\right]$ from Ref.\ \cite{Bazavov:2017dus}. (a) Entropy density. (b) Baryon charge density.}
\label{fig:thermofinitemuB2017}
\end{figure}

The results displayed in Fig.\ \ref{fig:thermomuB0} are not predictions of the holographic EMD model but rather, as discussed above, they were used to fix the free parameters of this bottom-up construction to generate black hole solutions which are now able to mimic equilibrium properties of the QGP at $\mu_B=0$. Indeed, with the parameters in Eq.\ \eqref{eq:EMDparameters} fixed by using lattice QCD inputs at $\mu_B=0$, one may employ the EMD model to obtain predictions for the physics of the QGP at finite $\mu_B$ and calculate a large set of physical observables across the $(T,\mu_B)$ phase diagram \cite{Rougemont:2015wca,Rougemont:2015ona,Finazzo:2015xwa}. 

In Ref.\ \cite{Rougemont:2015wca} this model's holographic prediction for the equation of state at finite baryon density was compared with lattice QCD results at $\mathcal{O}\left[(\mu_B/T)^2\right]$ from Ref.\ \cite{Borsanyi:2012cr}, and the results are shown in Fig.\ \ref{fig:thermofinitemuB2012}. We also present here for the first time, in Fig.\ \ref{fig:thermofinitemuB2017}, the comparison between the full holographic EMD equation of state at finite $\mu_B$ computed in this model and the very recent lattice QCD results truncated at $\mathcal{O}\left[(\mu_B/T)^6\right]$ from Ref.\ \cite{Bazavov:2017dus}. In both comparisons, one can see an overall reasonable agreement between the EMD and lattice QCD equations of state in the $(T,\mu_B)$ plane. By refining the holographic parameters used in the $\mu_B=0$ calculation one should be able to achieve a much better agreement with lattice results (and fix, for instance, the disagreement between our trace anomaly and the lattice). Such a refined (and time consuming) analysis is, however, beyond the scope of the current paper and it will be presented elsewhere in a separate study.

\section{Transport of conserved charges}
\label{sec:transport}

In this Section we discuss the transport of conserved charges in the hot and baryon rich QGP by computing the susceptibilities, conductivities, and diffusion rates for baryon charge, electric charge, and strangeness.

\subsection{Baryon sector}
\label{sec:baryon}

For the sake of completeness, we briefly review the results originally obtained in Ref.\ \cite{Rougemont:2015ona} for the baryon susceptibility, baryon conductivity, and baryon diffusion in the $(T,\mu_B)$ plane. For the present work we have improved our numerics, which allowed us to access a wider range of values in the $(T,\mu_B)$ plane than previously done in Ref.\ \cite{Rougemont:2015ona}.

\begin{figure}[htp!]
\center
\subfigure[]{\includegraphics[width=0.9\linewidth]{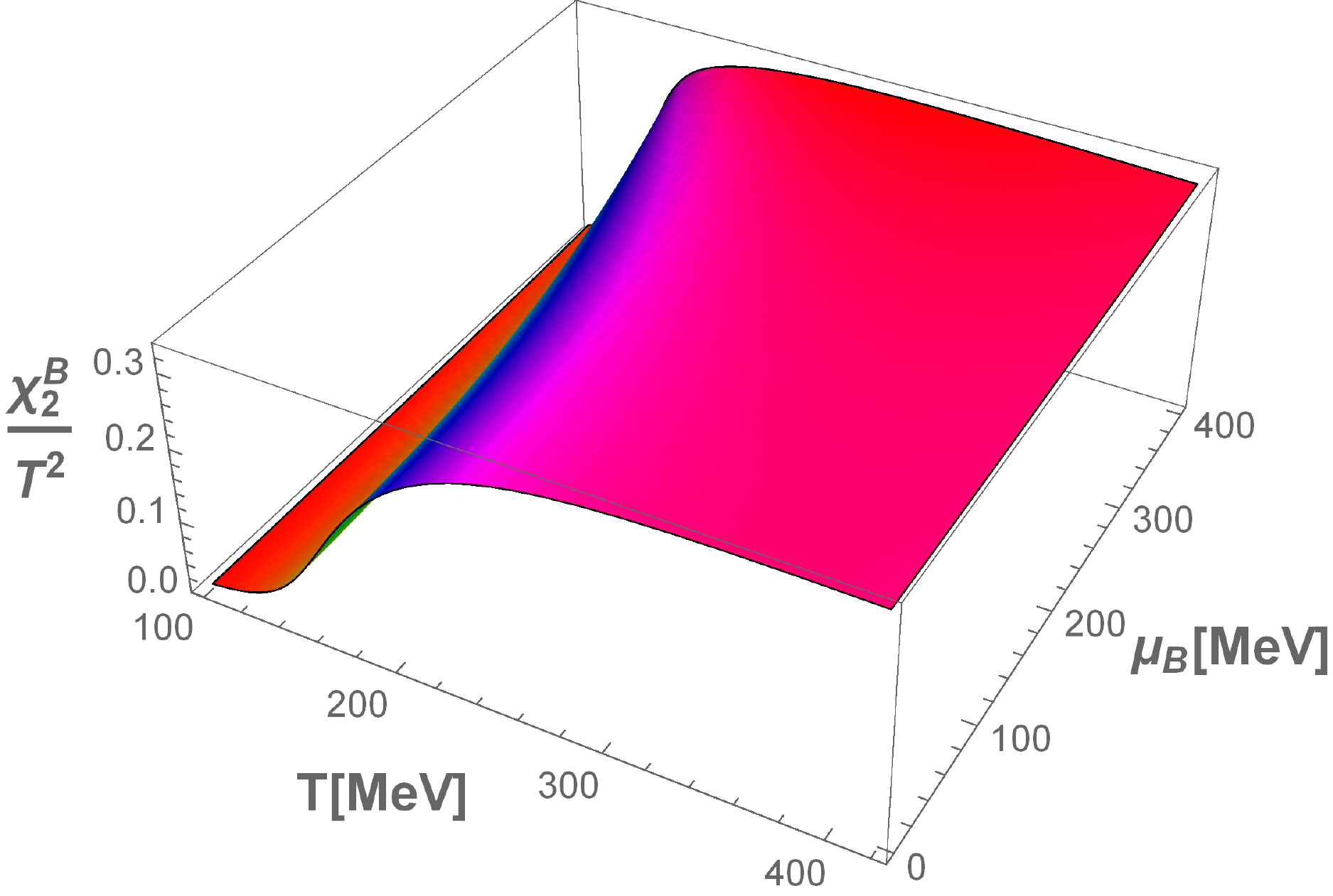}}
\qquad
\subfigure[]{\includegraphics[width=0.8\linewidth]{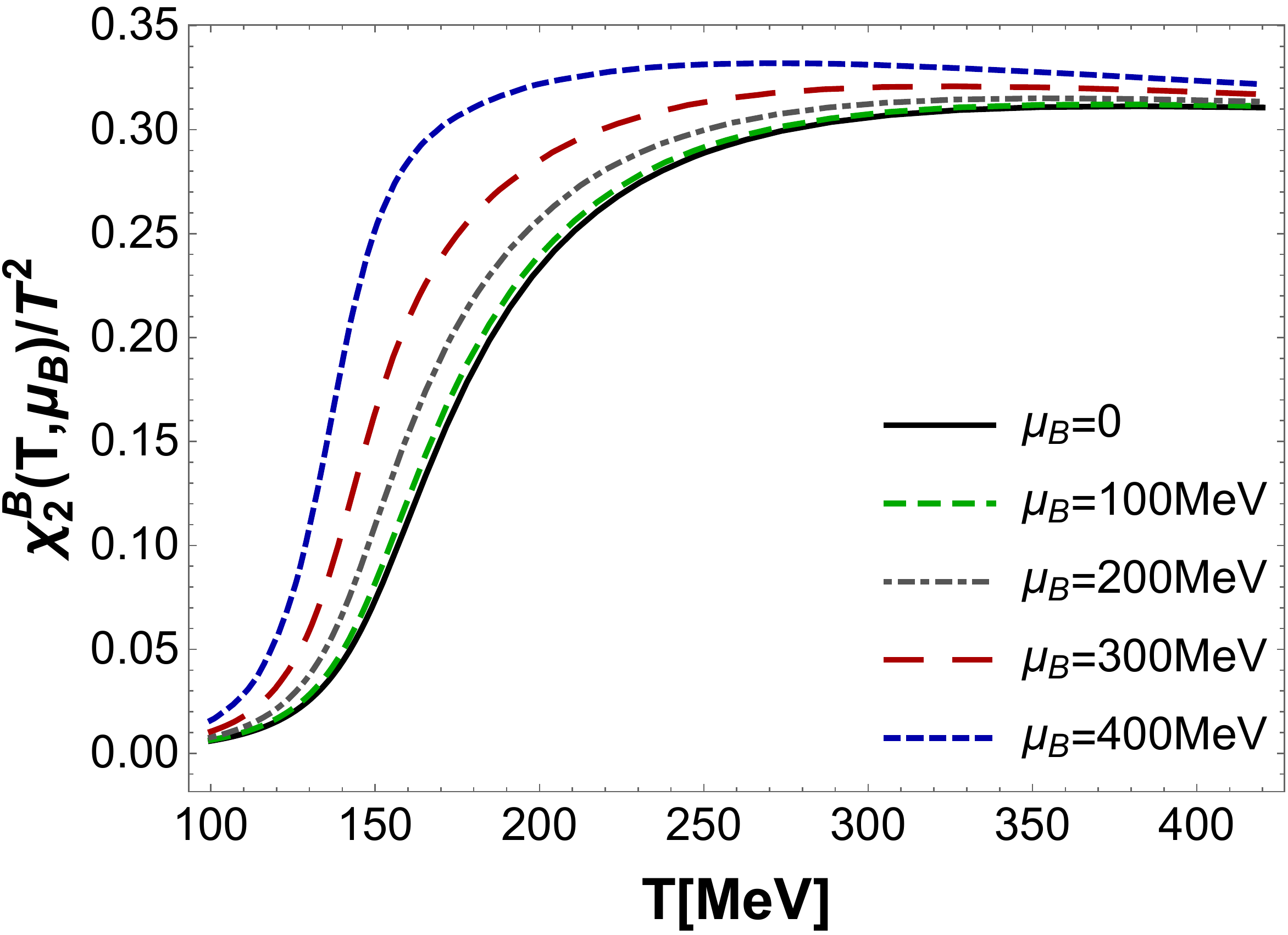}}
\caption{(Color online) Holographic baryon susceptibility. (a) Surface plot as a function of $T$ and $\mu_B$. (b) Curves as functions of $T$ for different values of $\mu_B$.}
\label{fig:chiBfinitemu}
\end{figure}

\begin{figure}[htp!]
\center
\subfigure[]{\includegraphics[width=0.9\linewidth]{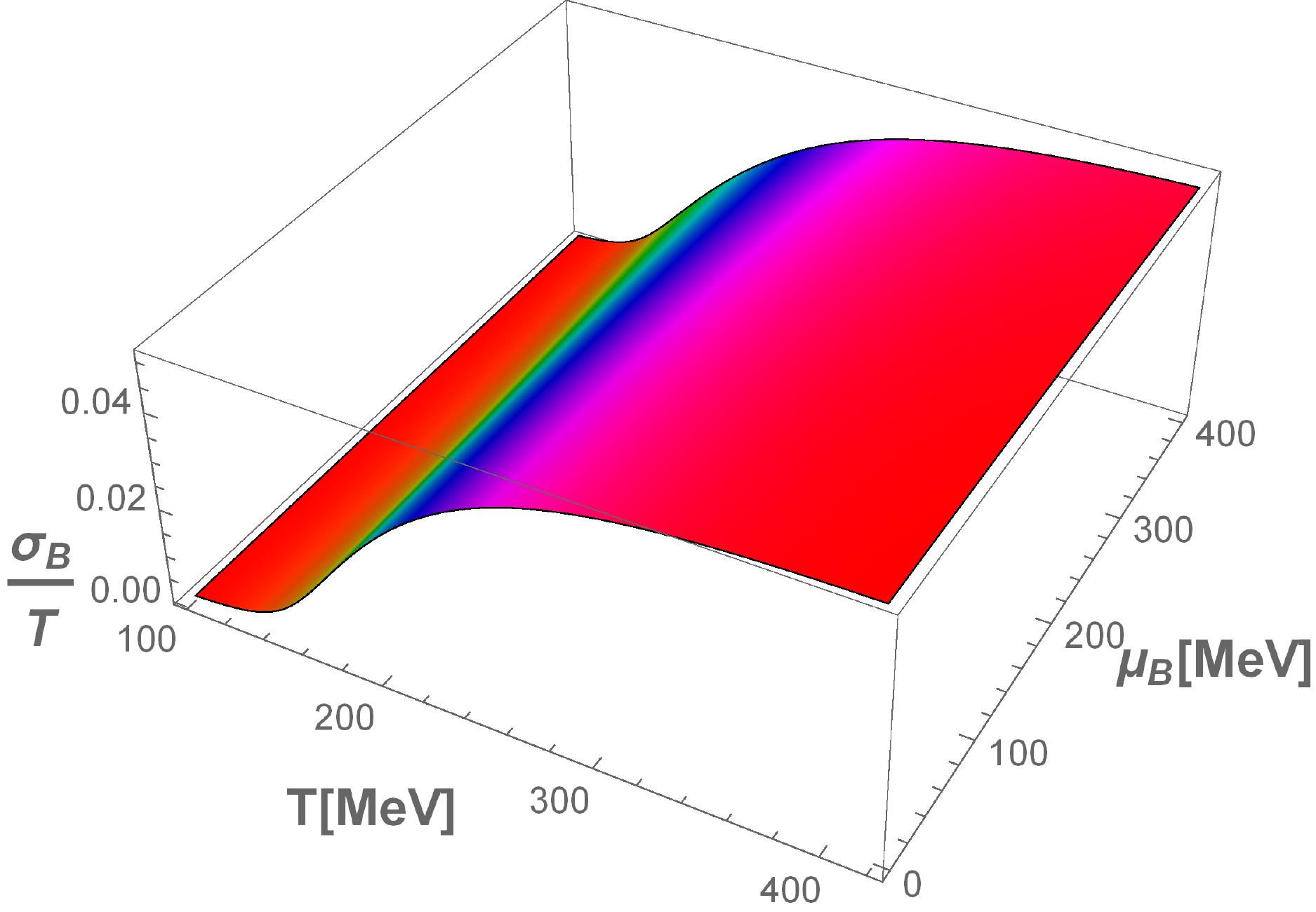}}
\qquad
\subfigure[]{\includegraphics[width=0.8\linewidth]{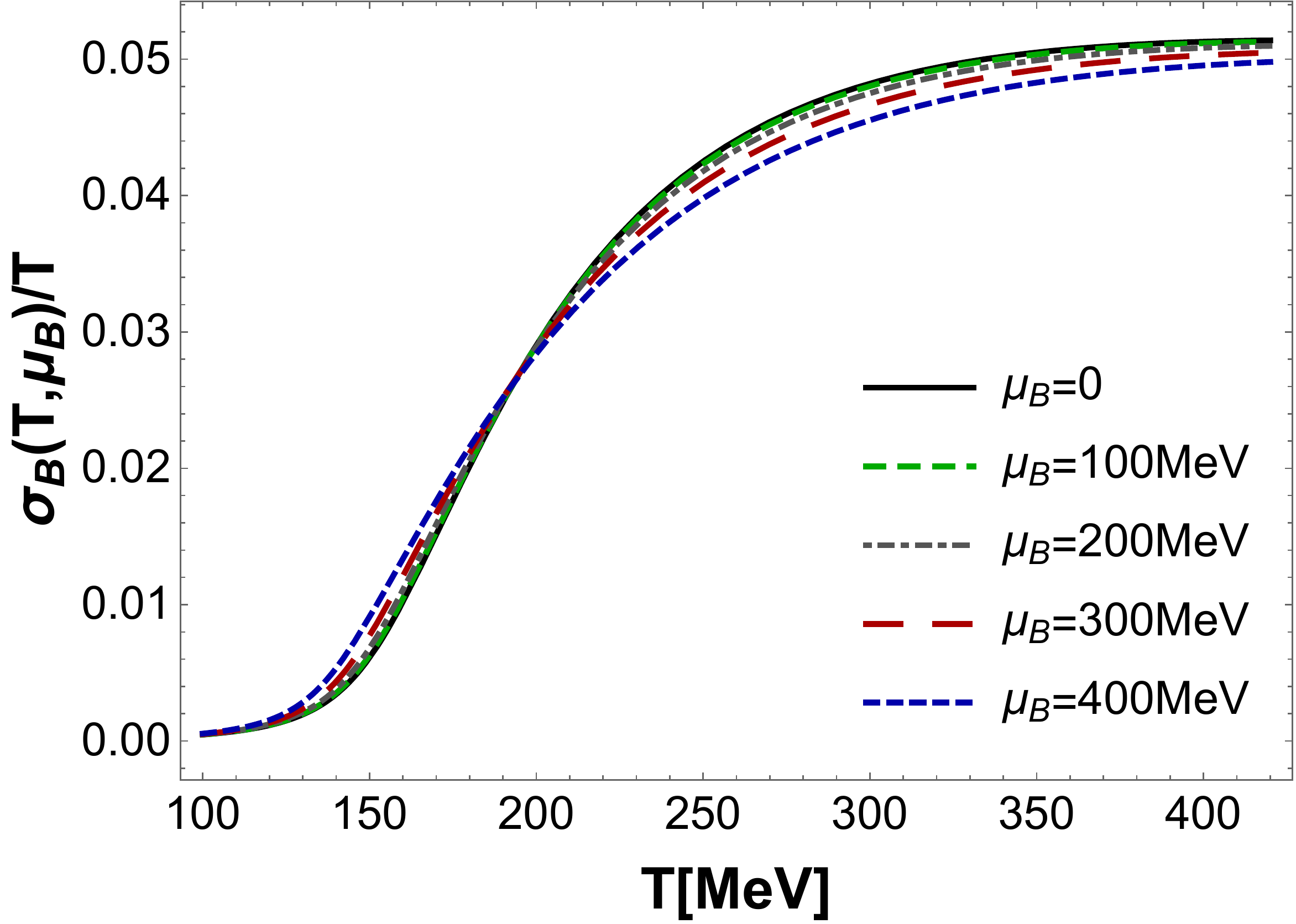}}
\caption{(Color online) Holographic baryon conductivity. (a) Surface plot as a function of $T$ and $\mu_B$. (b) Curves as functions of $T$ for different values of $\mu_B$.}
\label{fig:sigmaBfinitemu}
\end{figure}

\begin{figure}[htp!]
\center
\subfigure[]{\includegraphics[width=0.9\linewidth]{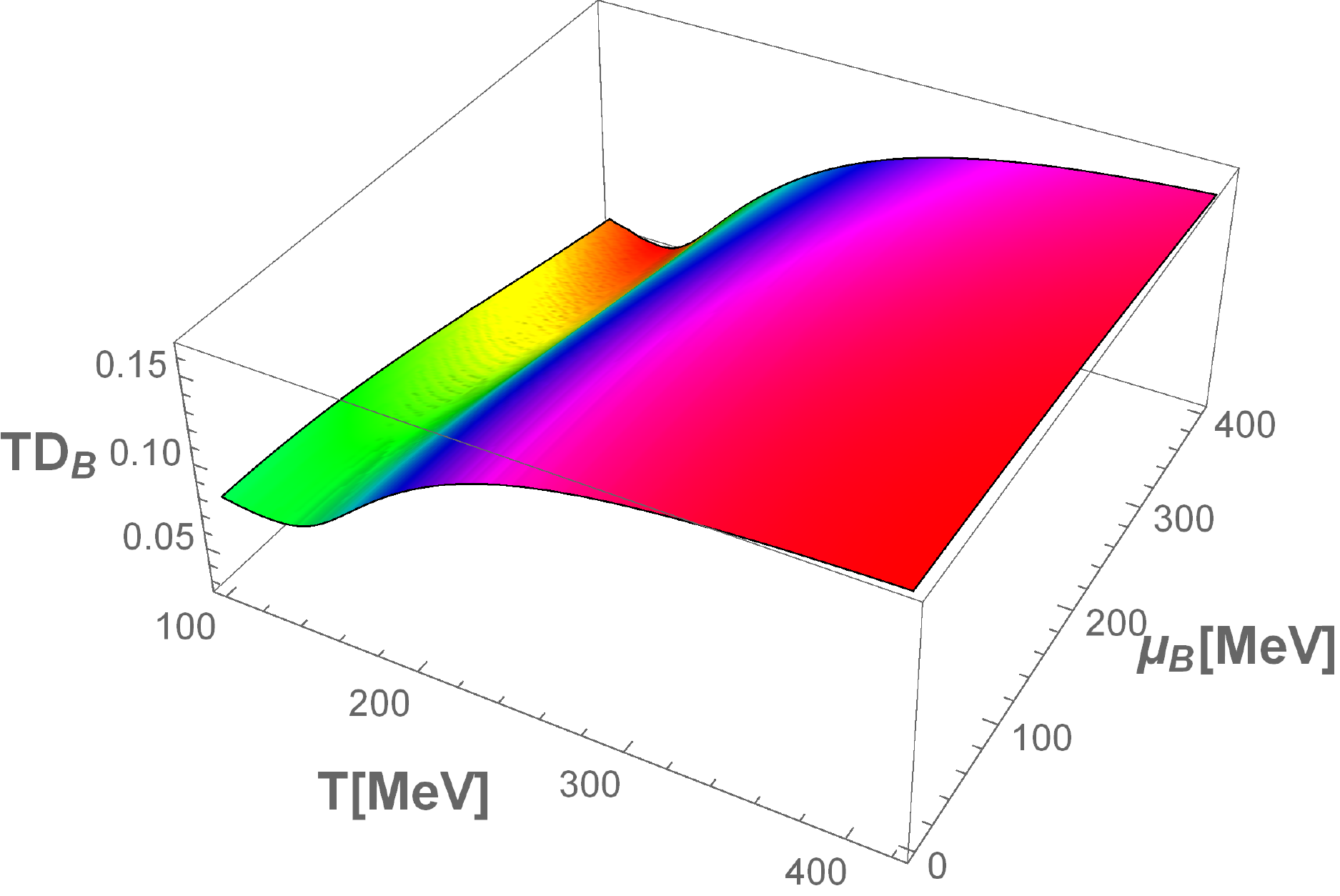}}
\qquad
\subfigure[]{\includegraphics[width=0.8\linewidth]{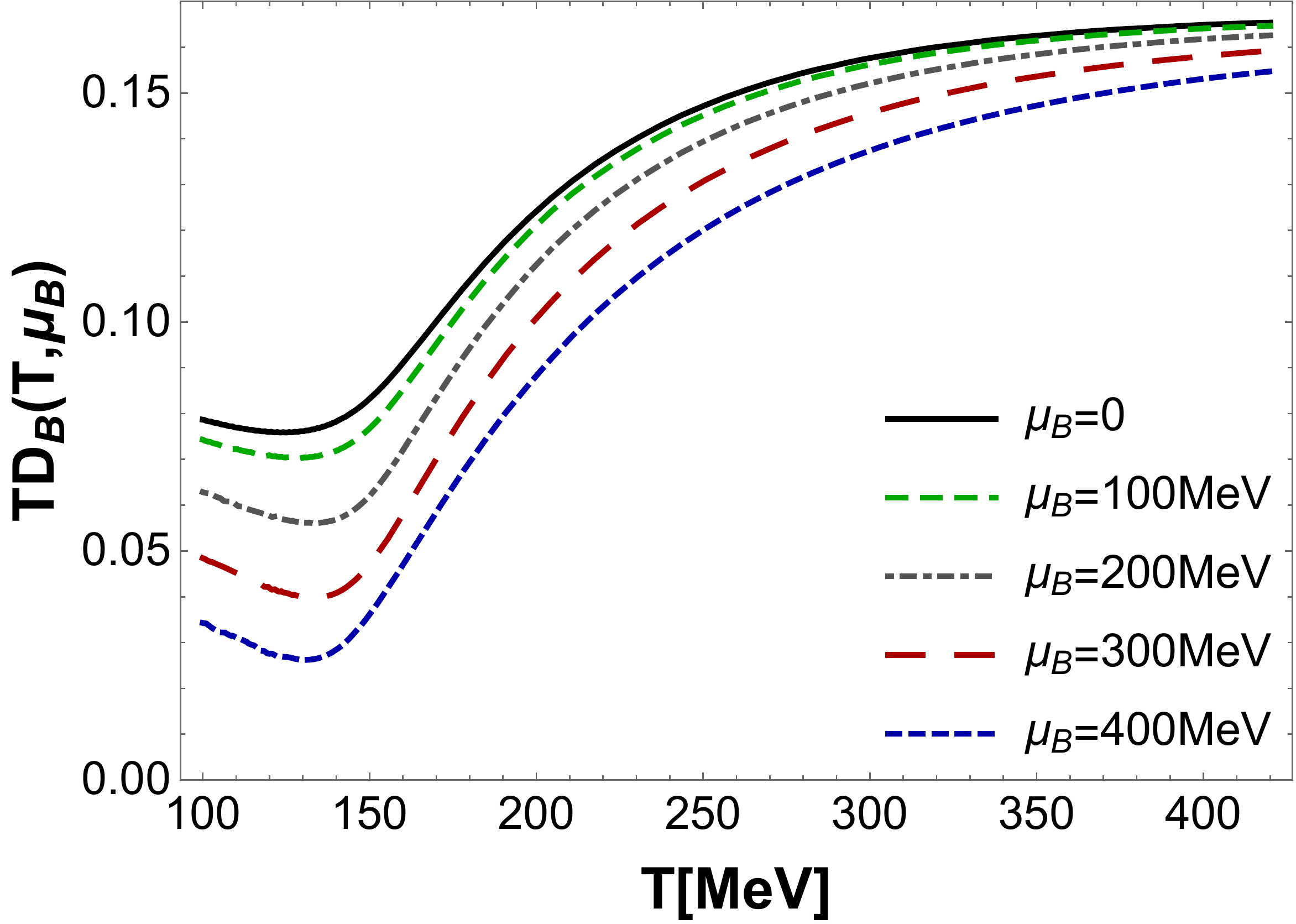}}
\caption{(Color online) Holographic baryon diffusion. (a) Surface plot as a function of $T$ and $\mu_B$. (b) Curves as functions of $T$ for different values of $\mu_B$.}
\label{fig:DBfinitemu}
\end{figure}

The generalized $n$-th order baryon susceptibility is given by,
\begin{align}
\chi_2^B=\frac{\partial^n p}{\partial\mu_B^n} = \frac{\partial^{n-1} \rho_B}{\partial\mu_B^{n-1}},
\label{eq:susc}
\end{align}
and we are interested here in the second order baryon susceptibility, whose results in the $(T,\mu_B)$ plane are displayed in Fig. \ref{fig:chiBfinitemu}.

In order to compute the baryon conductivity we follow the discussion in Ref.\ \cite{DeWolfe:2011ts} and consider linear disturbances of the EMD fields around the backgrounds in Eq.\ \eqref{eq:EMDansatz}. Since these finite temperature and baryon density backgrounds are isotropic and rotationally invariant, by taking a plane wave Ansatz for the EMD field perturbations with frequency $\omega$ and zero spatial momentum, $\vec{k}=\vec{0}$, the resulting EMD disturbances may be combined to construct gauge and diffeomorphism invariant variables which fall into different representations of the $SO(3)$ symmetry group of spatial rotations. There is a singlet, from which one may obtain the bulk viscosity ($\zeta$) of the dual plasma (as we are going to do in subsection \ref{sec:bulk}), a triplet, from which one can compute the baryon conductivity (as we do below), and a quintuplet, from which one obtains the shear viscosity ($\eta$).\footnote{In the case of the EMD construction discussed here, the holographic shear viscosity calculated in the traceless graviton channel, associated with the $SO(3)$ quintuplet, is given by the usual result $\eta/s=1/4\pi$. However, in Appendix \ref{sec:shear} we are going to use a different route to compute a ``phenomenological holographic'' shear viscosity, which provides in a simple way a temperature dependent profile for $\eta/s$ similar to what is expected to occur in the real-world QGP.} Since these gauge and diffeomorphism invariant combinations fall within different representations of $SO(3)$, they cannot mix at the linear level and each one of these perturbations will satisfy a decoupled equation of motion.\footnote{At finite spatial momentum, however, this would no longer be valid since in this case the perturbations would be only classified in terms of an $SO(2)$ symmetry group.}

As discussed in \cite{DeWolfe:2011ts}, due to $SO(3)$ symmetry, each spatial component of the Maxwell field perturbation, $\vec{a}$, satisfies the same decoupled differential equation and, therefore, we may take without loss of generality, $a\equiv a_x$, such that the corresponding equation of motion reads \cite{DeWolfe:2011ts},
\begin{align}
a''(r)&+\left[2A'(r)+\frac{h'(r)}{h(r)}+\frac{f_B'(\phi)}{f_B(\phi)}\phi'(r)\right]a'(r)\nonumber\\
&+\frac{e^{-2A(r)}}{h(r)}\left[\frac{\omega^2}{h(r)}-f_B(\phi)\Phi'(r)^2 \right]a(r)=0.
\label{eq:sigmaBeom}
\end{align}
One may then numerically solve Eq.\ \eqref{eq:sigmaBeom} over the EMD backgrounds with in-falling wave condition at the black hole horizon, normalizing the vector perturbation to unity at the boundary, and plug in the result into the following holographic Kubo formula for the baryon conductivity in the EMD model expressed in physical units \cite{DeWolfe:2011ts,Rougemont:2015ona} (discarding the usual delta function that appears in translationally invariant systems at finite density \cite{Hartnoll:2008kx}),
\begin{align}
\sigma_B= -\frac{\Lambda}{2\kappa_5^2\phi_A^{1/\nu}}\lim_{\omega\rightarrow 0}\frac{h(r)f_B(\phi)e^{2A(r)}\textrm{Im}\left[a^*(r,\omega) a'(r,\omega) \right]}{\omega},
\label{eq:sigmaB}
\end{align}
where $h f_B(\phi)e^{2A}\textrm{Im}\left[a^*a'\right]$ is a conserved flux in the radial direction, such that \eqref{eq:sigmaB} may be evaluated at any value of the holographic coordinate $r$. The results for the baryon conductivity as a function of $(T,\mu_B)$ are shown in Fig.\ \ref{fig:sigmaBfinitemu}.

Finally, we now consider the evaluation of the baryon diffusion, $D_B$, which controls the fluid response to inhomogeneities in the baryon charge density \cite{Kapusta:2014dja}. As discussed in Ref.\ \cite{Iqbal:2008by}, for black brane backgrounds such as ours the baryon diffusion may be calculated using Nernst-Einstein's relation,
\begin{align}
D_B=\frac{\sigma_B}{\chi_2^B}.
\label{eq:DB}
\end{align}
The holographic results for the baryon diffusion are shown in Fig.\ \ref{fig:DBfinitemu}, from which one can see that the diffusion of baryon charge is suppressed by the presence of a nontrivial baryon chemical potential. This is consistent with the presence of a CEP in the holographic model at higher baryon densities, since the baryon diffusion is expected to vanish at the CEP \cite{Son:2004iv}.

\subsection{Electric charge sector}
\label{sec:electric}

For the sake of completeness, we briefly review the results originally obtained in Ref.\ \cite{Finazzo:2015xwa} for the electric charge susceptibility, electric conductivity, and electric charge diffusion in the $(T,\mu_B)$ plane.

\begin{figure}[htp!]
\center
\subfigure[]{\includegraphics[width=0.8\linewidth]{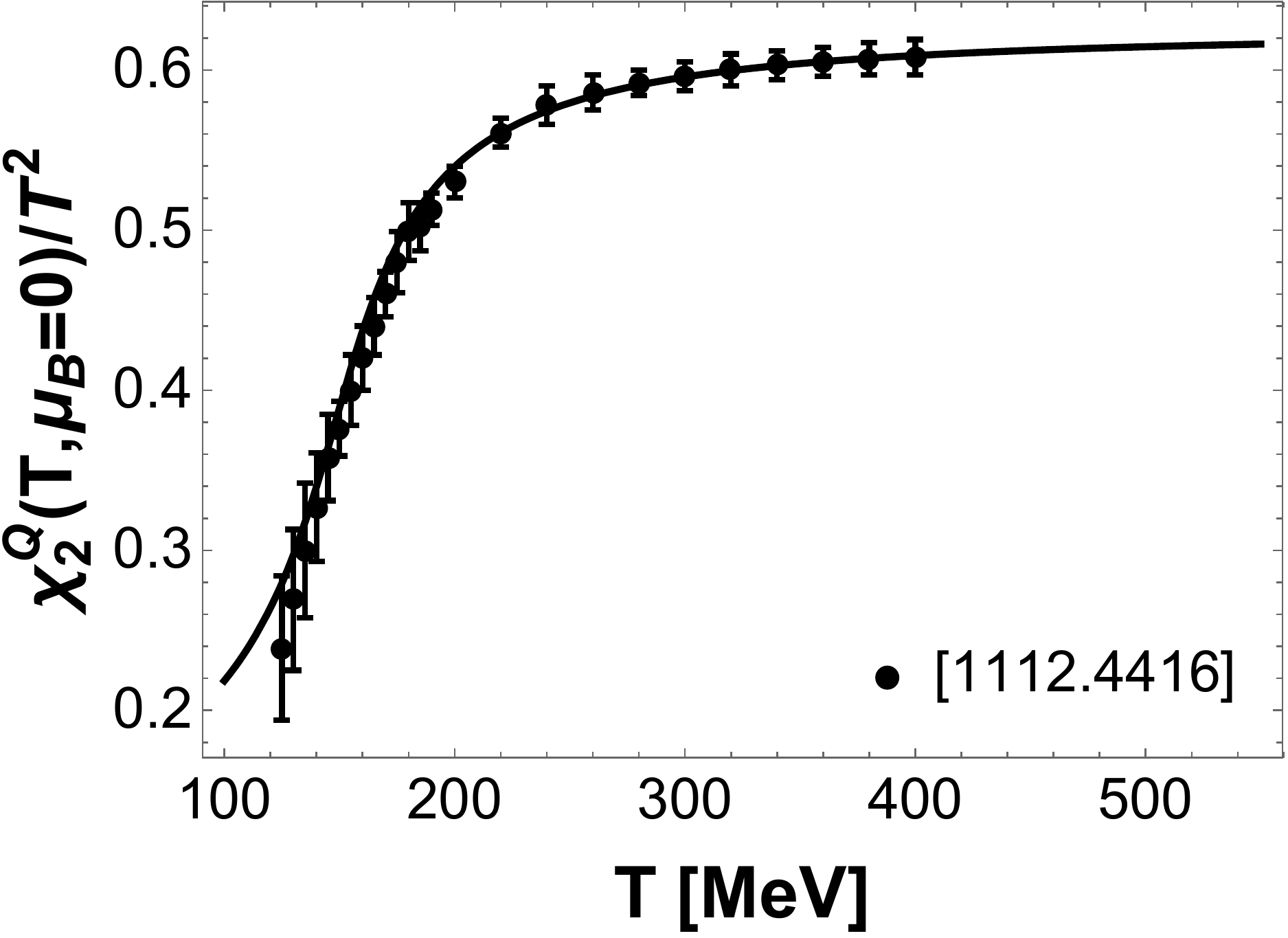}}
\qquad
\subfigure[]{\includegraphics[width=0.8\linewidth]{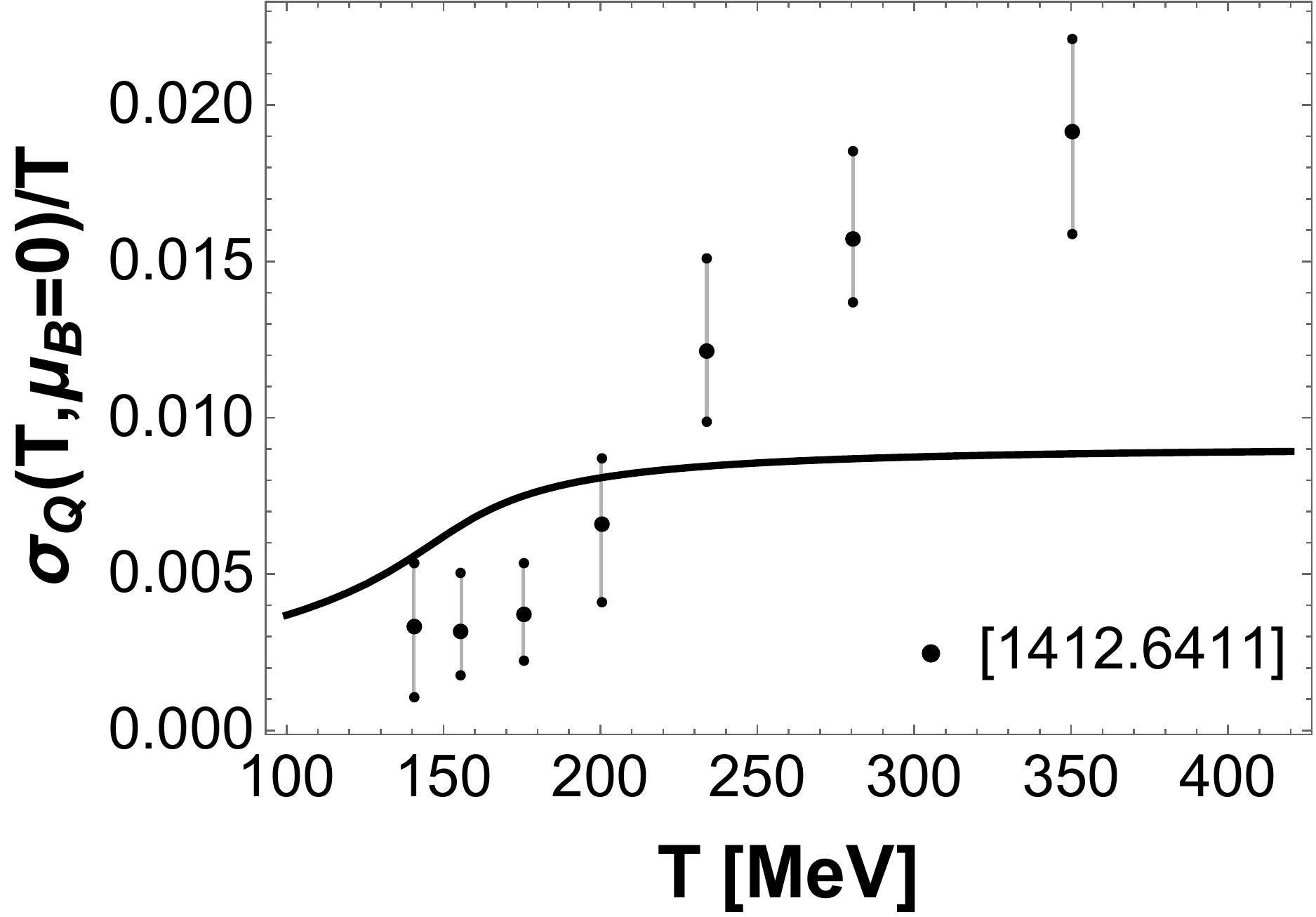}}
\qquad
\subfigure[]{\includegraphics[width=0.8\linewidth]{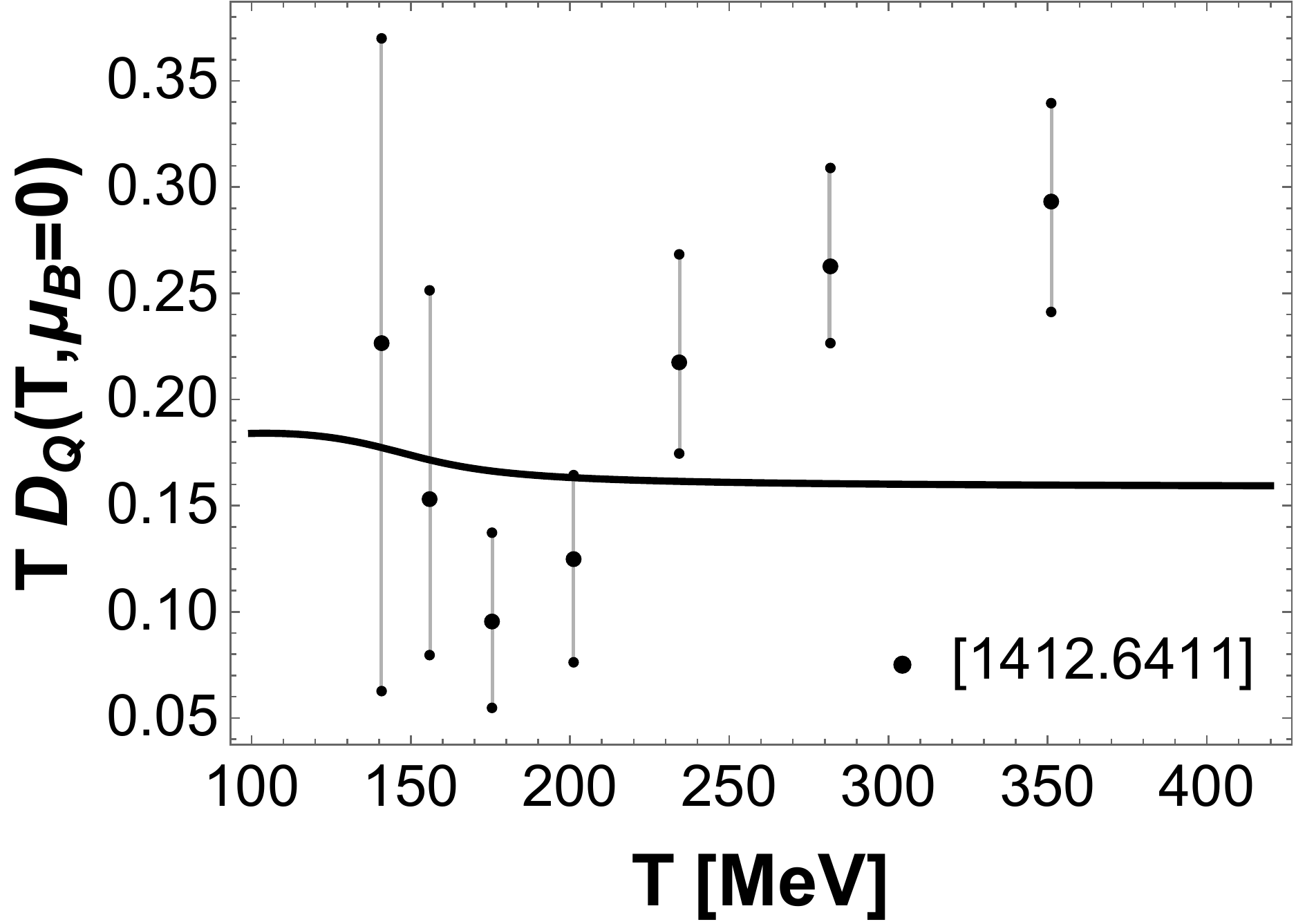}}
\caption{Holographic electric charge transport at $\mu_B=0$ compared to lattice results from Refs.\ \cite{Borsanyi:2011sw,Aarts:2014nba}. (a) Electric charge susceptibility. (b) Electric charge conductivity. (c) Electric charge diffusion.}
\label{fig:QtransportmuB0}
\end{figure}

\begin{figure}[htp!]
\center
\subfigure[]{\includegraphics[width=0.9\linewidth]{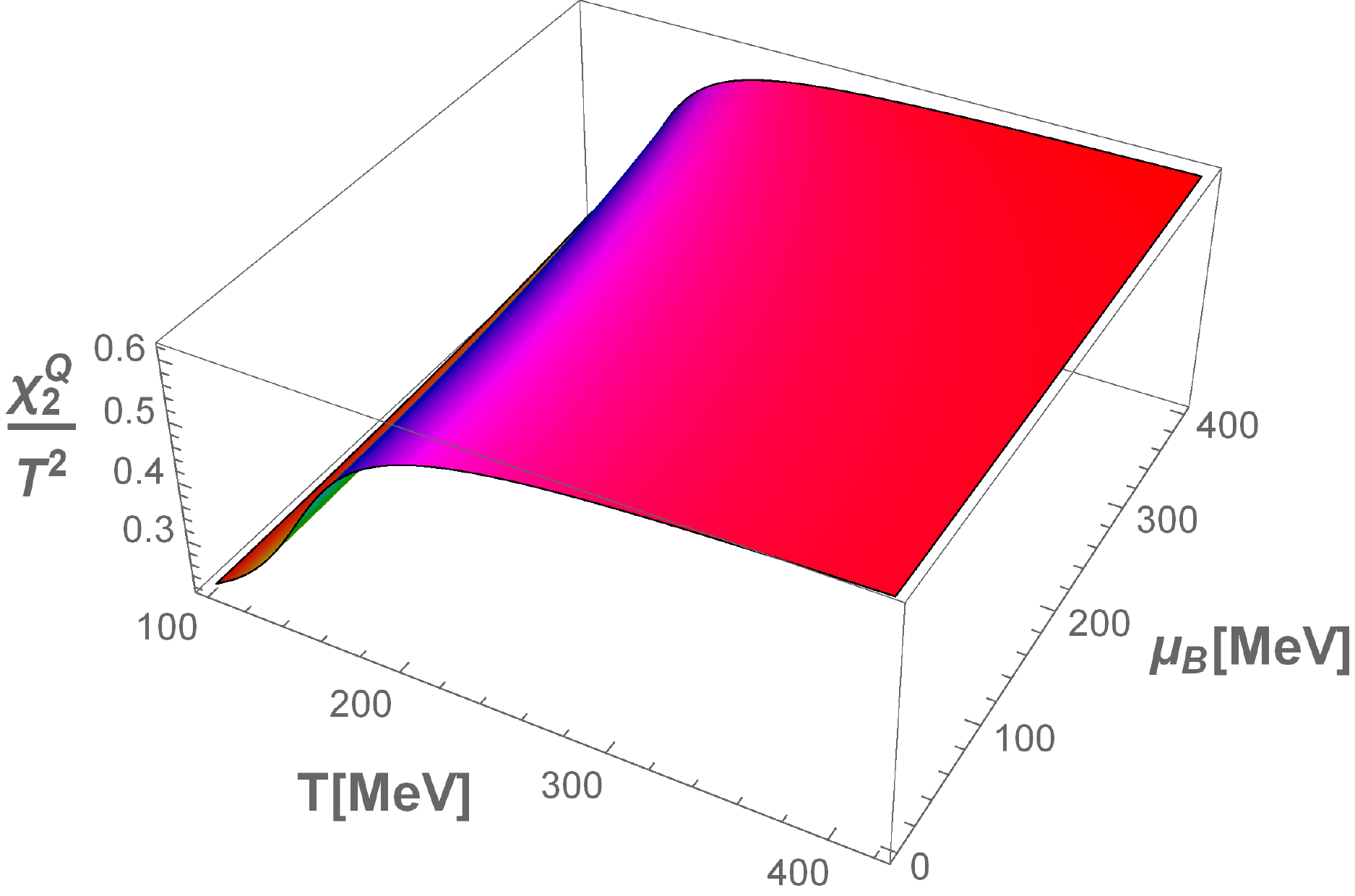}}
\qquad
\subfigure[]{\includegraphics[width=0.8\linewidth]{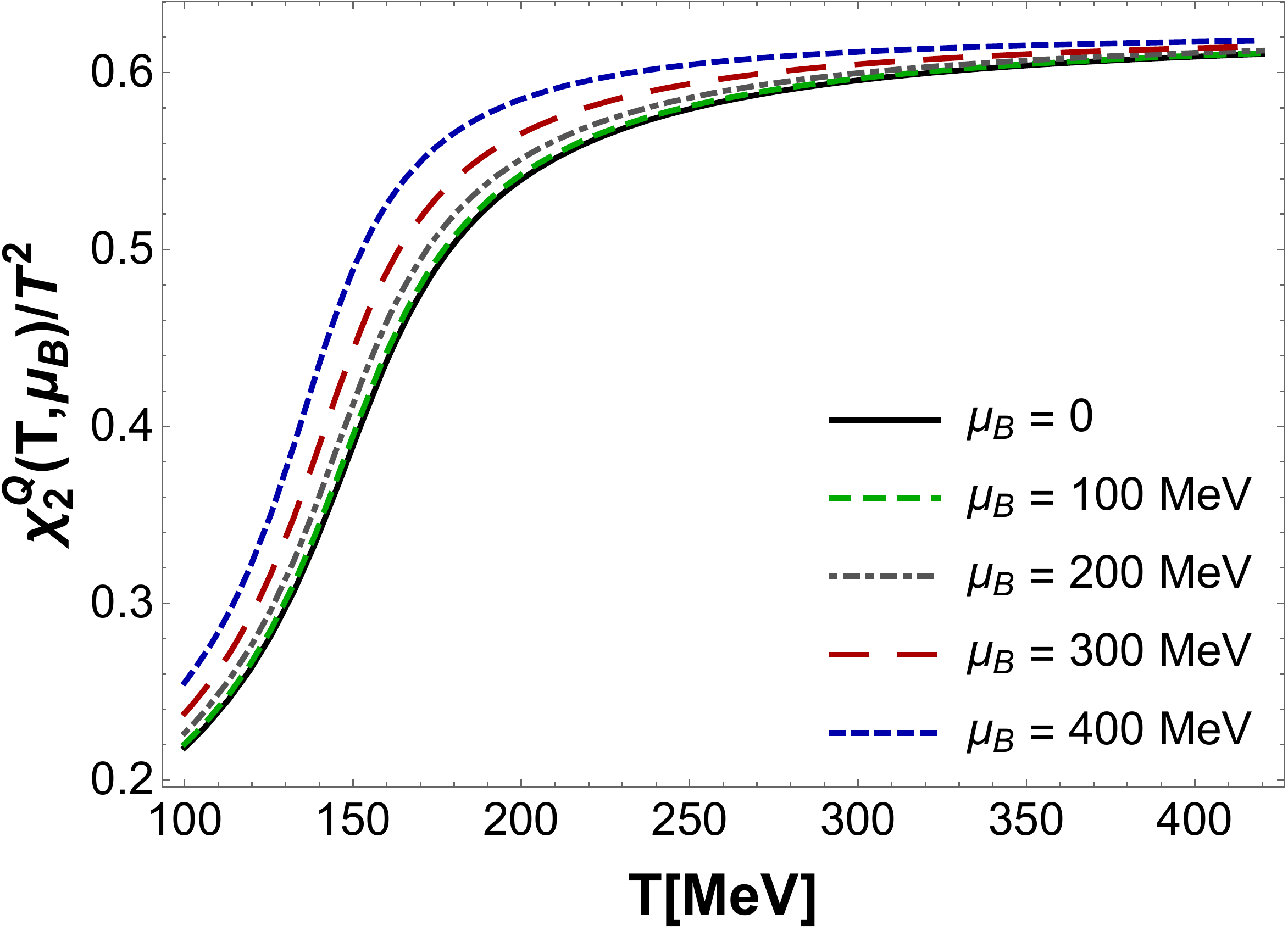}}
\caption{(Color online) Holographic electric charge susceptibility. (a) Surface plot as a function of $T$ and $\mu_B$. (b) Curves as functions of $T$ for different values of $\mu_B$.}
\label{fig:chiQfinitemu}
\end{figure}

\begin{figure}[htp!]
\center
\subfigure[]{\includegraphics[width=0.9\linewidth]{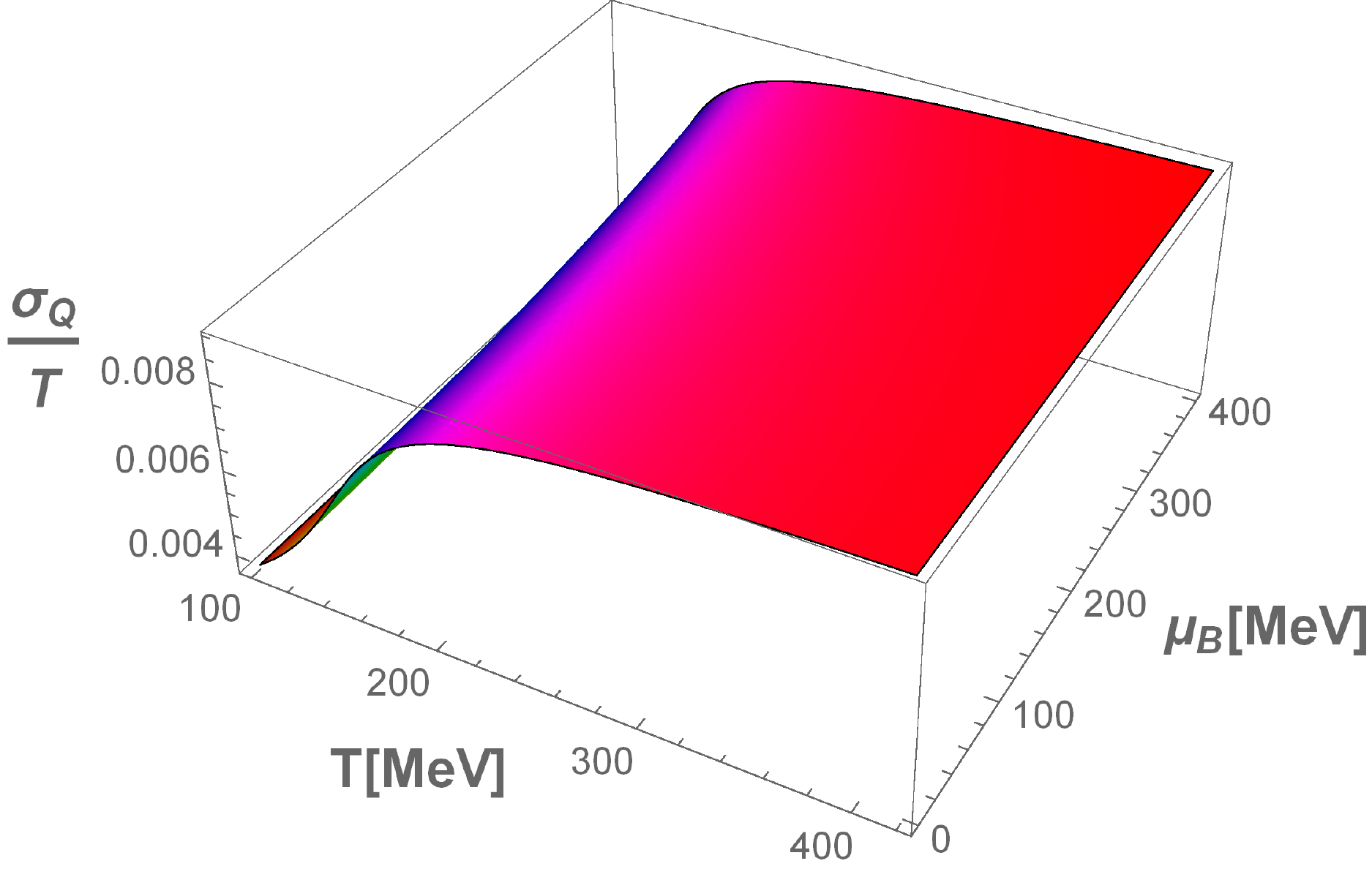}}
\qquad
\subfigure[]{\includegraphics[width=0.8\linewidth]{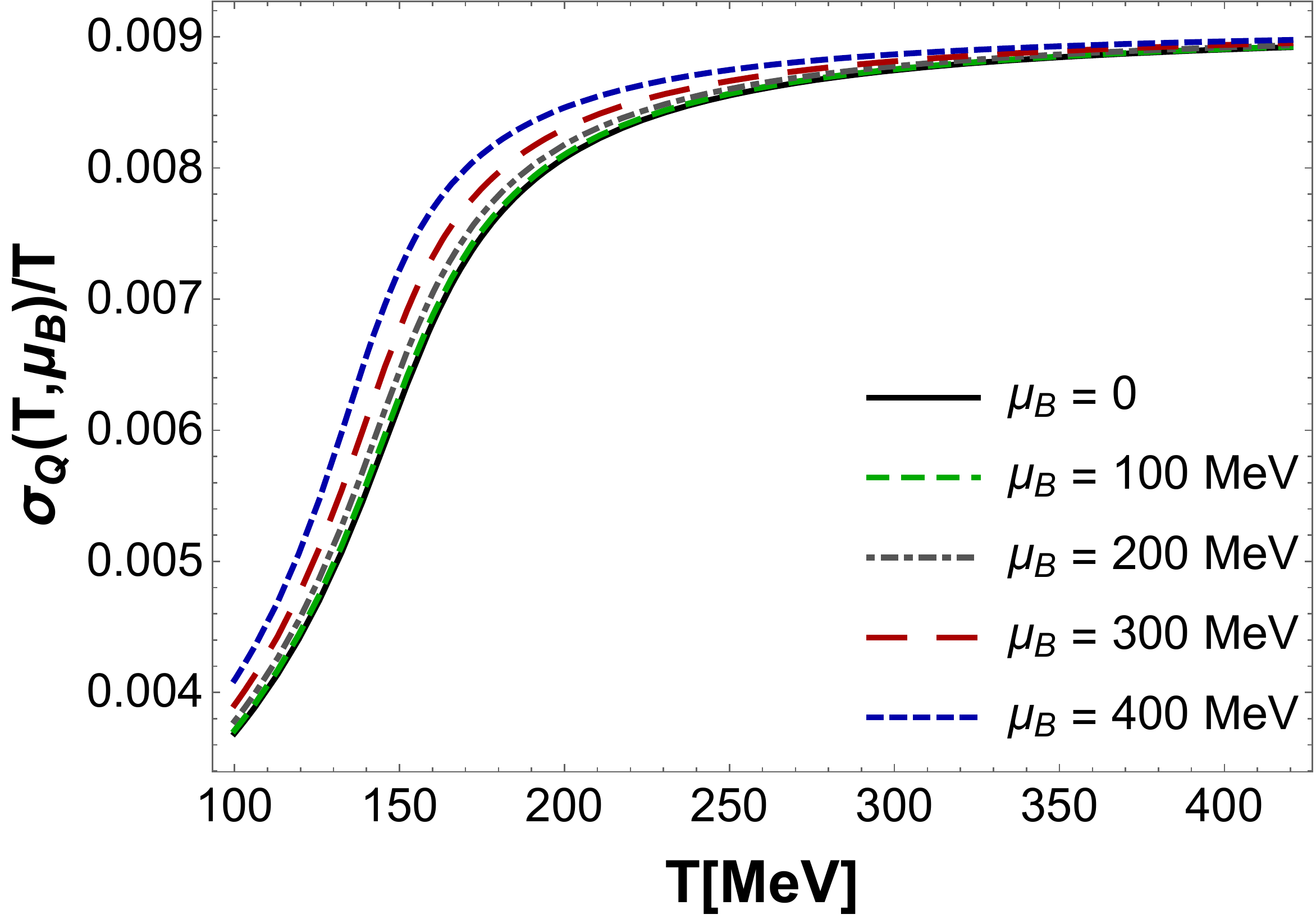}}
\caption{(Color online) Holographic electric charge conductivity. (a) Surface plot as a function of $T$ and $\mu_B$. (b) Curves as functions of $T$ for different values of $\mu_B$.}
\label{fig:sigmaQfinitemu}
\end{figure}

\begin{figure}[htp!]
\center
\subfigure[]{\includegraphics[width=0.9\linewidth]{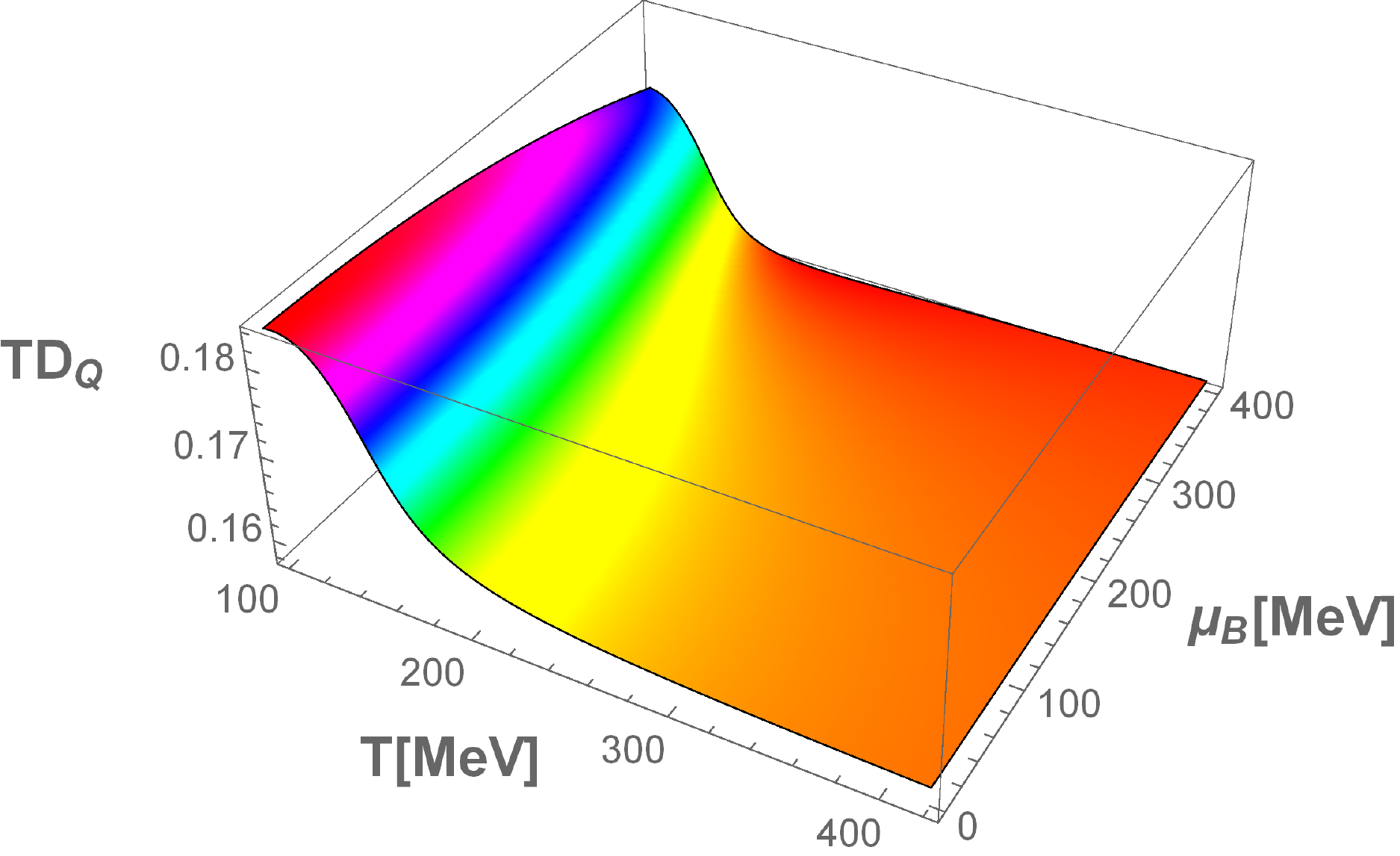}}
\qquad
\subfigure[]{\includegraphics[width=0.8\linewidth]{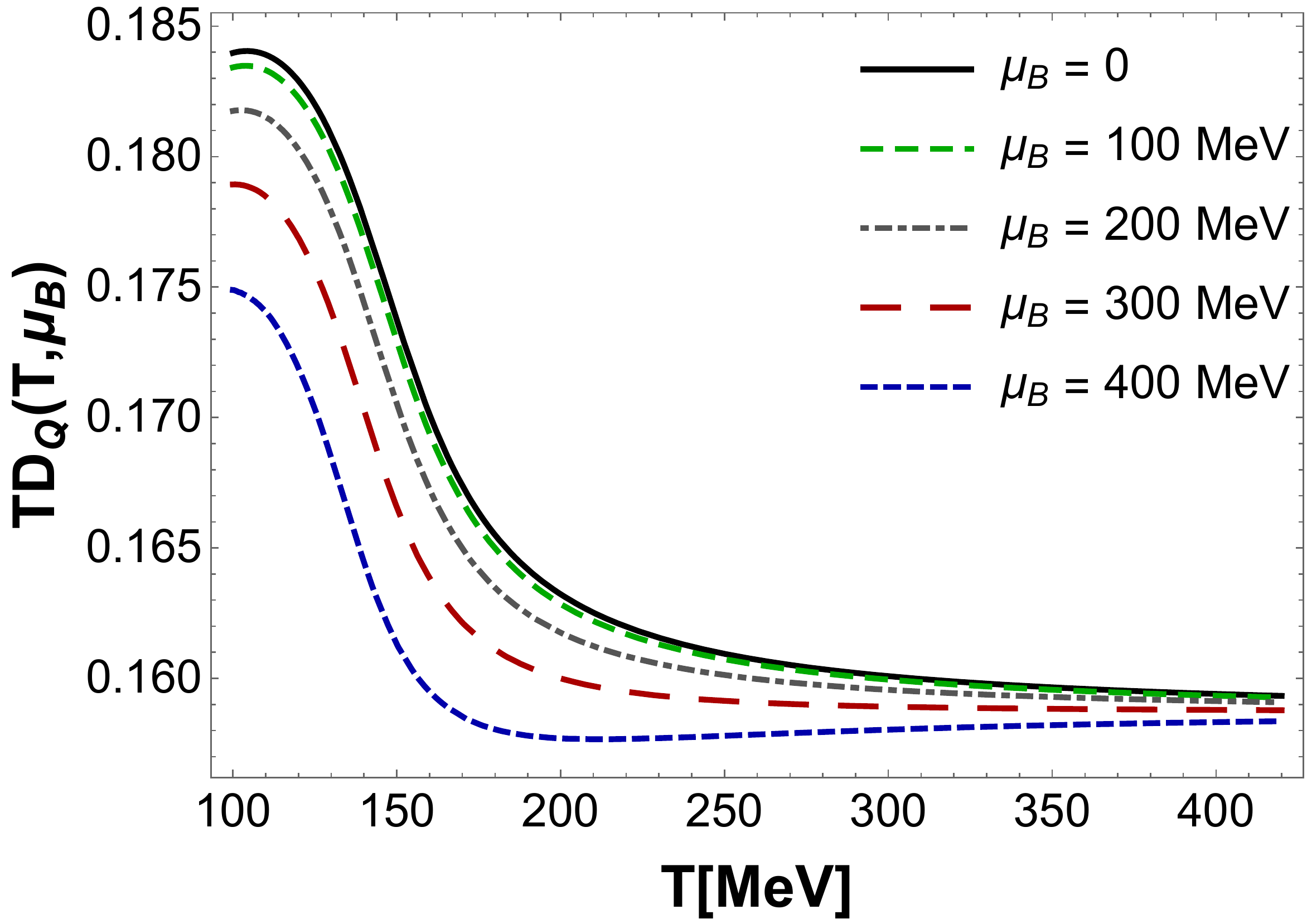}}
\caption{(Color online) Holographic electric charge diffusion. (a) Surface plot as a function of $T$ and $\mu_B$. (b) Curves as functions of $T$ for different values of $\mu_B$.}
\label{fig:DQfinitemu}
\end{figure}

As mentioned in the introduction, under the experimental conditions realized in heavy ion collisions, $\mu_B>\mu_S>\mu_Q$, and as a first approximation, we may consider the electric and strangeness sectors as probes over the finite $\mu_B$ black hole backgrounds discussed before. This means that we are going to take $\mu_S=\mu_Q=0$, which may be accomplished by considering the \emph{probe} Maxwell actions,
\begin{align}
S_X=-\frac{1}{8\kappa_5^2}\int d^5x\,\sqrt{-g}\,f_X(\phi)(F_{\mu\nu}^X)^2,
\label{eq:probes}
\end{align}
where $X=Q$ or $S$. In this subsection, we consider the electric charge sector as a probe on top of the finite $\mu_B$ backgrounds in Eq.\ \eqref{eq:EMDansatz}. Thus, the background value of the Maxwell field $A_\mu^Q$ vanishes and the probe action \eqref{eq:probes} with $X=Q$ is only nontrivial if we consider fluctuations of $A_\mu^Q$, which is all that is needed to compute the transport of electric charge.

In order to fix the electric coupling function $f_Q(\phi)$ in the probe action \eqref{eq:probes}, we evaluate the following integral on top of the EMD backgrounds \cite{DeWolfe:2010he,Finazzo:2015xwa},
\begin{align}
\frac{\chi_2^Q}{T^2}(\mu_Q = 0)=\frac{s/T^3}{16\pi^2 f_Q(0)\int_{r_H}^\infty dr\, e^{-2A(r)} f_Q^{-1}\left(\phi(r)\right)},
\label{eq:integralsusc}
\end{align}
with the requirement that it matches the lattice QCD results for the electric charge susceptibility at vanishing chemical potential from Ref.\ \cite{Borsanyi:2011sw}. With this, one may fix,
\begin{align}
f_Q(\phi)=0.0193\,\textrm{sech}(-100\,\phi)+0.0722\,\textrm{sech}(10^{-7}\,\phi),
\label{eq:fQ}
\end{align}
with the corresponding results shown in Fig.\ \ref{fig:QtransportmuB0} (a).

By using the membrane paradigm \cite{Iqbal:2008by}, one derives the following holographic Kubo formula for the electric conductivity in the EMD model (already expressed in physical units) \cite{Finazzo:2015xwa},
\begin{align}
\sigma_Q= \frac{f_Q\left(\phi(r_H)\right)}{2\kappa_5^2}\,\frac{e^{A(r_H)}}{\phi_A^{1/\nu}}\,\Lambda.
\label{eq:sigmaQ}
\end{align}
As before, the electric charge diffusion may be calculated using Nernst-Einstein's relation by dividing the electric conductivity by the electric susceptibility.

In Figs.\ \ref{fig:QtransportmuB0} (b) and (c) we show the EMD predictions for the electric conductivity and electric charge diffusion at $\mu_B=0$ compared to lattice QCD results from Ref.\ \cite{Aarts:2014nba}. These holographic results at vanishing baryon density constitute an update of the results first published in Ref.\ \cite{Finazzo:2013efa}, which were based on an older version of the holographic model (without chemical potential). We note that the EMD results for the electric conductivity of the QGP, which are predictions of the model, are much closer to the lattice QCD results from Ref.\ \cite{Aarts:2014nba} than many other models available in the literature; for comparisons between different models, see for instance Fig.\ 6 of Ref.\ \cite{Greif:2014oia} and Fig.\ 4 of Ref.\ \cite{Greif:2016skc}. We must also remark that there is room for further improvements in the agreement between the EMD predictions for the electric conductivity and charge diffusion and the corresponding lattice results from Ref. \cite{Aarts:2014nba}, once the latter are refined by taking the continuum limit and by considering physical quark masses (as in the case of the lattice inputs used to fix the free parameters of the EMD model).

In Figs.\ \ref{fig:chiQfinitemu}, \ref{fig:sigmaQfinitemu}, and \ref{fig:DQfinitemu} we show the EMD results for the electric charge susceptibility, conductivity, and diffusion at finite baryon chemical potential, respectively. As in the case of the baryon diffusion, also the diffusion of electric charge is found to be suppressed as one increases the baryon density of the medium.

\subsection{Strangeness sector}
\label{sec:strangeness}

We consider for the first time how the transport of strangeness in the QGP near the crossover transition is affected by a nonzero baryon chemical potential. The strangeness sector  is especially interesting in light of the $p/\pi$ puzzle at the LHC \cite{Floris:2014pta} where it has been suggested that there are possibly missing strange resonances when compared to Lattice QCD \cite{Bazavov:2014xya,Alba:2017mqu} and/or there could be a flavor hierarchy of chemical freeze-out temperatures \cite{Bellwied:2013cta,Noronha-Hostler:2016rpd}.  Recent work \cite{Takeuchi:2015ana} has also looked into the dynamics of strangeness and found that strange hadrons (with the exception of $\Lambda$'s) generally freeze-out sooner than light hadrons.  Thus, understanding how strongly strangeness diffuses throughout the Quark Gluon Plasma, especially at finite baryon densities, could help shed further light on such topics.

\begin{figure}[htp!]
\center
\includegraphics[width=0.8\linewidth]{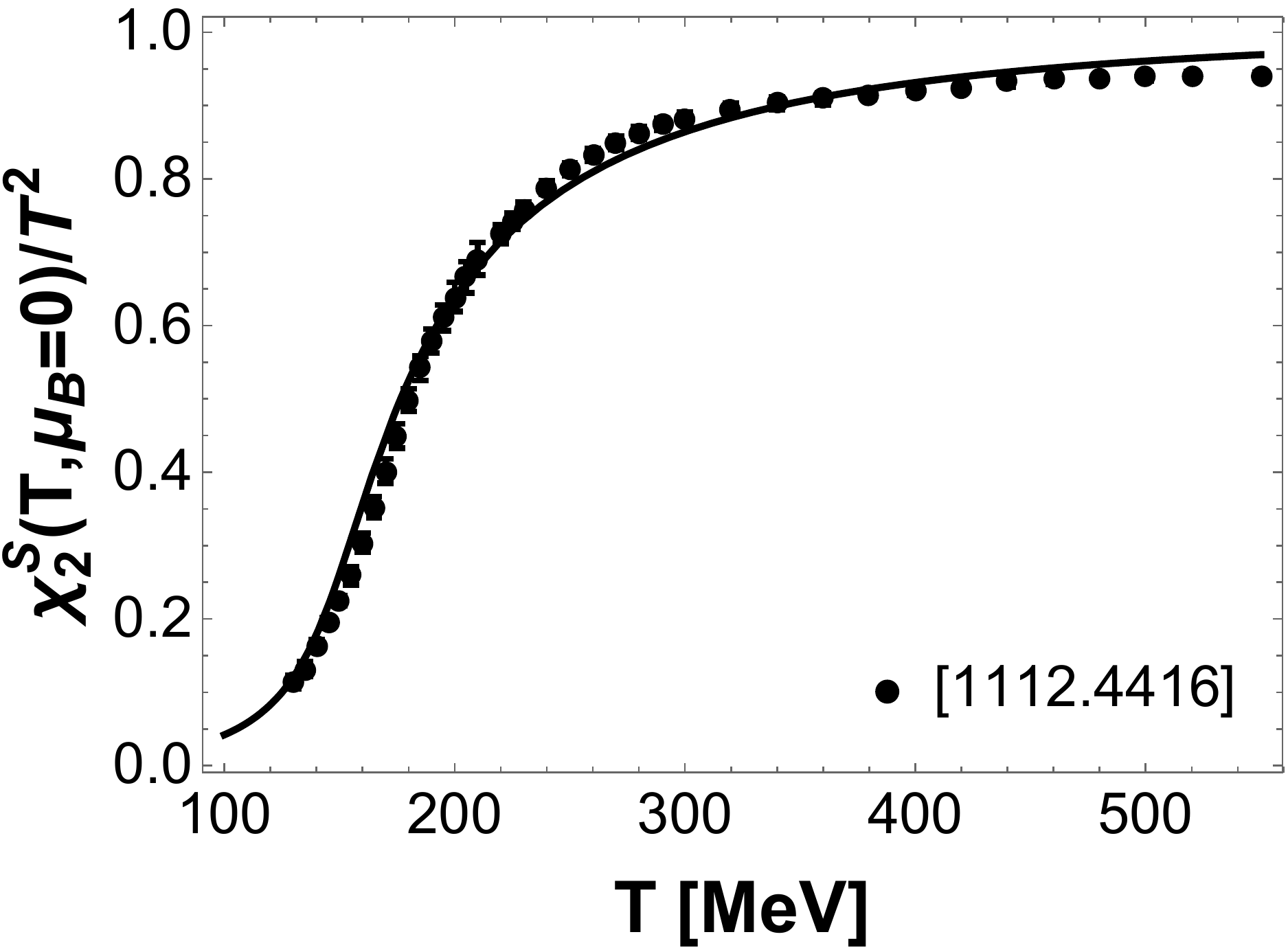}
\caption{Holographic strangeness susceptibility at $\mu_B=0$ compared to lattice results from Ref.\ \cite{Borsanyi:2011sw}.}
\label{fig:StransportmuB0}
\end{figure}

\begin{figure}[htp!]
\center
\subfigure[]{\includegraphics[width=0.9\linewidth]{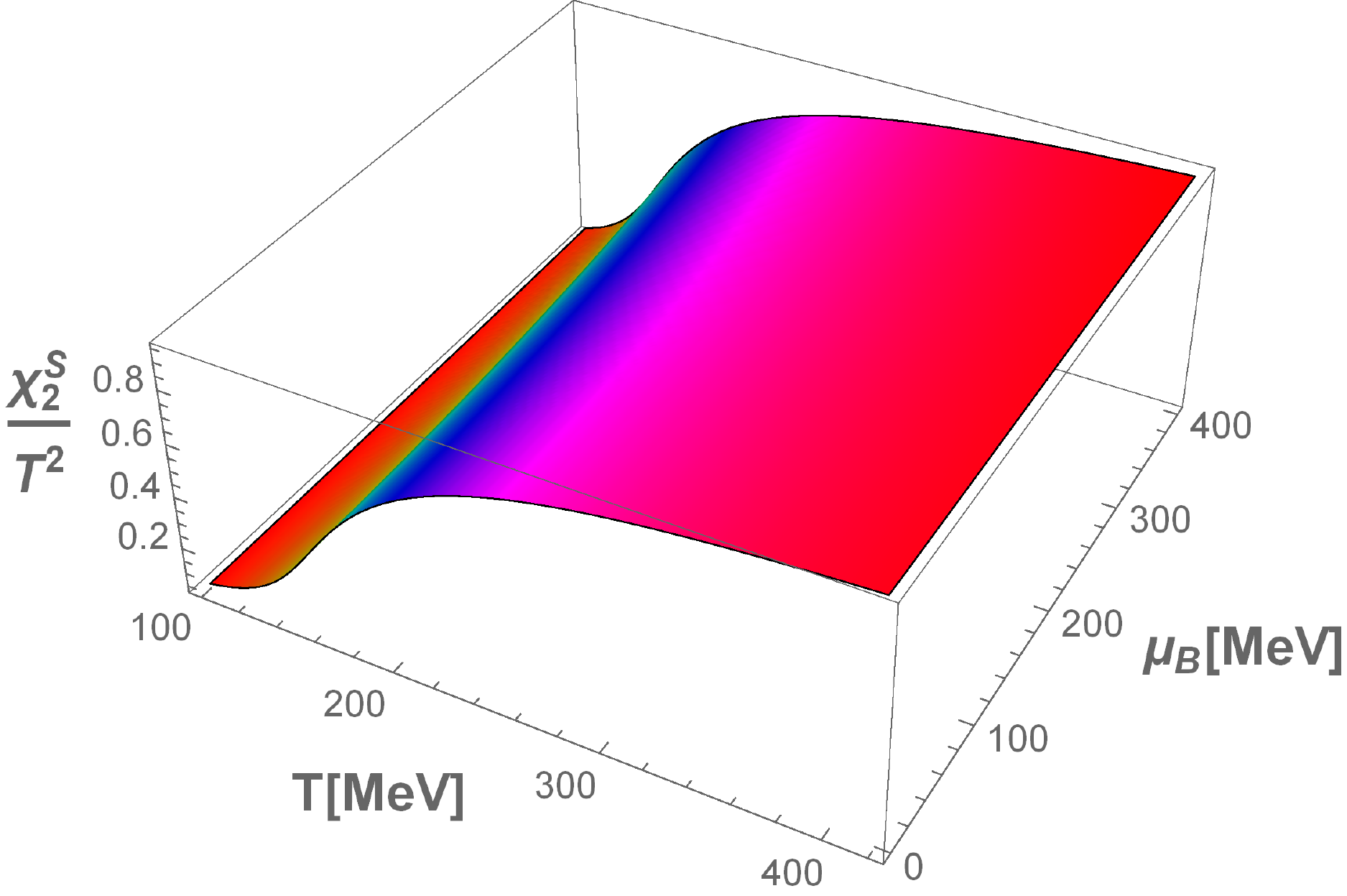}}
\qquad
\subfigure[]{\includegraphics[width=0.8\linewidth]{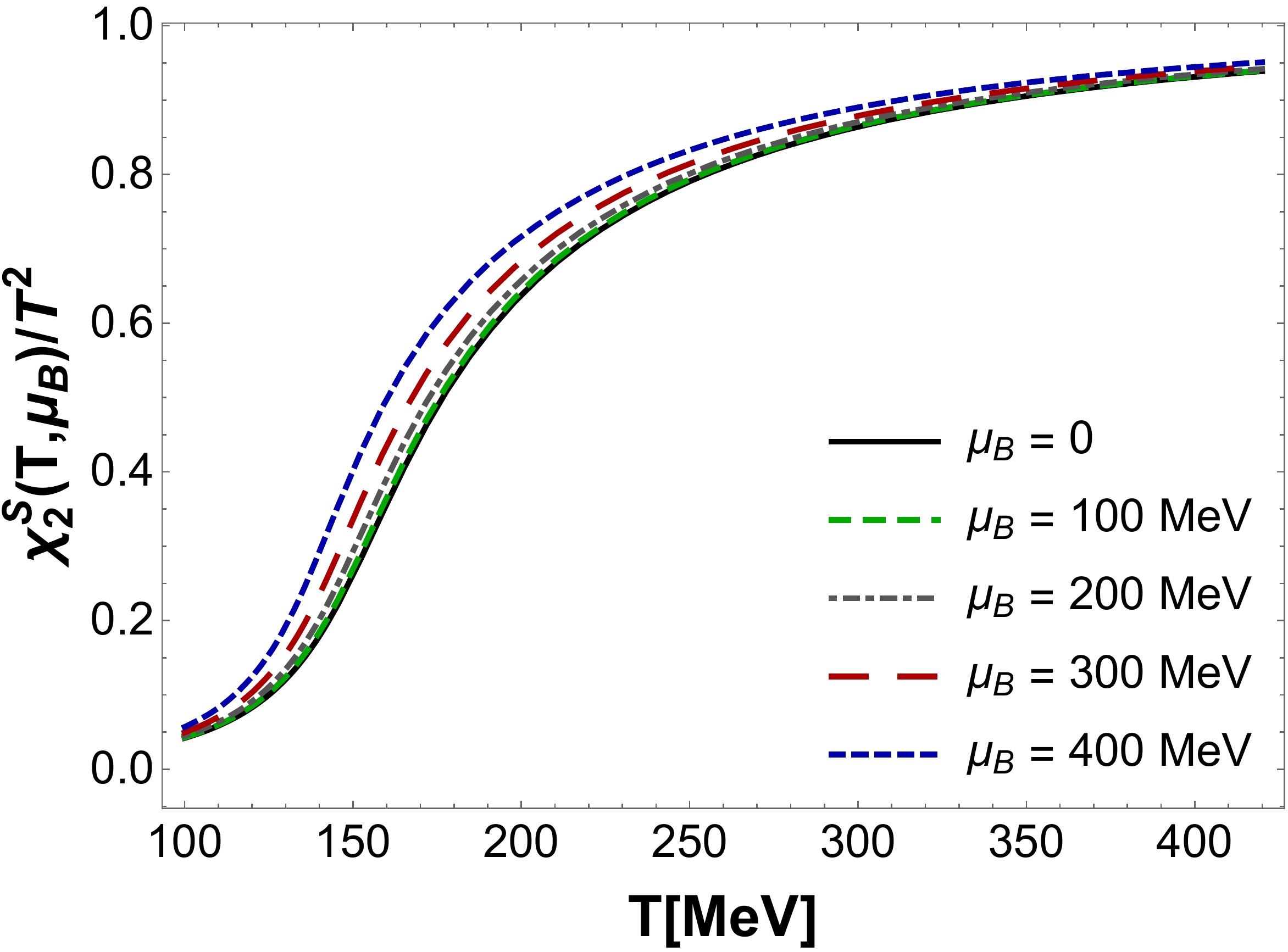}}
\caption{(Color online) Holographic strangeness susceptibility. (a) Surface plot as a function of $T$ and $\mu_B$. (b) Curves as functions of $T$ for different values of $\mu_B$.}
\label{fig:chiSfinitemu}
\end{figure}

\begin{figure}[htp!]
\center
\subfigure[]{\includegraphics[width=0.9\linewidth]{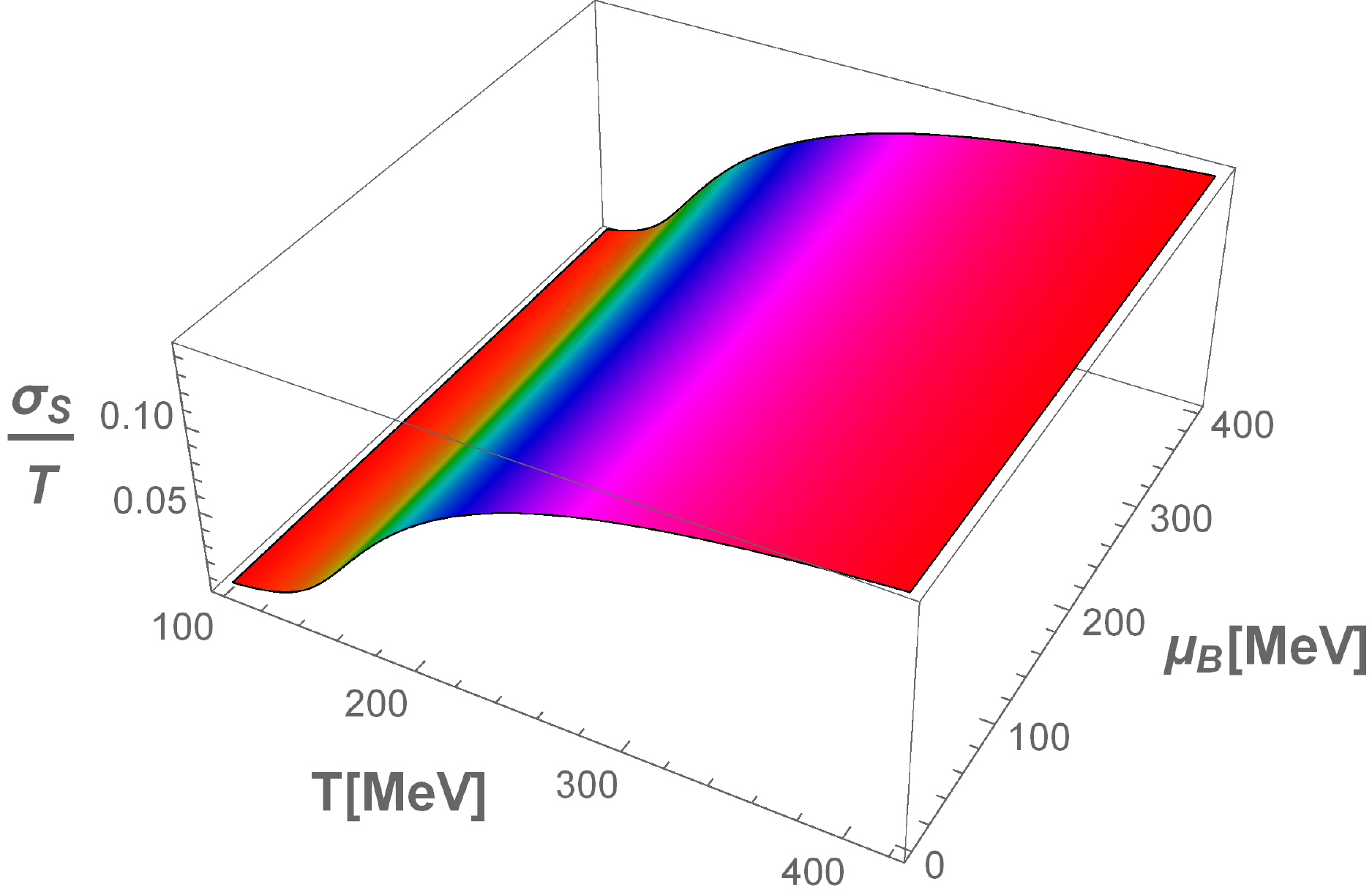}}
\qquad
\subfigure[]{\includegraphics[width=0.8\linewidth]{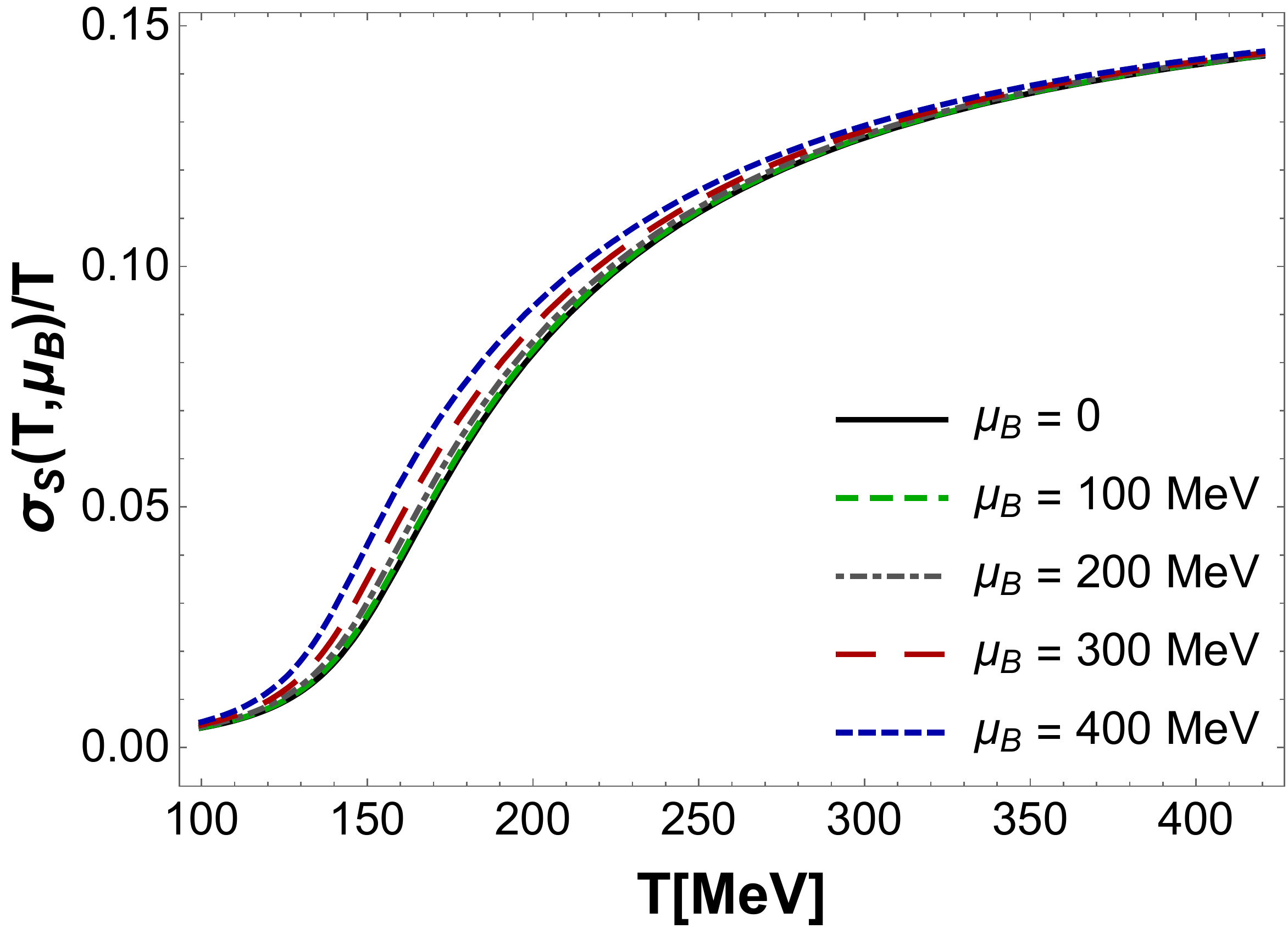}}
\caption{(Color online) Holographic strangeness conductivity. (a) Surface plot as a function of $T$ and $\mu_B$. (b) Curves as functions of $T$ for different values of $\mu_B$.}
\label{fig:sigmaSfinitemu}
\end{figure}

\begin{figure}[htp!]
\center
\subfigure[]{\includegraphics[width=0.9\linewidth]{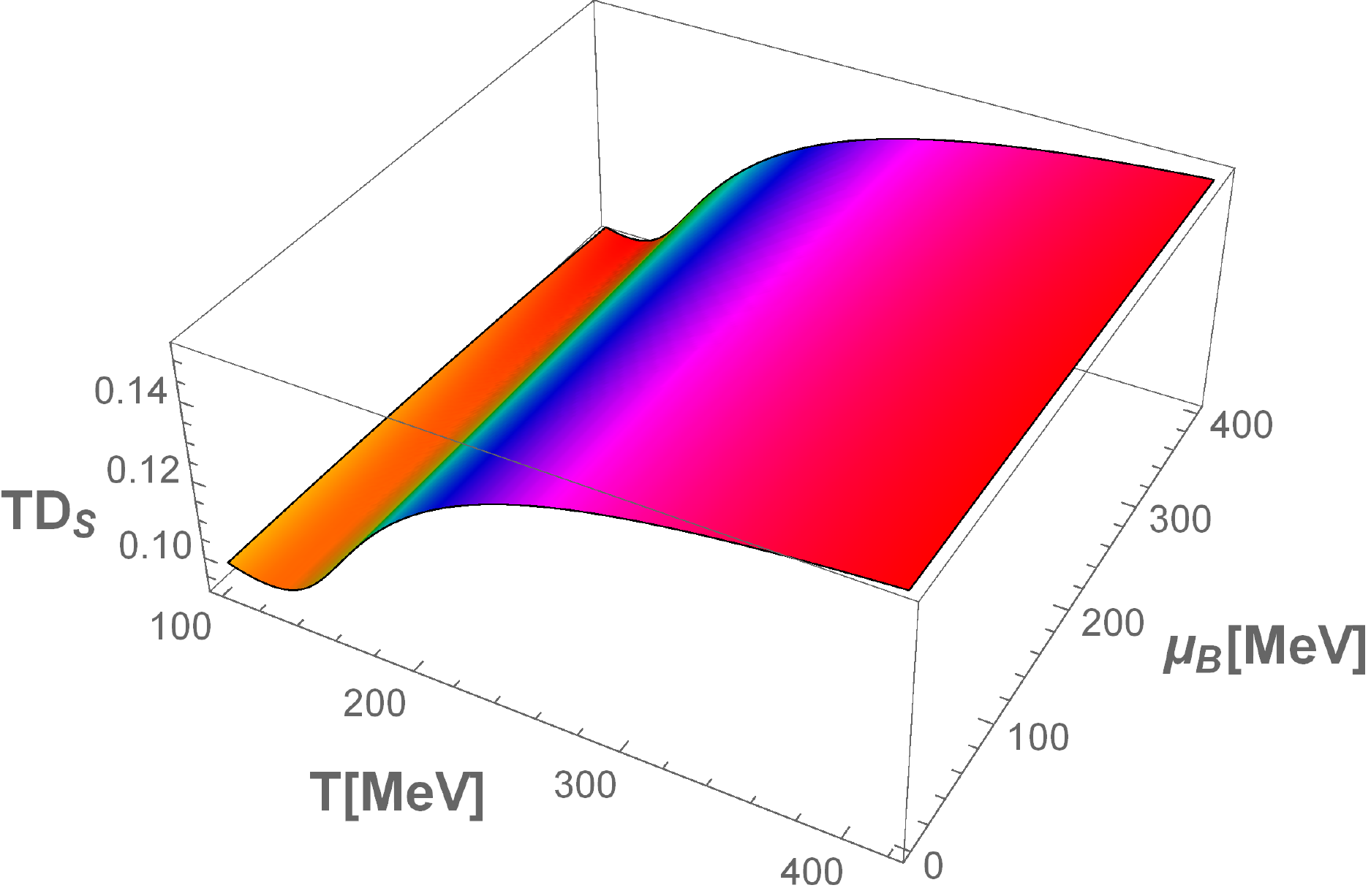}}
\qquad
\subfigure[]{\includegraphics[width=0.8\linewidth]{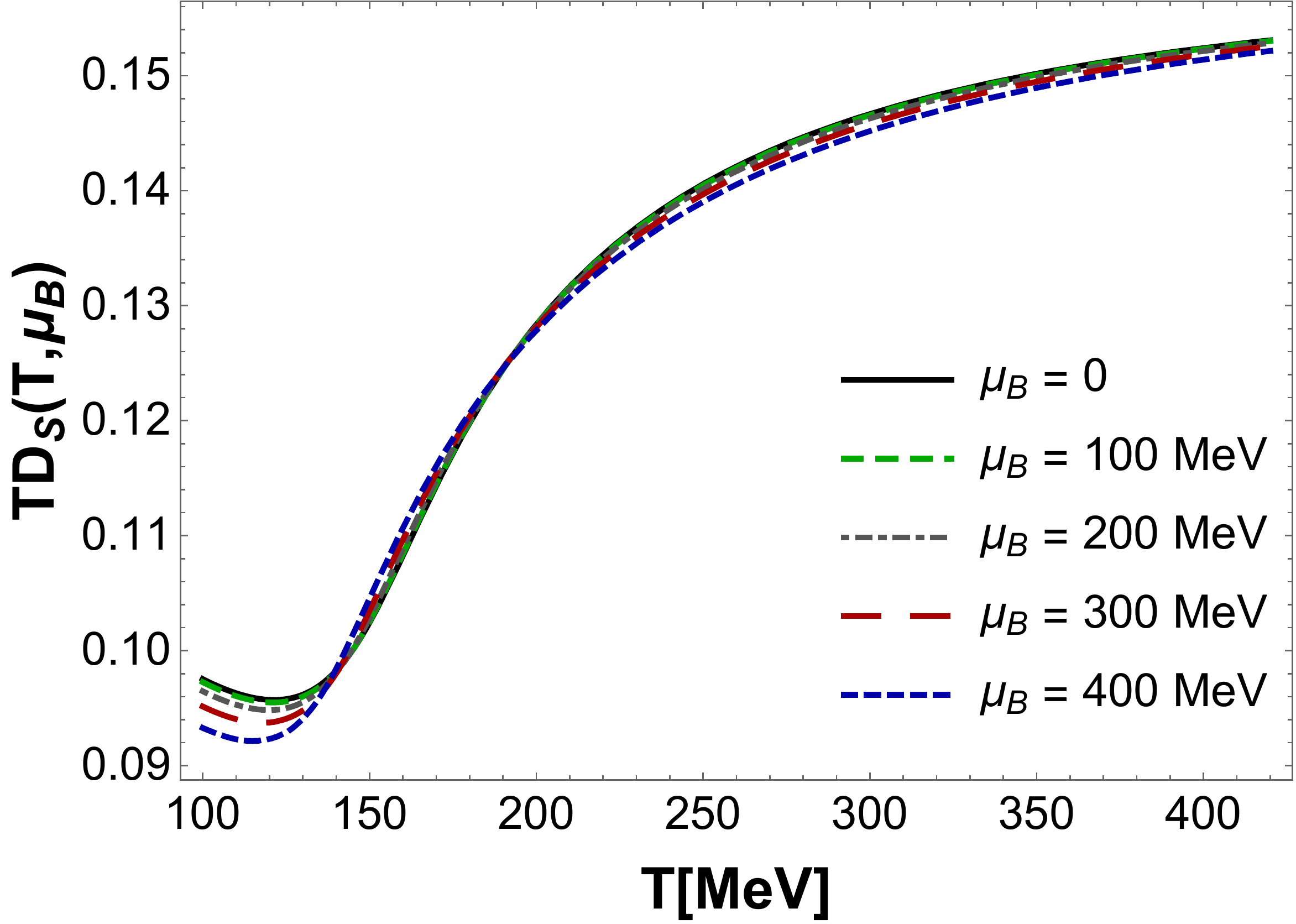}}
\caption{(Color online) Holographic strangeness diffusion. (a) Surface plot as a function of $T$ and $\mu_B$. (b) Curves as functions of $T$ for different values of $\mu_B$.}
\label{fig:DSfinitemu}
\end{figure}

As done in the previous section, we begin by fixing the coupling function $f_S(\phi)$ between the bulk fields $\phi$ and $A_\mu^S$ by solving the integral in Eq.\ \eqref{eq:integralsusc} (with the change $Q\to S$) with the constraint that the holographic result for the strangeness susceptibility at $\mu_B=0$ matches the corresponding lattice QCD result from Ref.\ \cite{Borsanyi:2011sw}. As in the case of electric charge, we remark that strangeness effects are only included here in the probe limit. By doing so, one can fix,\footnote{One consequence of the overall normalization chosen for the couplings $f_B(\phi)$, $f_Q(\phi)$, and $f_S(\phi)$ in Eqs.\ \eqref{eq:EMDparameters}, \eqref{eq:fQ}, and \eqref{eq:fS}, respectively, is that at high temperatures (compared to $\mu_B$) the diffusion rates for the conserved charges converge approximately to the same value obtained in the SYM plasma \cite{Policastro:2002se} (which is independent of the choice of $\kappa_5^2$).}
\begin{align}
f_S(\phi)= 1.282 \,\textrm{sech}(0.8 \,\phi) - 0.282 \,\textrm{sech}(50 \,\phi),
\label{eq:fS}
\end{align}
with the corresponding results displayed in Fig.\ \ref{fig:StransportmuB0}.

The calculation of the strangeness conductivity and diffusion proceeds as in the previous section by changing $f_Q(\phi)\to f_S(\phi)$, and the EMD predictions for the strangeness susceptibility, conductivity and diffusion at finite baryon density are shown in Figs.\ \ref{fig:chiSfinitemu}, \ref{fig:sigmaSfinitemu}, and \ref{fig:DSfinitemu}, respectively. We note that, compared to the baryon and electric charge sectors, the diffusion of strangeness is much more robust to the presence of a nontrivial baryon density, especially for temperatures above the crossover transition. However, also in this case the diffusion of conserved charge is suppressed by the baryon chemical potential, although the effect is very small. Overall, we observe a general suppression of diffusion of conserved charges through the medium as the baryon density of the plasma is increased, with such effect being stronger in the baryon sector and considerably weaker in the strangeness sector. As mentioned before, a more complete investigation of finite density effects on the transport of conserved charges would require to go beyond the probe limit in the electric and strangeness sectors. Such a study is much more challenging from a numerical point of view than the calculations performed here and it is left to a future investigation.

\subsection{Brief Summary of BSQ Conductivity and Diffusion}
\label{sec:summary}

A prime motivation for this work was to obtain the temperature and baryon chemical potential dependence of $\sigma_i/T$, where $i=B,S,Q$, which may then be used in future relativistic hydrodynamic models that include the effects of these three conserved charges in the evolution of the plasma. Such models would be instrumental to investigate, in a realistic manner, the observable consequences of the presence of the critical end point in the QCD phase diagram, which is one of the primary tasks of the Beam Energy Scan Theory (BEST) collaboration.

However, there are other questions that can be answered by studying these transport coefficients.  For instance, a larger $\sigma_i/T$ indicates that the current associated with the transport of this conserved charge is larger (also, in a gas, one may say that the mean free path between these interactions is larger). When one looks at the BSQ magnitudes of $\sigma_i/T$, one can clearly see that in our calculations there is a hierarchy between the three transport coefficients such that strangeness is the largest, followed by baryon number and electric charge. Though the magnitude of the strangeness conductivity has not yet been compared to Lattice QCD results, we do see that the electric conductivity is quite a bit smaller than the baryon conductivity found here. Additionally, in Sec.\ \ref{sec:Tcs} we will study the inflection points of the conductivities compared to their associated susceptibilities as a method of understanding the difference in transition temperatures between equilibrium and out-of-equilibrium properties.

\section{Bulk viscosity}
\label{sec:bulk}

In this section we turn our attention to the holographic calculation of the bulk viscosity in the hot and baryon dense QGP described by the EMD holographic model.

In \cite{Noronha-Hostler:2013gga,Noronha-Hostler:2014dqa}, effects from the bulk viscosity transport coefficient, $\zeta$, were included in event-by-event hydrodynamics and state-of-the-art hydrodynamic models \cite{Ryu:2015vwa,Bernhard:2016tnd} have shown that this coefficient is an important ingredient in the description of heavy ion experimental data. For instance, in the calculations performed in \cite{Ryu:2015vwa} a profile for the bulk viscosity peaking at the crossover allowed for a simultaneous agreement with experimental data for different physical observables. A similar profile was later used in the Bayesian analysis performed in \cite{Bernhard:2016tnd}, which confirmed the role played by bulk viscosity in comparisons to experimental data. However, at the moment it is not clear how the different assumptions regarding the many model parameters that enter in these complex simulations affect the magnitude and the location of the peak of $\zeta/s$ extracted from model comparisons to data. In fact, the small $\zeta/s$ found in the very recent Bayesian analysis done in \cite{jonah} is different than the result found in Ref.\ \cite{gabrielQM}, which made different assumptions regarding the location of the peak of the bulk viscosity.

In this subsection we calculate the $T$ and $\mu_B$ dependence of the bulk viscosity in the EMD model of Ref.\ \cite{Rougemont:2015wca} used here.\footnote{A previous estimate in an older version of the EMD model, constructed using as phenomenological inputs now outdated lattice results for the QCD equation of state at $\mu_B=0$, was originally presented in Ref.\ \cite{DeWolfe:2011ts}.}

\begin{figure}
\center
\subfigure[]{\includegraphics[width=0.9\linewidth]{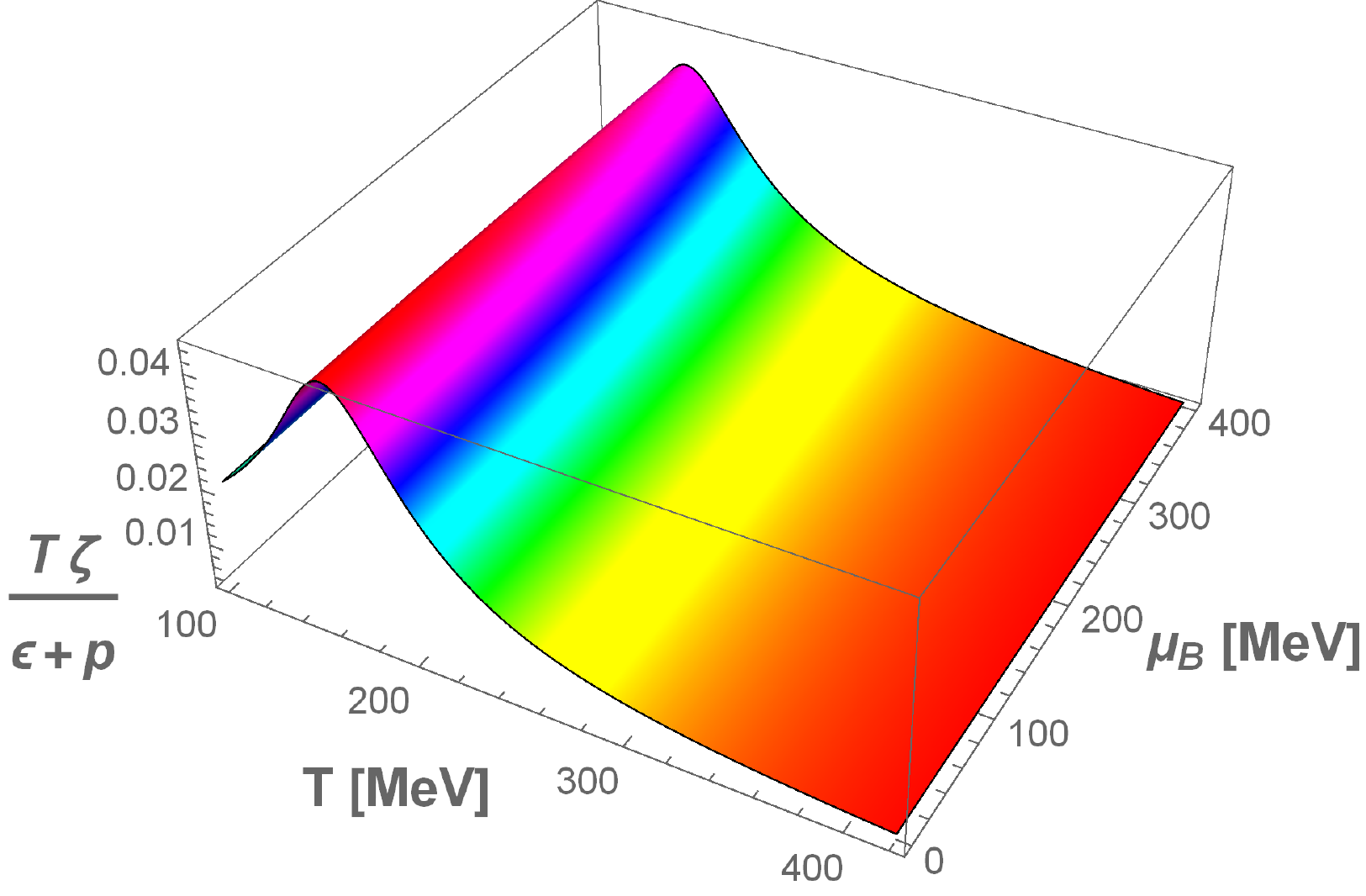}}
\qquad
\subfigure[]{\includegraphics[width=0.8\linewidth]{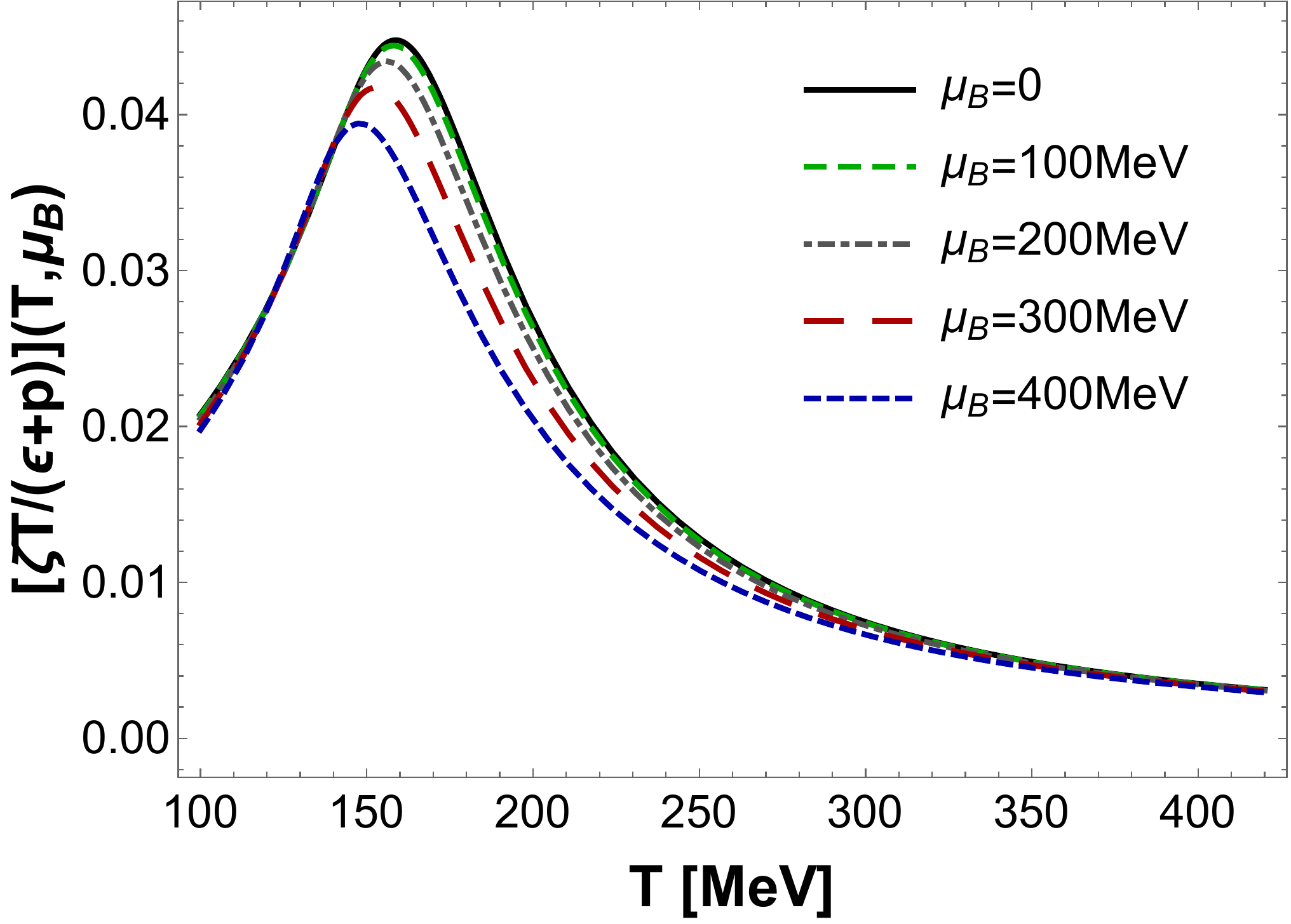}}
\caption{(Color online) Holographic bulk viscosity. (a) Surface plot as a function of $T$ and $\mu_B$. (b) Curves as functions of $T$ for different values of $\mu_B$.}
\label{fig:bulk}
\end{figure}

\begin{figure}[htp!]
\center
\includegraphics[width=0.8\linewidth]{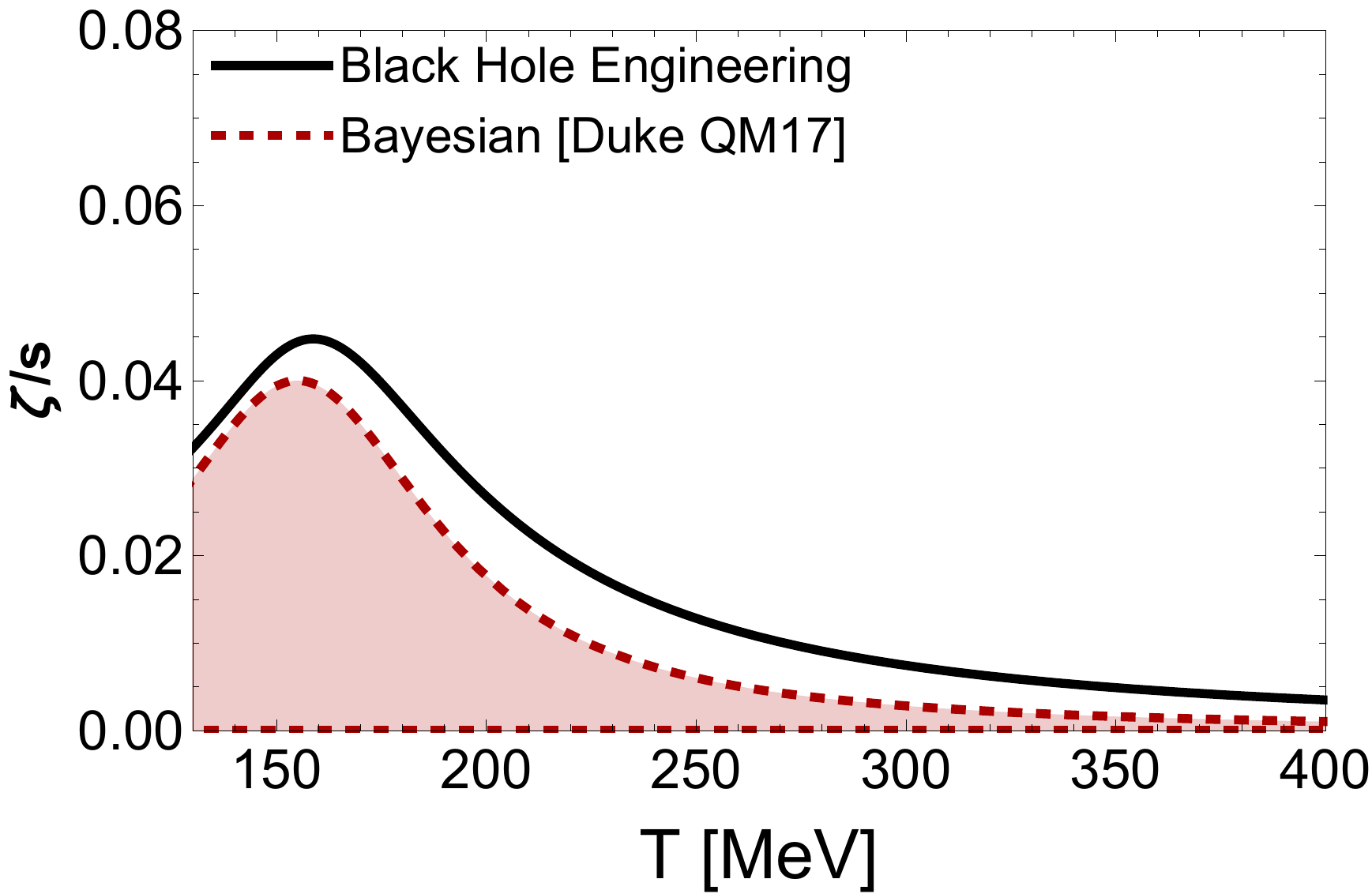}
\caption{(Color online) $\zeta/s(T)$ at $\mu_B=0$ compared to a recent study using Bayesian analysis \cite{jonah} to extract $\zeta/s(T)$ from comparisons of relativistic hydrodynamics calculations to experimental data.}
\label{fig:zetabay}
\end{figure}

In the holographic EMD setup at finite baryon density the bulk viscosity is associated with the gauge and diffeomorphism invariant combination of the EMD fields transforming as a singlet under $SO(3)$, which we denote by $\mathcal{H}$. The equation of motion for this scalar perturbation is given by \cite{DeWolfe:2011ts},
\begin{align}
&\mathcal{H}''(r)+\left[ 4A'(r)+\frac{h'(r)}{h(r)} +\frac{2\phi''(r)}{\phi'(r)} - \frac{2A''(r)}{A'(r)} \right]\mathcal{H}'(r)\nonumber\\
&+\left[ \frac{e^{-2A(r)}\omega^2}{h(r)^2} + \frac{h'(r)}{h(r)} \left( \frac{A''(r)}{A'(r)} - \frac{\phi''(r)}{\phi'(r)} \right) \right.\nonumber\\
&\left.+\frac{e^{-2A(r)}}{h(r)\phi'(r)}\left( 3A'(r)f'_B(\phi)-f_B(\phi)\phi'(r) \right)\Phi'(r)^2\right]\mathcal{H}(r)=0.
\label{eq:bulkeom}
\end{align}
One numerically solves Eq.\ \eqref{eq:bulkeom} over the EMD backgrounds with in-falling wave condition at the black hole horizon, normalizing the scalar perturbation to unity at the boundary. Then, one plugs in the solution into the following holographic Kubo formula for the dimensionless ratio between the bulk viscosity and the entropy density \cite{DeWolfe:2011ts},
\begin{align}
\frac{\zeta}{s}=-\frac{1}{36\pi}\lim_{\omega\to 0}\frac{e^{4A(r)}h(r)\phi'(r)^2\,\textrm{Im}\left[ \mathcal{H}^*(r)\mathcal{H}'(r) \right]}{\omega A'(r)^2},
\label{eq:bulk}
\end{align}
where $e^{4A}h\phi'\,^2\,\textrm{Im}\left[ \mathcal{H}^*\mathcal{H}' \right]/A'\,^2$ is a conserved flux in the radial direction such that \eqref{eq:bulk} may be evaluated at any value of $r$. At finite baryon density, the relevant combination entering in hydrodynamics is $\zeta T/(\epsilon + p)$, which reduces to the ratio $\zeta/s$ at $\mu_B=0$. The results for the bulk viscosity as a function of $(T,\mu_B)$ are shown in Fig.\ \ref{fig:bulk}. One notes that the bulk viscosity is reduced with increasing baryon density and, thus, one may say that there is an overall suppression of hydrodynamic viscosity coefficients in the medium due to a nonzero baryon chemical potential, since the same effect is observed in the shear viscosity (discussed in Appendix \ref{sec:shear}). The bulk viscosity develops a peak in the crossover (contrary to the shear viscosity which has a minimum there). Interestingly enough, we note that the magnitude of the bulk viscosity and its $T$ dependence is somewhat similar to the result of a recent Bayesian analysis \cite{jonah} shown in Fig.\ \ref{fig:zetabay}. The small value of $\zeta/s$ found in our calculations here is, however, not compatible with the profile used in previous model comparisons to data done in \cite{Ryu:2015vwa,Bernhard:2016tnd,gabrielQM}, which required an order of magnitude larger values of bulk viscosity.

\section{Transition temperatures}
\label{sec:Tcs}

The crossover transition observed at low baryon density in $(2+1)$-flavor lattice QCD calculations with physical quark masses is not a genuine phase transition since the derivatives of the pressure are continuous functions \cite{Aoki:2006we,Borsanyi:2016ksw,Bazavov:2017dus}. In other words, the change from hadronic degrees of freedom in the HRG phase to partonic degrees of freedom in the QGP phase proceeds  continuously, as a crossover, instead of a true phase transition. In the region of the QCD phase diagram where this qualitative change in the relevant degrees of freedom of the system takes place, some physical observables still display fast variations as functions of temperature, which may be characterized by inflection points or extrema. Consequently, instead of displaying a definite critical temperature that characterizes all the different observables (as it would be the case, for instance, in a first order phase transition), in the case of a crossover one may only define a band of transition temperatures and investigate how a set of observables, for instance the fluctuation of conserved charges, vary along that band. 

Recently, much attention has been paid to the equilibrium observables known as susceptibilities because strangeness has a higher transition temperature compared to light hadrons/quarks by at least $15$ MeV \cite{Bellwied:2013cta}, which may indicate that strange degrees of freedom not only hadronize at a higher temperature but could also reach chemical freeze-out earlier on than light hadrons.    

Within our holographic model, we are uniquely able to explore both the equilibrium observables (such as susceptibilities and other thermodynamic quantities) as well as the out-of-equilibrium observables i.e. the transport coefficients. Bayesian methods \cite{Pratt:2015zsa} have been employed in recent years to more thoroughly understand the temperature dependence and now finite baryon chemical potential dependence \cite{Auvinen:2016tgi} of shear and bulk viscosity.  However, there is no guidance from ab initio calculations to the placement of the transition temperatures of transport coefficients and how they should correlate with their relevant equilibrium observables. For instance, previous work has suggested that the temperature dependence of the bulk viscosity should be connected with the derivative of the trace anomaly \cite{Karsch:2007jc}.  For instance, here we are uniquely in the position to test exactly how closely connected the bulk viscosity is with the trace anomaly in a realistic manner and can also see if the BSQ conductivity transport coefficients are closely correlated with their corresponding BSQ susceptibilities determined in equilibrium.  

In the following, the type of transition temperature used depends on the behavior of the quantity of interest.  For instance, the position of the minimum of $\eta/s(T)$ (discussed in Appendix \ref{sec:shear}) is chosen as its transition temperature whereas $\zeta/s(T)$ has a clear peak that is chosen as its transition temperature.  Other quantities such as the susceptibilities and conductivities do not have a peak or minimum but do have a clear inflection point, which is then chosen as its corresponding transition temperature.  The one quantity where two separate points are taken is the trace anomaly that has a peak at a high temperature (larger than all other equilibrium transition temperatures) but also has an inflection point at a temperature in the midst of all other transition temperatures.

\begin{figure}[htp!]
\center
\subfigure[]{\includegraphics[width=0.8\linewidth]{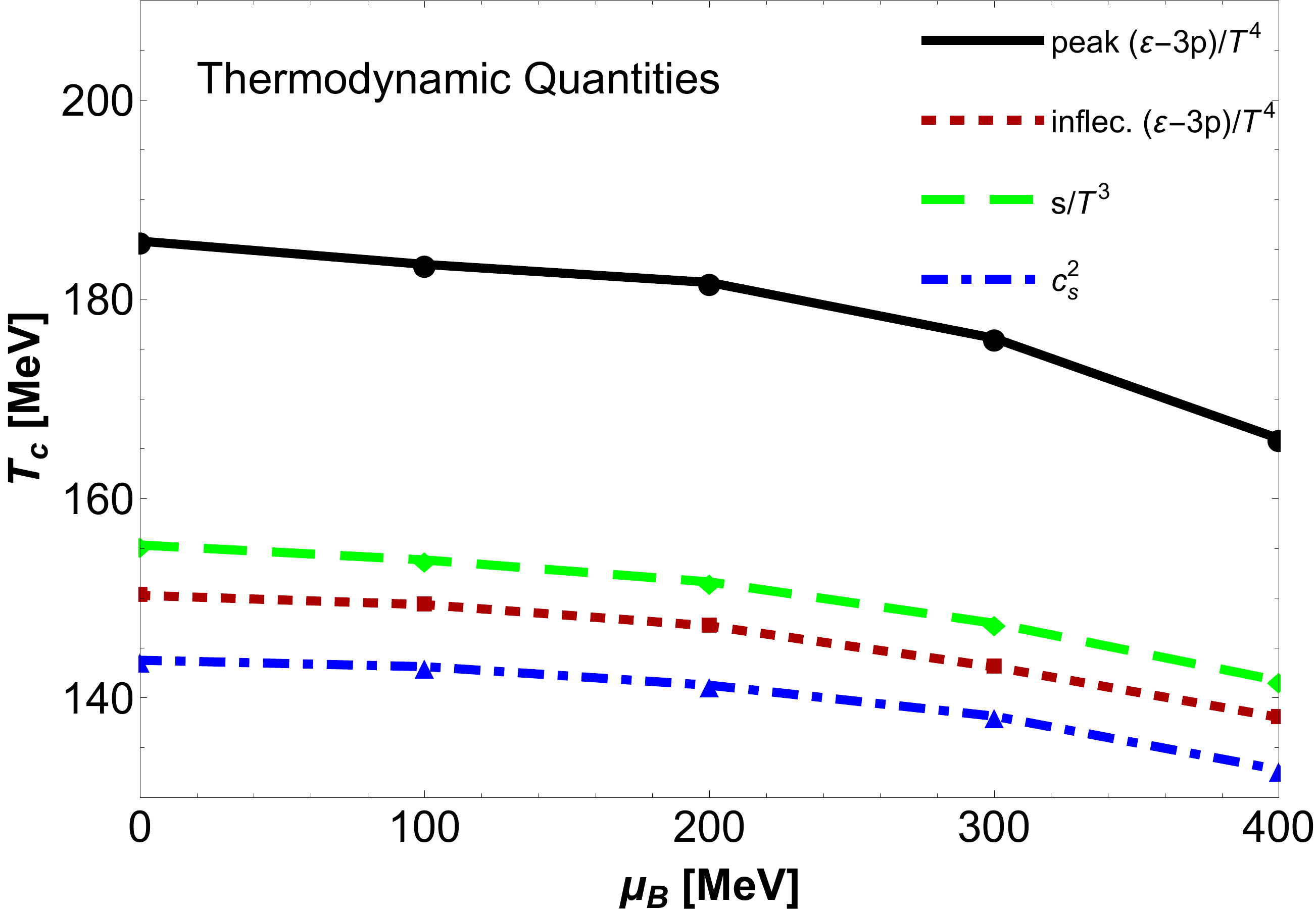}}
\qquad
\subfigure[]{\includegraphics[width=0.8\linewidth]{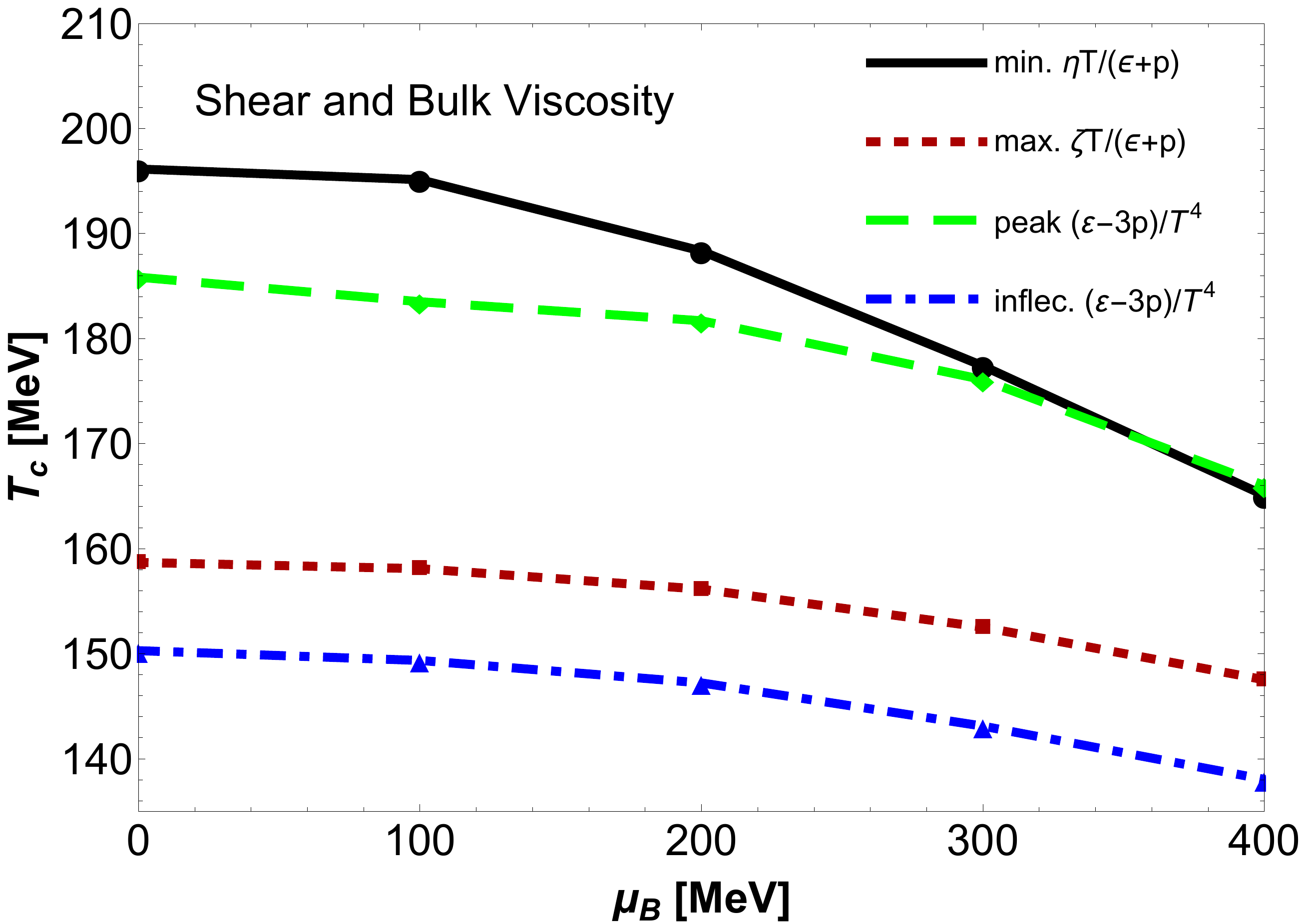}}
\qquad
\subfigure[]{\includegraphics[width=0.8\linewidth]{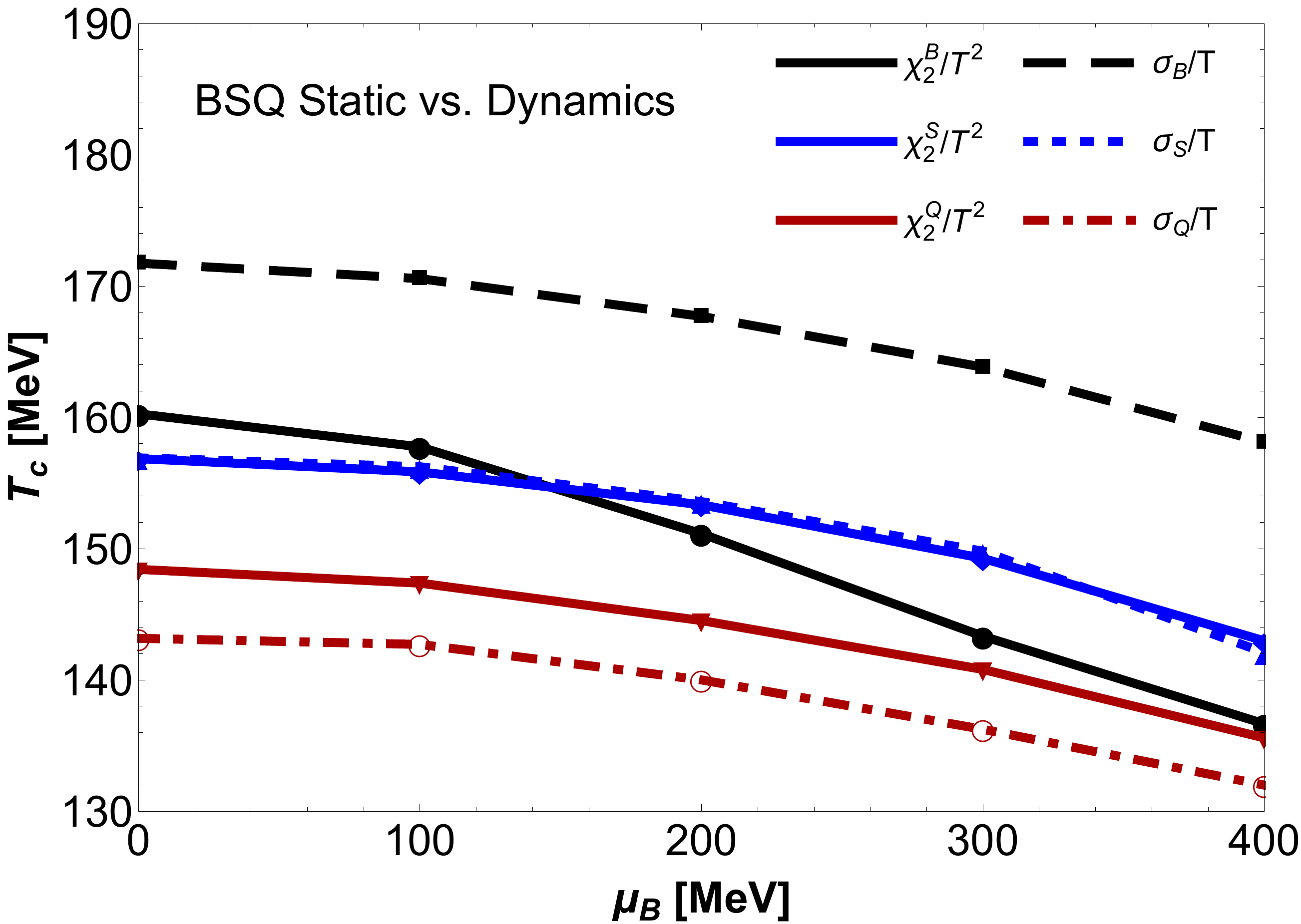}}
\caption{(Color online) (a) Transition temperatures associated with thermodynamic quantities as a function of $\mu_B$. (b) Transition temperatures associated with hydrodynamic viscosities as a function of $\mu_B$. (c) Comparison between the transition temperatures associated with conserved charges extracted from second order susceptibilities (equilibrium) and their respective conductivities (transport) as a function of $\mu_B$.}
\label{fig:pseudotemp}
\end{figure}

In Fig.\ \ref{fig:pseudotemp} we display the holographic EMD result for the crossover lines of several equilibrium and transport quantities, constructed using the values of their characteristic points (inflection points and extrema), as a function of $\mu_B$. In Fig.\ \ref{fig:pseudotemp} (a) one can see that the crossover lines constructed following the inflection point of $s/T^3$ and the dip in $c_s^2$ as a function of $\mu_B$ are much more similar to the curve traced out by the inflection point of the trace anomaly than by the curve associated with the peak of the same quantity. In Fig.\  \ref{fig:pseudotemp} (b) we show the crossover lines across $\mu_B$ associated with the minimum of the phenomenologically constructed shear viscosity and the peak of the bulk viscosity. 

One can see that the crossover line defined by the peak of the bulk viscosity is much more similar to the curve associated with the inflection point of the trace anomaly than the curve traced out by its maximum. However, it would be misleading to associate the peak of $\zeta/s$ directly with $(\varepsilon-3p)/T^4$ since the peak of $\zeta/s$ remains roughly $10$ MeV higher than the inflection point of $(\varepsilon-3p)/T^4$ while the peak of $\zeta/s$ is also roughly 20 MeV lower than the peak of $(\varepsilon-3p)/T^4$. These results strongly highlight the importance of overcoming Lattice QCD issues to directly calculate transport coefficients since some assumptions about the temperature dependence of the transport coefficients may prove to be wrong. Furthermore, they clearly demonstrate that Bayesian analysis techniques to extract transport coefficients should not assume one single temperature for the inflection point or extrema of transport coefficients but rather it appears that the bulk viscosity may be constrained within the (rather large) bounds given by the trace anomaly.  

On the other hand, it is interesting to note that the crossover line given by the minimum of $\eta T/(\epsilon+p)$ across $\mu_B$, extracted here from the jet quenching parameter, as explained in Appendix \ref{sec:shear}, involves much larger temperatures starting around 200 MeV at zero baryon density with a steep fall towards lower temperatures with increasing $\mu_B$. If one takes these estimates for the hydrodynamic viscosity crossover lines and applies them in hydrodynamic calculations, this would correspond to simulations in which the minimum of the shear viscosity at $\mu_B=0$ occurs far from the typical values of the switching temperature $T_{sw} \sim 150$ MeV \cite{Bernhard:2016tnd}. This points to an interesting interplay between shear and bulk viscosities during the hydrodynamic evolution of such a plasma: as the system cools down from the high temperatures achieved in the initial state towards $T\sim $ 200 MeV, the bulk viscosity remains small and $\eta T/(\epsilon+p) < 1/4\pi$ and, thus, the plasma in this regime displays nearly perfect fluid response to spatial inhomogeneities. Below $T \sim 160$ MeV, $\eta/s$ starts to increase and the particle number changing processes that take place within hadronization contribute to generate a peak in the bulk viscosity, which however still remains smaller than the shear viscosity in the model presented here.

Clearly, further study is needed to investigate the consequences of the results discussed above. Our results for the high transition temperature of $\eta T/(\epsilon+p)$ are especially surprising, which indicates either that the correlation between $\hat{q}$ and $\eta/s(T)$ is not as strong as previously thought or that this relationship is, indeed, correct but that the transition temperature of $\eta/s(T)$ is much higher than previously thought.  While the high transition temperature of $\eta/s(T)$ may be surprising we also see a large flat region through the entirety of the crossover as shown in Fig.\ \ref{fig:shearbay}, which seems to be consistent with current Bayesian analyses. Perhaps this is not surprising in light of the success of many hydrodynamic models that use a constant $\eta/s$. The holographic picture would then  provide an $\eta/s(T)$ dependence that only sees an increase of $\eta/s$ in the purely hadronic region as well as at very high temperatures. Otherwise $\eta/s(T)$ would be more or less flat throughout most of the temperature regime that current experiments can explore.

In Fig.\ \ref{fig:pseudotemp} (c) we show the crossover lines (obtained from inflection points) associated with the second order susceptibilities of conserved baryon, electric, and strangeness charges and compare them with their corresponding conductivities. Such a comparison provides a way to gauge how transport coefficients (which involve real time non-equilibrium dynamics) feel the change of degrees of freedom that takes place in the crossover region, which has been usually studied using only quantities defined in equilibrium such as the susceptibilities of conserved charges. In our calculations there is a clear split between the crossover line associated with $\chi_2^B/T^2$ and that of $\sigma_B/T$, while for the other conserved charges equilibrium and out-of-equilibrium crossover lines are not very separated. Moreover, in the baryon sector one can see that the transport coefficient $\sigma_B$ feels the change in degrees of freedom around the transition already at higher temperatures than its equilibrium counterpart $\chi_2^B$. This suggests that as the system cools down, baryon transport switches off earlier in the evolution than one would expect using as an estimate the longer time it takes for the system to achieve the lower $\chi_2^B$ transition temperature. The same does not occur for the other conserved charges where the crossover lines obtained from equilibrium quantities provide a good estimate for the electric and strangeness conductivity transition temperatures. A detailed analysis of the interplay between the $T$ and $\mu_B$ dependence of the susceptibilities and the conductivities of the QGP's conserved charges, and their potential effects on its evolution, may be performed using hydrodynamics augmented by the inclusion of the effects of B, S, and Q conserved currents. 

We note that we expect some differences to arise when calculations are further improved beyond only finite $\mu_B$ but also  simultaneously include nonzero $\mu_S$ and $\mu_Q$. While one expects that in heavy ion collisions $\mu_Q$ remains quite small even at large $\mu_B$, $\mu_Q>0$ could induce some small corrections on $\chi_2^Q$ and $\sigma_Q$. Additionally, $\mu_S$ should not be nearly as small as $\mu_Q$ and, thus, the most logical next step would be to include the backreaction of the strangeness conserved charge in our model. However, this presents significant numerical challenges that are beyond the scope of this paper. 

Finally, we note that all the lines that define the crossover region computed here bend towards lower temperatures and that the width of this band shrinks from $T \sim 140 - 200$ MeV to $T \sim 130-170$ MeV when $\mu_B$ goes from 0 to 400 MeV. It is expected that all the curves in Fig.\ \ref{fig:pseudotemp} will converge to the same point at the CEP of the model, which can only occur at much larger values of $\mu_B$ that go beyond the range considered in this paper.

\section{Discussion}
\label{sec:conclusion}

Employing a phenomenological bottom-up holographic EMD model we train over 10000 holographic black holes to reproduce Lattice QCD results at $\mu_B=0$. We began by showing that the predictions of the holographic EMD model for the QGP equation of state are in good agreement with the latest lattice results \cite{Bazavov:2017dus} for the QCD equation of state with $(2+1)$ flavors and physical quark masses at finite temperature and baryon density. For the first time we look not only at quantities sensitive to baryon charge but also explore strangeness and electric charge conservation through the calculation of electric and strangeness charge susceptibilities, conductivities, and diffusion coefficients. As we also pointed out, the EMD results for the electric conductivity of the QGP are still within reasonable agreement with the available lattice QCD results for this quantity.

Regarding the phenomenological applicability of the EMD model predictions for the QCD phase diagram, as it is well known, any holographic calculation performed in the classical (supergravity) limit of the gauge/string duality, as is the case of the EMD model, cannot describe the weakly coupled ultraviolet regime of QCD (gauge/gravity models have strongly coupled ultraviolet fixed points and do not display asymptotic freedom). Therefore, for phenomenological applications, our calculations should be restricted to the strongly coupled infrared regime of QCD, which is actually the case realized close to the crossover transition region and the main focus of the present work.

We observed an overall suppression of the diffusion of conserved charges and hydrodynamic viscosities in the plasma with increasing baryon density. Regarding the suppression of conserved charges, such an effect is more noticeable in the baryon sector while being very small for the strangeness sector. 

The lattice QCD results presented in \cite{Borsanyi:2011sw} show that the inflection point associated with $\chi_2^S$ is at a larger temperature than the one found from $\chi_2^B$. Given that $\chi_2^B$ is dominated by the light flavors, this suggests that the strangeness sector may hadronize and reach chemical freeze-out at a higher temperatures than the light sector \cite{Bellwied:2013cta}. Our results do indicate that the strangeness susceptibility generally has an inflection point at higher temperatures than the baryon susceptibility at finite densities, see Fig.\ \ref{fig:pseudotemp} (c).  However, close to $\mu_B=0$ the relationship is flipped, which may be due to the particular choice in the parameters used in $f_S(\phi)$ for the calculation of $\chi_2^S$\,\footnote{By comparing the curves in Figs.\ \ref{fig:thermomuB0} and \ref{fig:StransportmuB0}, one can see that in our black hole model $\chi_2^S$ has a softer slope compared to the Lattice QCD results than our corresponding calculation for $\chi_2^B$.}. This provides a new constraint to be considered in future work.  Also, it will be interesting to check  if the inflection point of the strangeness conductivity remains smaller than the inflection point associated with the baryon conductivity.

Furthermore, we studied how the hydrodynamic shear and bulk viscosities in this model vary with $T$ and $\mu_B$. By mixing a quasiparticle relation between the shear viscosity and the jet quenching parameter with a holographic calculation of the latter, we estimated the temperature and baryon chemical potential dependence of the shear viscosity transport coefficient in a phenomenological manner that fits within current expectations from Bayesian analysis. Our $\eta/s(T)$ may be useful in realistic hydrodynamic simulations where $\eta/s$ is not a constant and, in fact, the profile we found for this phenomenologically extracted shear viscosity displays a minimum at the crossover transition temperature, as expected on theoretical grounds, and is indeed also close in magnitude to some profiles already used in hydrodynamic simulations. The bulk viscosity, on the other hand, displays a (small) peak at the crossover and it remains smaller than the shear viscosity. Both hydrodynamic viscosities were found to be suppressed with increasing baryon density, which give support to the picture suggested in \cite{Denicol:2013nua} that the hot and baryon rich QGP formed in low energy heavy ion collisions at RHIC may be even closer to perfect fluidity than its $\mu_B=0$ counterpart.

One clear conclusion from this work is that there is a need for the development of new experimental observables that are sensitive to the BSQ conserved currents. Unlike the shear and bulk viscous transport coefficients that affect the entire QGP fluid as a whole, the BSQ conductivities directly affect a subset of variables within the fluid description that are related to the corresponding conserved charge (and some subsets are affected by multiple conductivities).  For instance, one could compare the current of neutral pions (that should be unaffected by these conductivities) to that of $\Xi$ baryons that carry electric charge, strangeness, and baryon number and look for  some systematic differences. It is important to note here that while the shear and bulk viscosities are suppressed at larger baryon chemical potentials, the magnitude of the conductivities either remain the same or are enhanced at finite $\mu_B$, which implies that not only the pressure gradients are being smoothed out by viscosity but that there is also a significant smoothening of the gradients of the chemical potentials associated with the conserved charges at finite density. The implications of this effect to the dynamical evolution of the hot and baryon rich QGP are now being investigated using state-of-the-art simulations (e.g., \cite{Shen:2017ruz}).  

We also computed the temperature crossover band as a function of $\mu_B$ from characteristic temperature points (inflections and extrema) of the physical observables (defined either in equilibrium such as susceptibilities or near equilibrium quantities such as conductivities) calculated in the present work. The crossover region bends to lower temperatures and its width shrinks as one increases the baryon density. The expectation is that the crossover band should shrink to a point at the CEP, at which a second order phase transition takes place. Other aspects related to critical phenomena in the QCD phase transition, computed in the context of a holographic EMD model, will be investigated in a future work.

Overall, our results establish a framework for studying both equilibrium and out-of-equilibrium quantities at finite baryon densities. A better understanding of the transition temperatures can provide deeper insight into the crossover nature of the QCD phase transition and provide a basis for further phenomenological studies. Clearly, some of the new questions brought up by the current work may only be completely answered using first principles calculations, which are still beyond reach due to the Fermi-sign problem.  Until then, the type of holographic black hole engineering procedure used here to match equilibrium Lattice QCD results at $\mu_B=0$ and then make predictions for observables at finite $\mu_B$, as well as for transport coefficients, may provide a useful alternative route to unravel some of the new properties of the hot and baryon dense QGP formed in heavy ion collisions.

\acknowledgments

R.R. acknowledges financial support by Funda\c{c}\~{a}o Norte Riograndense de Pesquisa e Cultura (FUNPEC). R.C. acknowledges financial support by the S\~{a}o Paulo Research Foundation (FAPESP) under FAPESP grant number 2016/09263-2. J.N. acknowledges financial support by FAPESP and Conselho Nacional de Desenvolvimento Cient\'{i}fico e Tecnol\'{o}gico (CNPq). This  material  is  based  upon  work  supported by the National Science Foundation under grant no.   PHY-1513864  and  by  the  U.S.  Department  of  Energy, Office of Science, Office of Nuclear Physics, within the framework of the Beam Energy Scan Theory (BEST) Topical Collaboration. The authors gratefully acknowledge the use of the Maxwell Cluster and the advanced support from the Center of Advanced Computing and Data Systems at the University of Houston.

\appendix

\section{Shear viscosity}
\label{sec:shear}

One of the biggest discoveries in the field of heavy-ion collisions is the nearly perfect fluid-like nature of the QGP. Early results derived within holography for the shear viscosity of strongly coupled plasmas \cite{Kovtun:2004de} propelled the field forward such that the first relativistic viscous hydrodynamical models were developed \cite{Romatschke:2007mq}. However, to this day a clear understanding of the temperature dependence of $\eta/s(T)$ has remained elusive.

In this appendix, with the aim to obtain a temperature dependent $\eta/s(T)$ in a simple way, we estimate the $T$ and $\mu_B$ dependence of the shear viscosity for the hot and baryon dense QGP by using a phenomenological hybrid approach that mixes a quasiparticle relation between the shear viscosity and the jet quenching parameter \cite{Majumder:2007zh} (see also \cite{Li:2014hja}) with EMD holography.

\begin{figure}[htp!]
\center
\subfigure[]{\includegraphics[width=0.9\linewidth]{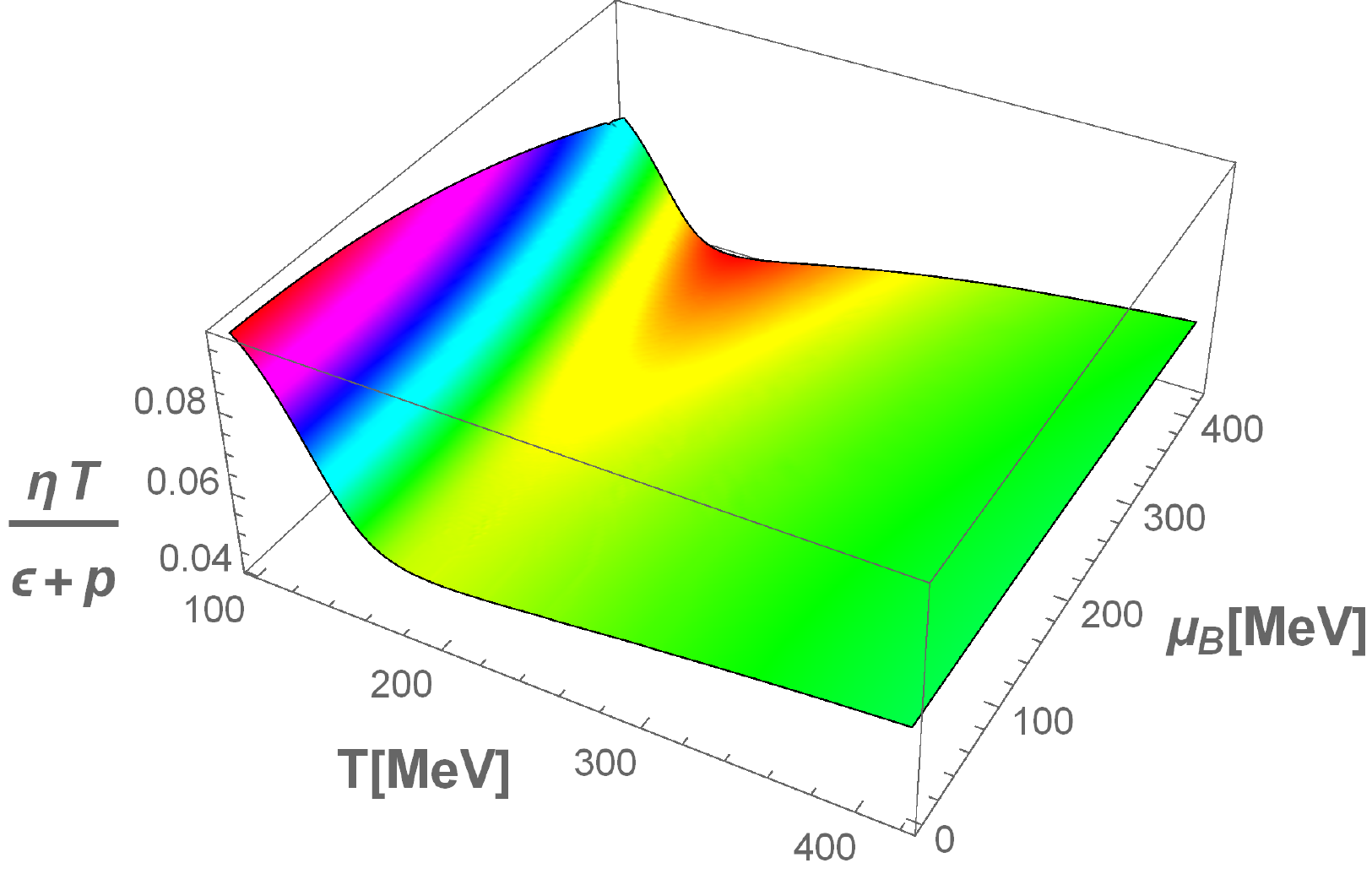}}
\qquad
\subfigure[]{\includegraphics[width=0.8\linewidth]{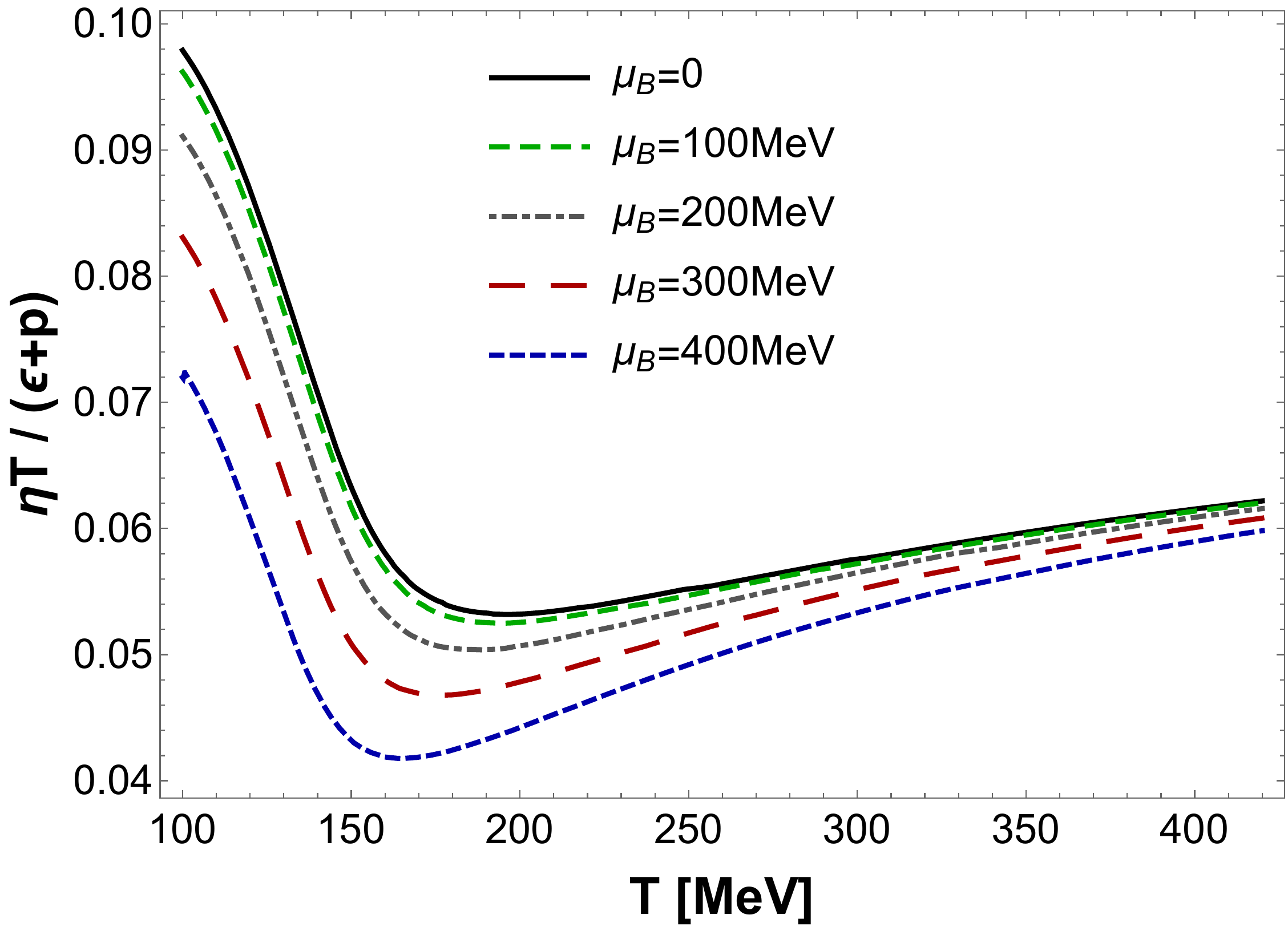}}
\caption{(Color online) ``Phenomenological" holographic shear viscosity ratio. (a) Surface plot as a function of $T$ and $\mu_B$. (b) Curves as functions of $T$ for different values of $\mu_B$.}
\label{fig:shear}
\end{figure}

Before doing that, let us remind the reader of the well-known fact that in isotropic, rotationally and translationally invariant holographic backgrounds with at most two derivatives in the gravity action the holographic shear viscosity ratio is given by $\eta/s=1/4\pi$ \cite{Policastro:2001yc,Buchel:2003tz,Kovtun:2004de}, which is known as the KSS value (higher order derivatives in the gravity action change this result, as shown in \cite{Kats:2007mq,Brigante:2007nu}). This small value (when compared to perturbative estimates \cite{Arnold:2000dr,Arnold:2003zc}) is remarkably close to the value used in recent hydrodynamic simulations of the spacetime evolution of the QGP that simultaneously match different experimental data from heavy ion collisions, $\eta/s \approx 0.095$ \cite{Ryu:2015vwa}. On the other hand, it is also clear that the shear viscosity in QCD cannot be a constant since the small value favored by hydrodynamic simulations in the crossover region must increase at higher temperatures, eventually converging to the perturbative QCD results \cite{Arnold:2000dr,Arnold:2003zc}. Moreover, at lower temperatures $\eta/s$  is expected to also increase and reach hadron resonance gas (HRG) model values \cite{NoronhaHostler:2008ju,NoronhaHostler:2012ug,Demir:2008tr,Denicol:2013nua}. Therefore, a realistic temperature dependent profile for $\eta/s$ in QCD should display a dip around the crossover transition temperature, $T_c$ \cite{Hirano:2005wx,Csernai:2006zz,NoronhaHostler:2008ju}.

In Ref.\ \cite{Cremonini:2012ny} it was shown that a nontrivial $T$ dependence for $\eta/s$ in the gauge-gravity correspondence may be obtained by considering higher order curvature corrections in a holographic setting with a dilaton profile breaking conformal symmetry in the infrared. While the dilaton field in the EMD model breaks conformal invariance in the infrared, we have not considered higher order curvature corrections in the present work, a task we postpone for the future due to the fact that it is still not clear how one should phenomenologically fix the new free function which would enter the EMD model once higher order curvature corrections are taken into account.

The route we are going to follow here to estimate a non-constant profile for the shear viscosity to entropy density ratio is considerably simpler and makes use of the quasiparticle relation discussed in Ref.\ \cite{Majumder:2007zh}, involving $\eta/s$ and the jet quenching parameter $\hat{q}$ \cite{Liu:2006ug}, on top of the holographic EMD black hole backgrounds discussed before. This quasiparticle relation may be written as follows \cite{Majumder:2007zh},
\begin{align}
\frac{\eta}{s}\sim\frac{T^3}{\hat{q}}.
\label{eq:quasiparticle}
\end{align}

The jet quenching parameter for the present EMD model at finite baryon density was computed already in Ref.\ \cite{Rougemont:2015wca} by evaluating the following integral on top of the numerical black hole backgrounds in Eq.\ \eqref{eq:EMDansatz},
\begin{align}
\frac{\hat{q}}{\sqrt{\lambda_t}\,T^3} = \frac{64\pi^2 h_0^{\textrm{far}}}{\int_{r_H}^\infty dr \,\frac{e^{-\sqrt{2/3}\phi(r)-3A(r)}}{\sqrt{h(r)\left[h_0^{\textrm{far}}-h(r)\right]}}},
\label{eq:qhat}
\end{align}
where $\lambda_t$ is 't Hooft coupling.

By using the suggested relation given in Eq.\ \eqref{eq:quasiparticle}, the following thermodynamic identity,
\begin{align}
\epsilon+p=Ts+\mu_B\rho_B,
\label{eq:thermo}
\end{align}
and by imposing that $\eta/s(T)$ should flow to the KSS value ($1/4\pi$) in the conformal limit of very high temperatures,\footnote{This is so because for the black hole backgrounds considered here the dilaton vanishes at the boundary. Note also that the gamma functions in Eq. \eqref{eq:shear} come from the conformal limit of the holographic jet quenching parameter \cite{Rougemont:2015wca,Liu:2006ug}.} we consider here the following hybrid ``phenomenological holographic'' formula for the shear viscosity at finite temperature and baryon density in the strongly coupled regime,
\begin{align}
\frac{\eta T}{\epsilon+p}= \frac{\pi^{1/2}\Gamma[3/4]}{4\Gamma[5/4]}\,\frac{\sqrt{\lambda_t}\,T^3/\hat{q}}{1+\frac{\mu_B\rho_B}{Ts}},
\label{eq:shear}
\end{align}
where the dimensionless combination $\sqrt{\lambda_t}\,T^3/\hat{q}$ is to be evaluated using Eq.\ \eqref{eq:qhat}. We remark that the specific hybrid formula \eqref{eq:shear} we proposed above lies within the broader conjecture done in \cite{Liu:2006ug} for the strongly coupled regime, since the 't Hooft coupling $\lambda_t$ in holographic calculations must be large.

Note that at finite baryon density the relevant observable entering in hydrodynamics is $\eta T/(\epsilon + p)$, which reduces to the ratio $\eta/s$ at $\mu_B=0$ \cite{Liao:2009gb,Denicol:2013nua}. The corresponding results for this ``phenomenological"\, holographic estimate of the shear viscosity of the QGP are shown in Fig.\ \ref{fig:shear}. It is interesting to note that the temperature profile estimated for the shear viscosity through this hybrid quasiparticle-holographic approach is actually similar to some profiles used in current hydrodynamic simulations (see, for instance, the orange curve in Fig.\ 1 of Ref.\ \cite{Denicol:2015nhu}). We also note that the ratio involving the shear viscosity is reduced with increasing baryon density, which indicates that the QGP becomes closer to its perfect fluid limit in the baryon dense regime (a similar conclusion was reached within a kinetic approach to a hadronic gas in \cite{Denicol:2013nua}). Additionally, this has an interesting consequence for the $v_2$ to $v_3$ relationship at the Beam Energy Scan and it favors a shorter lifetime of the QGP as argued in \cite{Auvinen:2013sba}.

\begin{figure}[htp!]
\center
\includegraphics[width=0.8\linewidth]{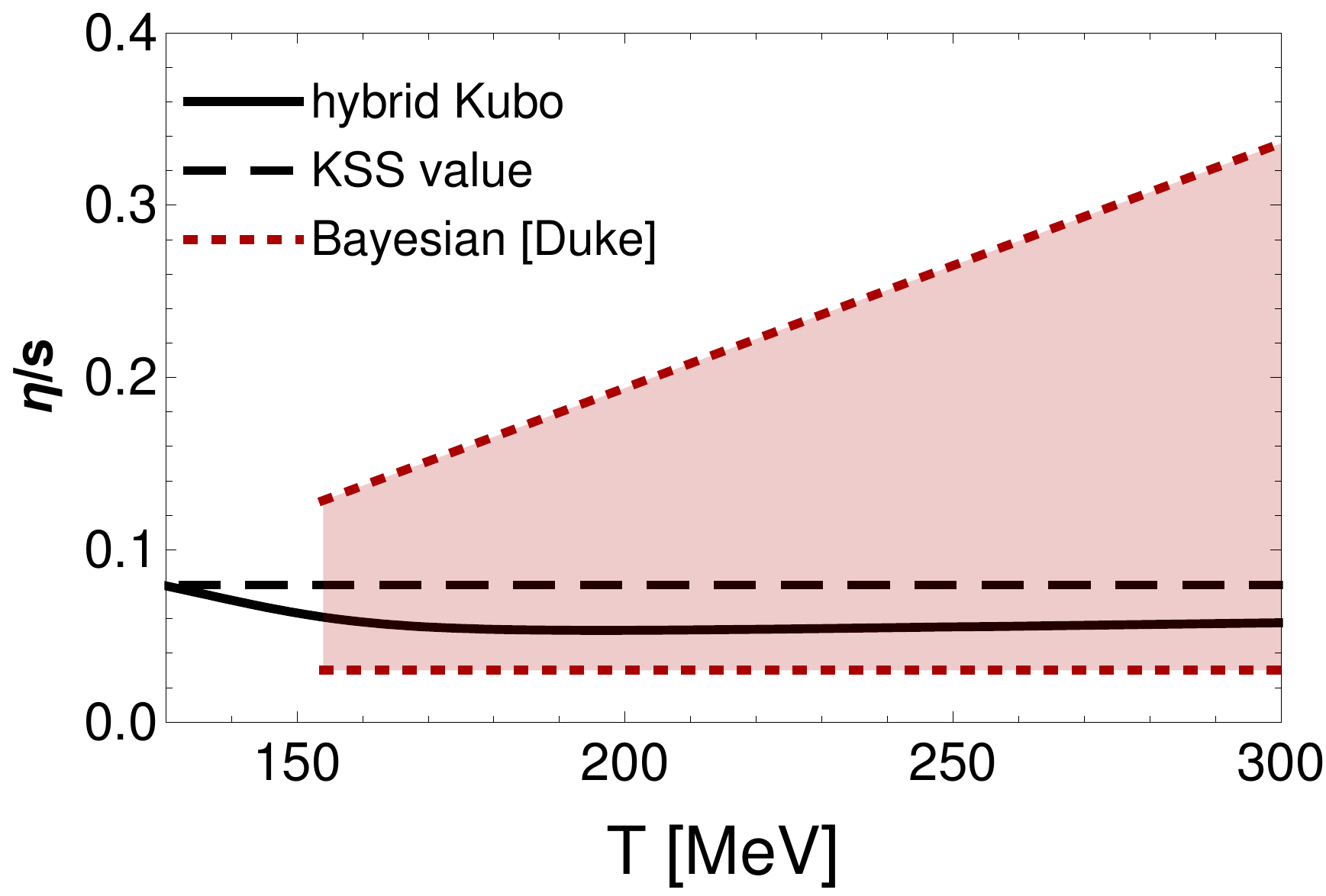}
\caption{(Color online) Estimate for $\eta/s(T)$ at $\mu_B=0$ (solid line) compared to the result of a Bayesian analysis \cite{Bernhard:2016tnd} involving comparisons of relativistic hydrodynamics calculations to experimental data.}
\label{fig:shearbay}
\end{figure}

A comparison of our results for $\eta/s(T)$ at $\mu_B=0$ to the extracted $\eta/s(T)$ parameters determined using a Bayesian analysis \cite{Bernhard:2016tnd} is shown in Fig.\ \ref{fig:shearbay}. One can see that  our results appear to be within their uncertainty band. Our $\eta/s(T)$ exhibits a relatively flat upward slope in the QGP phase but has a steeper slope in the hadron gas phase. Most relativistic hydrodynamic work demonstrates that the hadron gas region and the crossover region can be reasonably well-constrained by theory vs. experimental data comparisons. However, the high temperature region is exceedingly difficult to constrain in the same manner with current experimental data \cite{Niemi:2011ix,Niemi:2015qia}. 

Finally, we must remark that the dip obtained for $\eta/s(T)$ in our hybrid approach is not artificial, since the EMD model actually predicts that the dimensionless jet quenching parameter $\hat{q}/T^3$ displays a peak near the crossover. Rather, the minimum for this phenomenological calculation of $\eta/s(T)$ follows when one assumes that the quasiparticle relation \cite{Majumder:2007zh} between $\eta/s(T)$ and the jet quenching parameter can be extrapolated to the strong coupled regime. Even though it is not known at the moment how much such an extrapolation is justified, given our ignorance with respect to the actual values of transport coefficients in QCD, this kind of weak to strong coupling extrapolation is commonly employed in the field. Moreover, it may very well be that the simple hybrid formula proposed above in Eq. \eqref{eq:shear} does give a very similar profile for $\eta/s(T)$ as the one which we would obtain by considering higher order curvature corrections in the EMD setup, once the new free function which would appear in the model in this case is adequately fixed by some phenomenological input. There are hints that this may be in fact the case, since qualitatively similar profiles for $\eta/s(T)$ (displaying a dip and approaching the KSS result from below at very high $T$) have been already considered in \cite{Cremonini:2012ny} by taking into account higher order curvature corrections in an Einstein-dilaton model at zero baryon density.\footnote{In \cite{Cremonini:2012ny}, however, they considered a holographic model for a pure gluon plasma instead of a model for the QGP. More important, the extra free function comprised in the model due to the higher order curvature corrections was not fixed by any phenomenological inputs from QCD, instead, they considered many different trial profiles for this function and studied how they change the corresponding profile for the temperature dependent shear viscosity.} The explicit check of this claim, however, is left for a future work since, as mentioned before, at present we have no clear guide on how to fix this extra free function in a phenomenologically adequate way.


\end{document}